\documentclass[a4paper,11pt]{article}

\usepackage{jcappub}

\usepackage[T1]{fontenc}
\usepackage[utf8]{inputenc}

\usepackage{bm}
\usepackage{braket}

\usepackage{tensor}
\usepackage{amssymb,amsfonts,amsmath,amsthm}
\usepackage{mathtools}
\usepackage{relsize}
\usepackage{listings}

\renewcommand{\leq}{\leqslant}

\newcommand{\FouInd}[1]{\textbf{\textsf{#1}}} 
\newcommand{\FlipInd}[1]{\={#1}}

\newcommand{\semibold}[1]{{\fontseries{b}\selectfont{#1}}}
\newcommand{\para}[1]{\par\vspace{2mm}\noindent\semibold{{#1.}---}\ignorespaces} 

\newcommand{\packagefont}{\sffamily\fontseries{sbc}\selectfont}
\newcommand{\pytransport}{{\packagefont PyTransport}}
\newcommand{\cpptransport}{{\packagefont CppTransport}}
\newcommand{\mtransport}{\packagefont mTransport}

\newcommand{\fieldinf}{{\packagefont FieldInf}}
\newcommand{\modecode}{{\packagefont ModeCode}}
\newcommand{\multimodecode}{{\packagefont MultiModeCode}}
\newcommand{\pyflation}{{\packagefont PyFlation}}
\newcommand{\bingo}{{\packagefont BINGO}}
\newcommand{\sql}{{\packagefont SQL}}

\newcommand{\MPlanck}{M_{\mathrm{P}}}
\newcommand{\Hint}{H_{\text{int}}}
\newcommand{\vac}{\text{vac}}

\newcommand{\fNL}{f_{\text{NL}}}

\newcommand{\Nfields}{N}
\newcommand{\tinit}{t_{\text{init}}}

\newcommand{\Aslow}{A_{\text{slow}}}
\newcommand{\Afast}{A_{\text{fast}}}

\newcommand{\mgel}{m_{\text{gelaton}}}

\DeclareMathOperator{\TimeOrder}{T}
\DeclareMathOperator{\AntiTimeOrder}{\bar{T}}

\newcommand{\vect}[1]{\bm{\mathrm{{#1}}}}
\newcommand{\vectktot}{\vect{k}_{\text{tot}}}

\newcommand{\im}{\mathrm{i}}
\renewcommand{\d}{\mathrm{d}}
\newcommand{\e}[1]{\mathrm{e}^{{#1}}}

\newcommand\CC{C\nolinebreak\hspace{-.05em}\raisebox{.4ex}{\relsize{-3}{\textbf{+}}}\nolinebreak\hspace{-.10em}\raisebox{.4ex}{\relsize{-3}{\textbf{+}}}}

\makeatletter 
\renewcommand*\env@matrix[1][\arraystretch]{%
  \edef\arraystretch{#1}%
  \hskip -\arraycolsep
  \let\@ifnextchar\new@ifnextchar
  \array{*\c@MaxMatrixCols c}}
\makeatother

\title{Numerical evaluation of inflationary 3-point functions on curved field space\\
{\Large \it -- with the transport method \& CppTransport}
\vspace{5mm} \hrule
}

\author{Sean Butchers}
\author{and David Seery}
\affiliation{\vspace{2mm}\small
Astronomy Centre, Pevensey II Building, University of Sussex, Falmer, Brighton, BN1 9QH, UK \\
}

\emailAdd{smlb20@sussex.ac.uk}
\emailAdd{D.Seery@sussex.ac.uk}

\abstract{We extend the public
{\cpptransport} code to calculate the statistical properties of fluctuations
in multiple-field inflationary models with curved field space.
Our implementation accounts for all physical effects at tree-level in the `in--in' diagrammatic expansion.
This includes particle production due to time-varying masses, but excludes scenarios where the
curvature perturbation is generated by averaging over the decay of more than one particle.
We test our implementation by comparing results in Cartesian and polar field-space coordinates, showing excellent numerical
agreement and only minor degradation in compute time.
We compare our results with the {\pytransport} 2.0 code, which uses the same computational approach
but a different numerical implementation,
finding good agreement.
Finally, we use our tools to study a class of gelaton-like models which could produce an enhanced non-Gaussian signal
on equilateral configurations of the Fourier bispectrum.
We show this is difficult to achieve using hyperbolic field-space manifolds
and simple inflationary potentials.}

\begin{document}

\maketitle
\flushbottom

\newpage
\section{Introduction}
\label{sec:intro}
Inflation \cite{Guth:1980zm, Linde:1981mu, Albrecht:1982wi} has become established as a preferred framework
in which to describe the early universe.
In inflation, primordial quantum fluctuations are amplified, giving large-scale variations in energy density that are
inherited by later structure.
Recent ideas from theories of beyond-the-Standard-Model physics have
introduced multiple-field models yielding
2-point statistics consistent with measurement,
but which may be theoretically
preferable because their field values remain sub-Planckian.
In such theories the kinetic term
$X = -G_{IJ} \partial^a \phi^I \partial_a \phi^J / 2$
is often non-canonical and is expressed in terms of a
kinetic matrix $G_{IJ}(\bm{\phi})$.
(We define our notation more carefully below. Here, upper-case Latin
indices label the different
species of scalar fields, and lower-case indices label spacetime dimensions.)
The matrix $G_{IJ}$ is real, symmetric, and transforms as a covariant 2-tensor under
field redefinitions, so it may be interpreted as a metric.
The resulting `covariant' formalism constrains the ways in which $G_{IJ}$ can appear
in observable quantities and offers a convenient computational framework with the usual advantages
of tensor calculus.

Examples in this class include models descending from string theory or supergravity
where the kinetic matrix is inherited from a K\"{a}hler potential
$K(\phi^I, \phi^{I*})$~\cite{Lyth:1998xn,Baumann:2014nda}.
The $\alpha$-attractor scenario suggested by
Kallosh \& Linde is of this type~\cite{Kallosh:2013hoa,Ferrara:2013rsa,Kallosh:2013yoa},
including its multiple-field variants~\cite{Achucarro:2017ing}.
Also, a full description of the interesting Higgs inflation model,
including Goldstone modes, requires a noncanonical metric
that derives from the Goldstone sigma model~\cite{Greenwood:2012aj}.
Alternatively, the freedom to choose non-Cartesian coordinates on field
space may simply provide a more convenient option,
as with the `gelaton' and `quasi-single field inflation'
scenarios~\cite{Tolley:2009fg,Chen:2009we}.

\para{Numerical tools}
Whatever the origin of the noncanonical kinetic structure,
to constrain such models using modern datasets we require precise numerical predictions.
Numerical tools for performing inflationary calculations have existed for some time,
but their capabilities have been limited.
Ringeval et al. provided the early code
\href{http://cp3.irmp.ucl.ac.be/~ringeval/fieldinf.html}{{\fieldinf}},
which is capable of computing 2-point functions with an arbitrary choice of
metric $G_{IJ}$
\cite{Ringeval:2005yn, Martin:2006rs, Ringeval:2007am},
but many other tools restrict to the canonical case
$G_{IJ} = \delta_{IJ}$.
Major examples include
\href{http://modecode.org}{\modecode}/\href{http://modecode.org}{\multimodecode}~\cite{Mortonson:2010er,
Easther:2011yq,Norena:2012rs,Price:2014xpa},
\href{http://pyflation.ianhuston.net}{\pyflation}~\cite{Huston:2009ac,Huston:2011vt,Huston:2011fr}
and \href{https://sites.google.com/site/codecosmo/bingo}{\bingo}~\cite{Hazra:2012yn,Sreenath:2014nca}.
{\modecode}/{\multimodecode} and {\pyflation} are 2-point function solvers for canonical multiple-field
models, and {\bingo} is a 2- and 3-point function solver for single-field models.

All these are traditional codes that require customization by the user for each model of interest.
Recent developments in inflationary perturbation
theory~\cite{Mulryne:2009kh,Mulryne:2010rp,Seery:2012vj,Mulryne:2013uka}
have allowed the construction of
\emph{automated} tools~\cite{Dias:2015rca,Dias:2016rjq,Ronayne:2017qzn}.
These accept the specification of an inflationary model by its Lagrangian
and leverage symbolic algebra methods to produce custom code that
solves for the inflationary $n$-point functions.
We collectively refer to these as the \emph{transport tools}
(\href{http://transportmethod.com}{transportmethod.com}).
The suite contains three tools, all of which apply to multiple-field
models:
\begin{itemize}
	\item {\mtransport}~\cite{Dias:2015rca}
	is a 2-point function solver implemented in \emph{Mathematica}.
	It allows a nontrivial kinetic matrix and is suited to interactive model exploration.

	\item {\pytransport}~\cite{Mulryne:2016mzv,Ronayne:2017qzn}
	is a 2- and 3-point function solver implemented in Python.
	Version 1 (September 2016) restricted to canonical kinetic terms.
	Version 2 (September 2017) introduced support for an arbitrary kinetic matrix.
	Because it is implemented as a Python library it is well-suited to scripting or
	inclusion in other codes.

	\item {\cpptransport}~\cite{Seery:2016lko}
	is a 2- and 3-point function solver implemented in {\CC}.
	It has built-in functionality to parallelize computations and
	can postprocess correlation functions to produce inflationary observables.
	It manages storage of its data products as {\sql} databases.
	It is well-suited to larger calculations that benefit from its
	auto-parallelization or which produce significant data volumes,
	and performs well with `feature'
	models containing steps or kinks where its library of sophisticated steppers
	offers assistance.
	It is less easy (but still possible) to incorporate within
	larger codes than {\pytransport}.
	The original release 2016.3 restricted to canonical kinetic terms.
\end{itemize}

In this paper we describe a new release
of {\cpptransport}
(\href{https://zenodo.org/record/1183518}{2018.1})
that extends its functionality to nontrivial kinetic matrices.
We apply these new tools to a class of gelaton-like scenarios and show
that (at least in the scenarios we study) the parameter space available to
generate enhanced equilateral correlations is very small.
We compare our numerical results with the
independent {\mtransport} and {\pytransport} implementations, finding excellent agreement.%
    \footnote{Although the transport tools all use the same computational
    framework, their numerical implementations vary considerably in detail
    and therefore this constitutes a nontrivial check on numerical correctness.}

\para{Synopsis}
The necessary equations for computation of the inflationary two-point function
were given by Mulryne~\cite{Mulryne:2013uka}
and extended to a non-Euclidean field space by Dias et al.~\cite{Dias:2015rca}.
We have nothing novel to say about this part of the analysis.
The extension to three-point correlations was given in Ref.~\cite{Dias:2016rjq},
but this was limited to models with canonical kinetic terms.

This paper is divided into three principal parts.
First,
in \S\ref{sec:DifferenceCanonicalCase}, we highlight the key modifications
required to adapt
the analysis of Ref.~\cite{Dias:2016rjq}
for a nontrivial field-space metric.
A similar discussion has already been given by Ronayne et al.~\cite{Ronayne:2017qzn}.
We briefly review the field-space covariant formulation of inflationary perturbations
in~\S\ref{subsec:FieldCovarFormalism},
and use this to derive the covariant cubic Hamiltonian in~\S\ref{subsec:Hamiltonian}.
In~\S\ref{subsec:NewInitialConds} we discuss the computation of initial
conditions for each correlation function.
We formulate the covariant transport hierarchy in~\S\ref{subsec:CovarTransEqs}
and explain how to relate covariant correlation functions to the
curvature perturbation in~\S\ref{subsec:GaugeTrans_Short}.

Second, in~\S\ref{sec:NumRes} we present a selection of numerical results.
For those wishing to replicate our numerics,
we explain how to obtain {\cpptransport}
in~\S\ref{subsec:ObtainUsingCodes}.
In~\S\S\ref{subsec:CanoCompQsfiGel}--\ref{subsec:Quas2FldInflat}
we validate our numerical implementation by comparing results computed using
polar field-space coordinates with known results in Cartesian coordinates.
In~\S\ref{subsec:Gelaton} we apply our method to the `gelaton' model proposed by
Tolley \& Wyman~\cite{Tolley:2009fg}.
In this scenario a light degree of freedom is `dressed' by the interactions of a noncanonical
heavy mode, obtaining a subluminal phase velocity
and potentially enhanced correlations on equilateral Fourier configurations.
Our numerical tools successfully reproduce the features of the scenario,
but we show that (at least
for the range of potentials we consider) it is difficult to find suitable parameters
that allow both sufficient inflation \emph{and} large enhancement of the equilateral modes.
We conclude in~\S\ref{sec:Concs}.

Third, we include a large amount of supplementary information
in Appendix~\ref{app:DetailedCalcs}.
This includes more detailed computations
of the transport hierarchy
given in~\S\ref{sec:DifferenceCanonicalCase},
together with a selection of intermediate
results not discussed in the main text.

\para{Obtaining {\cpptransport}}
The latest builds of {\cpptransport} and {\pytransport} are available from the
website \href{https://transportmethod.com}{transportmethod.com}.
Alternatively, both {\cpptransport}
and {\pytransport} are permanently deposited at
\href{http://zenodo.org}{zenodo.org};
at the time of writing, the current releases are
\href{https://zenodo.org/record/1183518}{2018.1 for {\cpptransport}}
and
\href{https://zenodo.org/record/848220}{2.0 for {\pytransport}}.

\para{Notation}
We use natural units where $c = \hbar = 1$.
The reduced Planck mass is $\MPlanck^2 = (8 \pi G)^{-1}$.
We use the metric signature $(-,+,+,+)$. Greek indices $(\mu, \nu, ...)$
label space-time indices, whereas lower-case Roman indices
from the middle of the alphabet, $(i, j, ...)$, label spatial indices.
Upper-case Roman indices $(I, J, ...)$ label field-space coordinates.
We employ a compressed Fourier notation defined in Eq.~\eqref{eq:FourierConvention1_paper}
in which these labels appear in a bold, sans-serif font: $(\FouInd{I}, \FouInd{J}, ...)$.
For phase-space coordinates, we use Roman letters from the start of the alphabet, $(a, b, ...)$.

\section{Differences from the canonical case}
\label{sec:DifferenceCanonicalCase}

To accommodate a non-Euclidean field-space metric we require a covariantization
(with respect to the metric $G_{IJ}$)
of the formalism developed in Ref.~\cite{Dias:2016rjq} for the Euclidean case.
The advantage of a covariant formalism is that it naturally packages additional terms
arising
from the metric as Christoffel and Riemann contributions in the same way as spacetime
covariance.
Its construction entails the replacement of ordinary derivatives by covariant derivatives
and contraction of all indices with $G_{IJ}$.
However, detailed computations show that
Riemann terms also appear, meaning that the resulting formalism is not
`minimally coupled' to the field-space curvature.
The details of this covariantization were given in Gong \& Tanaka~\cite{Gong:2011uw}
and Elliston et al.~\cite{Elliston:2012ab}.

Dias et al.~\cite{Dias:2015rca} applied these ideas to find a covariant formulation
of the transport equations for the two-point function.
In this section we briefly review this construction and extend it to the
three-point function.
A more detailed discussion is given in Appendices~\ref{subsec:PertAction}--\ref{subsec:GaugeTransCurvs}.

\subsection{Field-covariant formalism}
\label{subsec:FieldCovarFormalism}

\para{Perturbation series}
In a covariant formalism our aim is to construct correlation functions that
transform tensorially under field redefinitions. These are coordinate transformations
in field space.
Correlation functions of the field perturbations
$\delta\phi^I \equiv \phi^I(\vect{x},t) - \phi^I(t)$ do
\emph{not} have this property, because the
coordinates $\phi^I$ do not themselves transform tensorially
(despite the species label `$I$').

A suitable alternative was given by Gong \& Tanaka~\cite{Gong:2011uw}, who observed
that in a normal neighbourhood of $\phi^I(t)$ we can associate $\phi^I(\vect{x}, t)$ with
the geodesic that connects it to $\phi^I(t)$. The geodesic is uniquely determined by its
tangent vector $Q^I$ at $\phi^I(t)$. By construction $Q^I$ is field-space covariant
and is defined in the unperturbed spacetime. It is therefore a candidate
to appear in correlation functions of the form
$\langle Q^I \rangle$,
$\langle Q^I Q^J \rangle$,
\ldots,
$\langle Q^I Q^J \cdots Q^K \rangle$,
each of which will inherit a tensorial transformation law from $Q^I$.
See Refs.~\cite{Gong:2011uw,Elliston:2012ab,Dias:2015rca}
for further details.

\para{Correlation functions}
After quantization,
our intention is to compute 2- and 3-point correlation functions of the
Heisenberg-picture fields $Q^I$ together
with their canonical momenta $P^J \equiv D_t Q^J$,
where $D_t \equiv \dot{\phi}^I \nabla_I$ is the covariant time derivative
and $\dot{\phi^I} = \d \phi^I / \d t$.
As usual, in order to use time-dependent perturbation theory, we split the
Hamiltonian into free and interacting parts corresponding to the quadratic and
cubic (or higher) terms~\cite{Dias:2016rjq}.
Notice that, with this definition, all mass terms are included in the free
Hamiltonian.
Finally, we define interaction-picture fields $q^I$ and $p^J$ that are related
to the Heisenberg-picture fields by a similarity transformation
$q^I = F^\dag Q^I F$,
$p^J = F^\dag P^J F$,
where $F$ is the unitary operator
\begin{equation} \label{eq:Fsolution_paper}
	F = \AntiTimeOrder \exp \left( \im \int^t_{-\infty^+} \Hint(t') \; \d t' \right),
\end{equation}
and $\AntiTimeOrder$ is the anti-time ordering operator that arranges the
fields in its argument
in order of increasing time.
The interacting part of the Hamiltonian is $\Hint$.
The notation `$-\infty^+$' indicates that the integral is to be performed over a contour
deformed away from the real axis into the positive complex plane in the distant past,
with analytic continuation used to define the integrand.
This can be regarded as the theorem of Gell-Mann \& Low in the present
context~\cite{GellMann:1951rw,Weinberg:2005vy}.

We frequently collect the phase-space coordinates $Q^I$, $P^I$ into a single
vector $X^a = (Q^I, P^J)$, and likewise for the interaction picture fields
$x^a = (q^I, p^J)$. Latin indices $a$, $b$, \ldots, from the early part of the
alphabet run over the dimensions of phase space, on which
the metric should be taken to have block-diagonal form
 \begin{equation}
  G_{ab} =
  \begin{pmatrix}
    G_{IJ} & 0 \\
    0 & G_{KL}  
  \end{pmatrix}
  .
\end{equation}
The vacuum expectation value of any (possibly composite)
Heisenberg-picture operator $\mathcal{O}(X)$
can be written in terms of $F$, $F^\dag$ and the interaction picture
fields using
\begin{equation} \label{eq:VEVsolution_paper}
\begin{split}
	\braket{\mathcal{O}(X)} & =
	\left\langle 0 \left| F \mathcal{O}(x) F^\dag \right| 0 \right\rangle \\
	& =
	\left\langle 0 \left| \AntiTimeOrder \exp \left( \im \int^t_{-\infty^{+}} \Hint(t') \; \d t' \right) \mathcal{O}(x) \TimeOrder \exp \left( -\im \int^t_{-\infty^{-}} \Hint (t'') \; \d t'' \right) \right| 0 \right\rangle ,
\end{split}
\end{equation}
where $|0\rangle$ is the vacuum of the free Hamiltonian.
We describe Eq.~\eqref{eq:VEVsolution_paper} as the `in--in' formula, use it compute all
correlation functions of cosmological perturbations in our field-covariant formalism.
For further details, see Appendix~\ref{subsubsec:FieldPerts}
for the definition of the covariant variable $Q^I$,
and Appendix~\ref{subsubsec:CorrFuncs}
for the definition of correlation functions.

\subsection{Hamiltonian}
\label{subsec:Hamiltonian}

In addition to the change from $\delta\phi^I$ to $Q^I$, the Hamiltonian
acquires new terms generated by derivatives of the metric.
The procedure to calculate these follows
Maldacena~\cite{Maldacena:2002vr,Seery:2005wm,Seery:2005gb}.
We minimally couple $\Nfields$ fields to gravity,
allowing a nontrivial kinetic matrix and a potential $V$,
and use the ADM decomposition to integrate out the Hamiltonian
and momentum constraints.
Finally the result is expanded to the desired order in perturbations.
The computation to third order in $Q^I$
was done by Elliston et al.~\cite{Elliston:2012ab},
or see Appendix~\ref{subsubsec:ConstaintFourierAction}
for further details.

\para{Summation convention}
To write the results, we use a compact notation in which
repeated index labels imply both summation over species labels
and integration over Fourier wavenumbers.
We indicate that this convention is in use
by writing the species indices in bold sans-serif.
Specifically, such contractions should be interpreted to mean
\begin{equation} \label{eq:FourierConvention1_paper}
	A_{\FouInd{I}}B^{\FouInd{I}} = \sum_I \int \frac{\d^3 k_I}{(2 \pi)^3} A_{I} (\vect{k}_I) B^{I} (\vect{k}_I),
\end{equation}
where, as always, the field metric $G_{IJ}$ is used to raise and lower indices.
In some manipulations
a $\delta$-function can be produced that changes the sign of a momentum label.
We indicate this by placing a bar on each index for which the sign of the momentum
should be reversed, eg.,
\begin{equation} \label{eq:FourierConvention2_paper}
	A_{\FouInd{I}}B^{\FouInd{\FlipInd{I}}} = \sum_I \int \frac{\d^3 k_I}{(2 \pi)^3} A_{I} (\vect{k}_I) B^{I} (- \vect{k}_I).
\end{equation}

\para{Second- and third-order kernels}
To third order, the result can be written
\begin{align}
\begin{split} \label{eq:PerturbedAction_Paper}
	S_\phi = \frac{1}{2} \int \d t \ a^3 \Big\{ &G_{\FouInd{IJ}} D_t Q^\FouInd{I} D_t Q^\FouInd{J} + M_{\FouInd{IJ}} Q^\FouInd{I} Q^\FouInd{J} + \\
	&A_{\FouInd{IJK}}Q^\FouInd{I} Q^\FouInd{J} Q^\FouInd{K} + B_{\FouInd{IJK}} Q^\FouInd{I} Q^\FouInd{J} D_t Q^\FouInd{K} + C_{\FouInd{IJK}} D_t Q^\FouInd{I} D_t Q^\FouInd{J} Q^\FouInd{K} \Big\},
\end{split}
\end{align}
where the second-order kernels
$G_{\FouInd{IJ}}$
and
$M_{\FouInd{IJ}}$
are defined as
\begin{align}
	\begin{split} \label{eq:Second-Order_Kernels}
	G_{\FouInd{IJ}} & \equiv (2\pi)^3 G_{IJ} \delta(\vect{k}_1 + \vect{k}_2), \\
	M_{\FouInd{IJ}} & \equiv (2\pi)^3 \delta(\vect{k}_1 + \vect{k}_2) \left( \frac{\vect{k}_1 \cdot \vect{k}_2}{a^2} G_{IJ} - m_{IJ} \right),
	\end{split}
\end{align}
and the mass-matrix $m_{IJ}$ satisfies
\begin{equation} \label{eq:m_IJ_paper}
	m_{IJ} \equiv V_{;IJ} - R_{KIJL}\dot{\phi}^K \dot{\phi}^L - \frac{1}{a^3 \MPlanck^2} D_t \bigg( \frac{a^3 \dot{\phi}_I \dot{\phi}_J}{H} \bigg).
\end{equation}
Then the third-order kernels
$A_{\FouInd{IJK}}$,
$B_{\FouInd{IJK}}$ and $C_{\FouInd{IJK}}$
are given by
\begin{subequations}
\begin{align}
\begin{split} \label{eq:Akernel_paper}
	A_{\FouInd{IJK}} & \equiv (2\pi)^3 \delta(\vect{k}_1 + \vect{k}_2 + \vect{k}_3) A_{IJK},
\end{split}\\
\begin{split} \label{eq:Bkernel_paper}
	B_{\FouInd{IJK}} & \equiv (2\pi)^3 \delta(\vect{k}_1 + \vect{k}_2 + \vect{k}_3) B_{IJK},
\end{split}\\
\begin{split} \label{eq:Ckernel_paper}
	C_{\FouInd{IJK}} & \equiv (2\pi)^3 \delta(\vect{k}_1 + \vect{k}_2 + \vect{k}_3) C_{IJK},
\end{split}
\end{align}
\end{subequations}
and the corresponding `species tensors'
are
\begin{subequations}
\begin{align}
\begin{split} \label{eq:ATensor_paper}
	A_{IJK}
	\equiv \mbox{} &
	-\frac{1}{3} V_{;IJK}
	- \frac{\dot{\phi}_I V_{;JK}}{2H \MPlanck^2}
	+ \frac{\dot{\phi}_I \dot{\phi}_J Z_K}{4H^2 \MPlanck^4}
	+ \frac{\dot{\phi}_I Z_J Z_K}{8H^3 \MPlanck^4}
	   \left(
	       1 - \frac{(\vect{k}_2 \cdot \vect{k}_3)^2}{k_2^2 k_3^2}
	   \right)
	\\ 
	&
	+ \frac{\dot{\phi}_I \dot{\phi}_J \dot{\phi}_K}{8H^3 \MPlanck^6}
	   (6H^2 \MPlanck^2 - \dot{\phi}^2 )
	- \frac{\dot{\phi}_K \dot{\phi}^L \dot{\phi}^M}{2H \MPlanck^2} R_{L(IJ)M}
	+ \frac{1}{3}R_{(I|LM|J;K)}\dot{\phi}^L \dot{\phi}^M \\
	&
	+ \frac{\dot{\phi}_I G_{JK}}{2H \MPlanck^2} \frac{\vect{k}_2 \cdot \vect{k}_3}{a^2}
	,
\end{split}\\
\begin{split} \label{eq:BTensor_paper}
	B_{IJK}
	\equiv \mbox{} &
	\frac{4}{3} R_{K(IJ)L} \dot{\phi}^L
	- \frac{\dot{\phi}_I Z_J \dot{\phi}_K}{4H^3 \MPlanck^4}
	   \left(
	       1 - \frac{(\vect{k}_2 \cdot \vect{k}_3)^2}{k_2^2 k_3^2}
	   \right)
	+ \frac{\dot{\phi}_I \dot{\phi}_J \dot{\phi}_K}{4H^2 \MPlanck^4}
	- \frac{Z_I G_{JK}}{H \MPlanck^2} \frac{\vect{k}_1 \cdot \vect{k}_2}{k_1^2},
\end{split}\\
\begin{split} \label{eq:CTensor_paper}
	C_{IJK}
	\equiv \mbox{} &
	- \frac{G_{IJ} \dot{\phi}_K}{2H \MPlanck^2}
	+ \frac{\dot{\phi}_I \dot{\phi}_J \dot{\phi}_K}{8H^3 \MPlanck^4}
	   \left(
	       1 - \frac{(\vect{k}_1 \cdot \vect{k}_2)^2}{k_1^2k_2^2}
	   \right)
	+ \frac{\dot{\phi}_I G_{JK}}{H \MPlanck^2} \frac{\vect{k}_1 \cdot \vect{k}_3}{k_1^2} .
\end{split}
\end{align}
\end{subequations}
The brackets surrounding indices in the Riemann terms indicate that
the enclosed indices should be symmetrized with weight unity, except for
indices between vertical bars $|$ which are excluded.
Further, note that the tensor $A_{IJK}$ should be symmetrized over all three indices
$IJK$ with weight unity,
and
$B_{IJK}$, $C_{IJK}$ should be symmetrized over $IJ$ with weight unity.
The indices on the
momentum labels $\vect{k}_1$, $\vect{k}_2$, $\vect{k}_3$
correspond to field-space labels as $1\rightarrow I$, $2\rightarrow J$ and $3\rightarrow K$,
and should be permuted during symmetrization.
The quantity $Z_I$ is defined by
\begin{equation} \label{eq:Z_I_paper}
	Z_I \equiv D_t \dot{\phi}_I + \frac{\dot{\phi}_I \dot{\phi}_J \dot{\phi}^J}{2H \MPlanck^2}.
\end{equation}
From these expressions it is simple to calculate the Hamiltonian using a Legendre transformation.
We define the canonical momentum $P_\FouInd{I}$ to satisfy
\begin{equation} \label{eq:ConjugateMomenta_paper}
	P_\FouInd{I} (t) \equiv \frac{\delta S_\phi}{\delta (D_t Q^\FouInd{I})},
\end{equation}
where the variational derivative can be computed using the rule
\begin{equation} \label{eq:VarDerivDefinition_paper}
	\frac{\delta [ Q^\FouInd{I} (\vect{k}_I, t) ]}{\delta [ Q^\FouInd{J} (\vect{k}_J, t') ]}
	= \delta_J^I (2\pi)^3 \delta(t - t') \delta(\vect{k}_I + \vect{k}_J) = \delta_\FouInd{J}^\FouInd{I} \delta(t - t') .
\end{equation}
To compute the Hamiltonian we require the relation
$H = \int \d t \, [ P^\FouInd{I} (D_t Q_{\FouInd{\FlipInd{I}}}) - L]$
which should be regarded as a function of $Q^I$ and $P^I$.
Finally, for convenience, we rescale the momentum by a factor $a^3$, viz.
$P_\FouInd{I} \rightarrow a^3 P_\FouInd{I}$, to obtain the final third-order Hamiltonian,
\begin{align}
	\begin{split} \label{eq:HamiltonianSolution_paper}
		H = \frac{1}{2} \int \d t \ a^3 \Big( &G_{\FouInd{IJ}} P^\FouInd{I} P^\FouInd{J} - M_{\FouInd{IJ}} Q^\FouInd{I} Q^\FouInd{J} - \\
		&A_{\FouInd{IJK}} Q^\FouInd{I} Q^\FouInd{J} Q^\FouInd{K} - B_{\FouInd{IJK}} Q^\FouInd{I} Q^\FouInd{J} P^\FouInd{K} - C_{\FouInd{IJK}} P^\FouInd{I} P^\FouInd{J} Q^\FouInd{K} \Big) .
	\end{split}
\end{align}
The second-order terms on the first line and represent the free part of the Hamiltonian $H_0$,
and the third-order terms on the second line represent the interacting part of the
Hamiltonian $\Hint$. The new contributions introduced by
derivatives of the nontrivial field-space
metric are given by the Riemann terms found in
the
$M_{IJ}$, $A_{IJK}$ and $B_{IJK}$ tensors.

\subsection{Initial conditions}
\label{subsec:NewInitialConds}
We will require suitable initial conditions for each correlation function on subhorizon
scales. To compute these we use Eq.~\eqref{eq:VEVsolution_paper}
to compute each correlation function at sufficiently early times---%
normally between four and ten e-folds inside the horizon, although the
precise numbers are model-dependent; see Ref.~\cite{Dias:2015rca}---%
that all species can be approximated as massless.
Such a time can normally be found, provided all masses remain bounded,
because the physical wavenumber $k/a$ corresponding to a fixed comoving wavenumber $k$
is pushed into the ultraviolet at early times, making each mode kinetically dominated
for sufficiently small $a$.
The outcome is that we can compute \emph{universal} initial conditions
applicable to any model, no matter what mass spectrum or interactions it contains,
provided the computation of its correlation functions begins sufficiently far inside
the horizon~\cite{Dias:2015rca,Dias:2016rjq}.
For more details see {\S}3 of Ref.~\cite{Dias:2015rca}
and {\S}6 of Ref.~\cite{Dias:2016rjq}.

\para{Two-point function}
A suitable initial condition for the covariant equal-time
2-point function was computed
by Dias et al.~\cite{Dias:2015rca},
following Elliston et al.~\cite{Elliston:2012ab}.
In our notation their results can be written
\begin{subequations}
	\begin{align}
		\label{eq:QQinit_conds_paper} \braket{Q^I(\vect{k}_1) Q^{J}(\vect{k}_2)}_{\mathrm{init}} &= (2\pi)^3 \delta(\vect{k}_1 + \vect{k}_2) G^{IJ} \left( \frac{1}{2ka^2} + \frac{H^2}{2k^3} \right) , \\
		\label{eq:QPinit_conds_paper} \braket{Q^I(\vect{k}_1) P^{J}(\vect{k}_2)}_{\mathrm{init}} &= (2\pi)^3 \delta(\vect{k}_1 + \vect{k}_2) G^{IJ} \left( - \frac{H}{2 k a^2} + \frac{\im}{2 a^3} \right), \\
		\label{eq:PQinit_conds_paper} \braket{P^I(\vect{k}_1) Q^{J}(\vect{k}_2)}_{\mathrm{init}} &= (2\pi)^3 \delta(\vect{k}_1 + \vect{k}_2) G^{IJ} \left( - \frac{H}{2 k a^2} - \frac{\im}{2 a^3} \right), \\
		\label{eq:PPinit_conds_paper} \braket{P^I(\vect{k}_1) P^{J}(\vect{k}_2)}_{\mathrm{init}} &= (2\pi)^3 \delta(\vect{k}_1 + \vect{k}_2) G^{IJ} \left( \frac{k}{2 a^4} \right) ,
	\end{align}
\end{subequations}
where the time-dependent quantities $H$, $a$ and $G_{IJ}$ appearing on the right-hand sides
should be evaluated at the initial time $\tinit$,
indicated by the subscript `init' attached to each correlation function.

Eqs.~\eqref{eq:QQinit_conds_paper}--\eqref{eq:PPinit_conds_paper}
are effectively the same as those found in the canonical case~\cite{Dias:2016rjq}
except that the Euclidean kinetic matrix
$\delta^{IJ}$ is replaced by the metric $G^{IJ}$.
For further details of the computation see Appendix~\ref{subsubsec:2pntCorrFns}.

\para{Three-point function}
Initial conditions for the 3-point functions require the in--in formula~\eqref{eq:VEVsolution_paper}.
The lowest-order nonzero contribution
to each correlator is given by
\begin{equation} \label{eq:3pntCorrFn_WickRotate_paper}
\begin{split}
	\braket{X^I X^J X^K} \subseteq
	\im
	\int_{-\infty^+}^\eta \d\tau \, H_{\FouInd{LMN}}
	\int &
	\Big(
	   \prod_{n=1}^3 \frac{\d^3 q_n}{(2\pi)^3}
	\Big)
	(2\pi)^3 \delta \Big( \sum_{i=1}^3 \vect{q}_i \Big)
	\\
	&
	\times \left\{ 
        \braket{X^{\FouInd{L}}_{q_1} X_{k_1}^I} \braket{X^\FouInd{M}_{q_2} X_{k_2}^J} \braket{X^\FouInd{N}_{q_3} X_{k_3}^K}
        + \text{perms}
    \right\}
    + \text{c.c.}
    ,
\end{split}
\end{equation}
where `perms' indicates a sum over permutations of the pairing
between `external' indices $IJK$ and the `internal' indices
$\FouInd{LMN}$,
`c.c.' indicates the complex conjugate of the preceding term,
and $H_{\FouInd{LMN}}$
contains all the cubic terms found in Eq.~\eqref{eq:HamiltonianSolution_paper}.
For further details we refer to Appendix~\ref{subsubsec:3pntCorrFns}.

To express the results we require some extra notation.
First, we divide $A_{IJK}$ into `fast' terms, which involve the scale factor $a$
and evolve exponentially fast in e-folds,
and `slow' terms, which evolve on slow-roll timescales,
\begin{equation}
    A^{IJK} \equiv
    \Aslow^{IJK} + \Afast^{IJK}
    =
    \Aslow^{IJK}
	+ \frac{\dot{\phi}^I G^{JK}}{2H \MPlanck^2} \frac{\vect{k}_2 \cdot \vect{k}_3}{a^2}
\end{equation}
The fast term grows rapidly on subhorizon scales and is always relevant
when computing initial conditions.
In Ref.~\cite{Dias:2016rjq} it was explained that the slow terms can also be relevant
in scenarios with enhanced three-body interactions such as a QSFI model.

Second, we introduce the quantities
$\vectktot \equiv \vect{k}_1 + \vect{k}_2 + \vect{k}_3$,
$k_t \equiv k_1 + k_2 + k_3$
and
$K \equiv k_1 k_2 + k_1 k_3 + k_2 k_3$.
The results for each correlation function are%
	\footnote{Eq.~\eqref{eq:PPPinitconds_paper}
	corrects a minor typo in v1 and v2 of the arXiv
	version of Ref.~\cite{Dias:2016rjq}. This typo was corrected
	in the arXiv v3.}
\begin{subequations}
\begin{align}
	\begin{split} \label{eq:QQQinitconds_paper}
		\braket{Q^I Q^J Q^K}_{\mathrm{init}}
		=
		\frac{(2\pi)^3 \delta (\vectktot)}{4a^4 k_1 k_2 k_3 k_t}
		\bigg\{
		  &
		  \frac{\dot{\phi}^I G^{JK}}{4H \MPlanck^2} \vect{k}_2 \cdot \vect{k}_3
		  + \frac{a^2}{2} \Aslow^{IJK}
		  - C^{IJK} \frac{k_1 k_2}{2}
		  \\
    	  & \mbox{}
		  + \frac{a^2 H}{2} B^{IJK}
		      \bigg[
		          \frac{(k_1 + k_2)k_3}{k_1 k_2}
		          - \frac{K^2}{k_1 k_2}
		      \bigg]
		  + \text{5 perms}
		\bigg\}
		,
	\end{split}
\end{align}
\begin{align}
	\begin{split} \label{eq:PQQinitconds_paper}
		\braket{P^I Q^J Q^K}_{\mathrm{init}}
		& =
		\frac{(2\pi)^3 \delta (\vectktot)}{4 a^4 (k_1 k_2 k_3)^2 k_t}
		\\
		& \hspace{-1.7cm}\times
		\bigg\{
		  k_1^2 (k_2 + k_3)
		  \bigg[
		      \frac{\dot{\phi}^I G^{JK}}{4H \MPlanck^2} \vect{k}_2 \cdot \vect{k}_3
		      + \frac{a^2}{2} \Aslow^{IJK}
		      - C^{IJK} \frac{k_1 k_2}{2}
		      + \text{5 perms}
		  \bigg]
		\\
		& \hspace{-1cm} +
		  k_1
		  \bigg[
		      {-\frac{\dot{\phi}^I G^{JK}}{4H \MPlanck^2}} \vect{k}_2 \cdot \vect{k}_3 
		      \Big(
		          K^2 + \frac{k_1 k_2 k_3}{k_t}
		      \Big)
		      - \frac{a^2}{2} \Aslow^{IJK}
		      \Big(
		          K^2 - \frac{k_1 k_2 k_3}{k_t}
		      \Big)
		\\
		& +
		      B^{IJK} \frac{k_1 k_2 k_3^2}{2H}
		      + C^{IJK} \frac{k_1^2 k_2^2}{2}
		      \Big(
		          1 + \frac{k_3}{k_t}
		      \Big)
		      + \text{5 perms}
		  \bigg]
        \bigg\}
        ,
	\end{split}
\end{align}
\begin{align}
	\begin{split} \label{eq:PPQinitconds_paper}
		\braket{P^I P^J Q^K}_{\mathrm{init}} 
		& =
		\frac{(2\pi)^3 \delta (\vectktot)}{4 a^6 H^2 (k_1 k_2 k_3)^2 k_t}
		\\
		& \hspace{-1.7cm}\times
		\bigg\{
		  k_1^2 k_2^2 k_3 
		  \bigg[
		      {-\frac{\dot{\phi}^I G^{JK}}{4H \MPlanck^2}} \vect{k}_2 \cdot \vect{k}_3
		      - \frac{a^2}{2} \Aslow^{IJK}
		      + C^{IJK} \frac{k_1 k_2}{2}
		      - \frac{a^2 H}{2} B^{IJK} \frac{(k_1 + k_2) k_3}{k_1 k_2}
		\\
		      & \hspace{3mm} + \text{5 perms}
		  \bigg]
		  +
		  k_1^2 k_2^2
		  \bigg[
		      \frac{a^2 H}{2} B^{IJK} k_3
		      + \text{5 perms}
		  \bigg]
		\bigg\}
		,
	\end{split}
\end{align}
\begin{align}
	\begin{split} \label{eq:PPPinitconds_paper}
		\braket{P^I P^J P^K}_{\mathrm{init}}
		=
		\frac{(2\pi)^3 \delta (\vectktot)}{4 a^6 H^2 k_1 k_2 k_3 k_t}
		\bigg\{
		  & 
		  \frac{\dot{\phi}^I G^{JK}}{4H \MPlanck^2} \vect{k}_2 \cdot \vect{k}_3
		  \Big(
		      K^2 + \frac{k_1 k_2 k_3}{k_t}
		  \Big)
		  + \frac{a^2}{2} \Aslow^{IJK}
		  \Big(
		      K^2 - \frac{k_1 k_2 k_3}{k_t}
		  \Big) \\
		  & \mbox{}
		  - B^{IJK} \frac{k_1 k_2 k_3^2}{2H}
		  - C^{IJK} \frac{k_1^2 k_2^2}{2}
		  \Big(
		      1 + \frac{k_3}{k_t}
		  \Big)
		  + \text{5 perms}
		\bigg\}
		.
	\end{split}
\end{align}
\end{subequations}
All time-dependent quantities on the right-hand side are to be evaluated at the
initial time $\tinit$,
and the tangent-space indices $I$, $J$, $K$, \ldots,
live in the tangent space associated with this time.

Where permutations are specified, they should be carried out
\emph{only within the bracket} in which the instruction to sum over permutations
is given.
(Notice that these means some momentum factors, such as those multiplying the square-bracket
terms in Eqs.~\eqref{eq:PQQinitconds_paper} and~\eqref{eq:PPQinitconds_paper},
are not symmetrized. This is correct because these momentum factors arise from wavefunctions
associated with the external fields, and these are not symmetric.)
Each permutation should be formed by simultaneous exchange of the species labels $I$, $J$, $K$
and their partner momenta $\vect{k}_1$, $\vect{k}_2$, $\vect{k}_3$.

The form of these equations matches the canonical case~\cite{Dias:2016rjq}, except for the
Riemann terms embedded in $\Aslow^{IJK}$ and $B^{IJK}$.
For further details of the calculation, see Appendix~\ref{subsubsec:3pntCorrFns}.

\subsection{Covariant transport equations}
\label{subsec:CovarTransEqs}
Next, we require differential equations to evolve each correlation function
from its initial value to any time of interest.
These equations were derived in the superhorizon limit by
Mulryne et al.~\cite{Mulryne:2009kh,Mulryne:2010rp,Elliston:2012ab}
and later extended to cover the subhorizon era~\cite{Mulryne:2013uka,Dias:2015rca}.

The procedure to derive these evolution equations matches that of
Dias et al.~\cite{Dias:2016rjq}.
We begin from the Hamiltonian~\eqref{eq:HamiltonianSolution_paper},
which can be written in the generic form
\begin{equation} \label{eq:HamiltFourierExpans_Paper}
	H = \frac{1}{2!} H_{\FouInd{ab}} X^\FouInd{a} X^\FouInd{b} 
	+ \frac{1}{3!} H_{\FouInd{abc}} X^\FouInd{a} X^\FouInd{b} X^\FouInd{c} + \cdots .
\end{equation}
The corresponding covariant evolution equation is
\begin{equation} \label{eq:HeisenbergEOMs_utensordefs}
	D_t X^\FouInd{a} = u\indices{ ^{\FouInd{a}} _{\FouInd{b}} } X^\FouInd{b}
	+ \frac{1}{2!} u\indices{ ^{\FouInd{a}} _{\FouInd{bc}}} X^\FouInd{b} X^\FouInd{c} + \cdots ,
\end{equation}
where $u\indices{ ^{\FouInd{a}} _{\FouInd{b}} }$ and
$u\indices{ ^{\FouInd{a}} _{\FouInd{bc}}}$
are phase-space tensors
that can be expressed in terms of $H_{\FouInd{ab}}$
and $H_{\FouInd{abc}}$~\cite{Dias:2016rjq}.
The derivative $D_t$ should be taken to act in phase space with a block-diagonal connexion.
For example, acting on contra- and covariant indices this produces
\begin{equation}
    D_t {X^a}_b = \frac{\d}{\d t} X^a + \Gamma^a_c {X^c}_b - \Gamma^c_b {X^a}_c ,
\end{equation}
where $\Gamma^a_b$ is the block matrix
\begin{equation}
    \Gamma^a_b =
    \begin{pmatrix}
        \dot{\phi}^K \Gamma^I_{JK} & 0 \\
        0 & \dot{\phi}^K \Gamma^I_{JK}
    \end{pmatrix}
    .
\end{equation}
In each block $I$ represents the species label associated with the
phase-space label $a$, and $J$ represents the species label associated with $b$.

A similar equation can be found for the fields in the interaction picture.
Using Eq.~\eqref{eq:VEVsolution_paper} to deduce tree-level expressions for the
2- and 3-point functions in terms of interaction-picture fields, it follows that
evolution equations can be derived by direct differentiation
and use of the interaction-picture equations of motion to rewrite
time derivatives.
The results are
\begin{subequations}
	\begin{align}
		\label{eq:2pointTransportEquation}
		D_t \Sigma^{ab} &= u\indices{^a_c} \Sigma^{cb} + u\indices{^b_c} \Sigma^{ac} , \\
		\label{eq:3pointTransportEquation}
		D_t \alpha^{abc} &= u\indices{^a_d} \alpha^{dbc} + u\indices{^a_{de}} \Sigma^{db} \Sigma^{ec} + 2 \ \mathrm{cyclic} \ (a\rightarrow b \rightarrow c),
	\end{align}
\end{subequations}
where
we have written the phase-space 2- and 3-point functions
as
\begin{subequations}
	\begin{align}
   	\braket{X^{\FouInd{a}} X^{\FouInd{b}}} & \equiv (2\pi)^3 \delta(\vect{k}_a + \vect{k}_b) \Sigma^{ab} , \\
   	\braket{X^{\FouInd{a}} X^{\FouInd{b}} X^{\FouInd{c}}} & \equiv (2\pi)^3 \delta(\vect{k}_a + \vect{k}_b + \vect{k}_c) \alpha^{abc} .
	\end{align}
\end{subequations}
These equations match those in the canonical case except that the time
derivative $D_t \equiv \dot{\phi}^I \nabla_I$ is now
covariant and will introduce terms involving the connexion components.
A more detailed derivation of these equations can be found in Appendix~\ref{subsubsec:EvoEqs}.

For practical calculations we need explicit expressions for the
$u$-tensors.
They are~\cite{Dias:2016rjq}
\begin{subequations}
\begin{align}
	\label{eq:u1_tensor_paper}
	u\indices{^a_b} =&
 	\begin{pmatrix}
  		0 & \delta^{I}_J \\
  		M\indices{^I_J} & -3H\delta_J^I
 	\end{pmatrix}, \\
 	\label{eq:u2_tensor_paper}
 	u\indices{^a_{bc}} =&
	\begin{Bmatrix}
		\begin{pmatrix}[1.1]
			- B\indices{_{JK}^I} & - C\indices{^I_{JK}} \\
			3 A\indices{^I_{JK}} & B\indices{^I_{KJ}}
		\end{pmatrix} \\
		\\
		\begin{pmatrix}
			-C\indices{^I_{KJ}} & 0 \\
			B\indices{^I_{JK}} & C\indices{_{KJ}^I}
		\end{pmatrix}
	\end{Bmatrix},
\end{align}
\end{subequations}
in which the index $a$ labels rows of the top-level matrix.
For ${u^a}_b$, the index $b$ labels the remaining columns;
for ${u^a}_{bc}$, the indices $bc$ label rows and columns of each submatrix.
As above, the field-space labels $I$, $J$, $K$
represent the species associated with the phase-space labels $a$, $b$ and $c$.

Further details of the calculation, including the Heisenberg equations of motion for $Q^I$ and $P^I$,
can be found in Appendix~\ref{subsubsec:CalcUtensors}.

\subsection{Gauge transformation}
\label{subsec:GaugeTrans_Short}
Although the formalism of covariant correlation functions is
computationally convenient, the covariant perturbations $Q^I$
and their statistical properties are not directly measurable.
The final step is therefore to express correlation functions
of measurable quantities such as the curvature perturbation $\zeta$
in terms of the covariant correlation functions.
This is a covariantization of the gauge transformation from
the spatially flat gauge to the uniform density gauge~\cite{Dias:2014msa,Dias:2015rca,Dias:2016rjq}.

Using the methods of Ref.~\cite{Dias:2014msa} we find that the
density fluctuation on spatially flat slices can be written
in terms of the covariant perturbations $Q^I$,
\begin{align} \label{eq:RhoPertUnsimplified_paper}
	\begin{split}
		\delta \rho & = \dot{\phi}^I D_t Q_I + V_I Q^I + \frac{1}{2} \left( 3\alpha_1^2 - 2\alpha_2 - 2\alpha_1 \right) \dot{\phi}^I \dot{\phi}_I \\
		& \quad + \frac{1}{2} V_{IJ} Q^I Q^J + \frac{1}{2} D_t Q^I D_t Q_I - 2\alpha_1 \dot{\phi}^I D_t Q_I + \frac{1}{2} R_{IJKL} Q^I \dot{\phi}^J \dot{\phi}^K Q^L
		,
	\end{split}
\end{align}
where $\alpha_1$ and $\alpha_2$, respectively, are the first- and second-order perturbations to the lapse.
We have neglected spatial gradients that become negligible on superhorizon scales.

Eq.~\eqref{eq:RhoPertUnsimplified_paper}
is superficially different to the canonical case due to the final term involving the
Riemann tensor.
However, the same term appears in the Hamiltonian constraint
(see Eq.~\eqref{eq:HamiltonianConstraint}), and after using this constraint
to simplify~\eqref{eq:RhoPertUnsimplified_paper}
the result matches the na\"{\i}ve covariantization
of the canonical formula~\cite{Dias:2014msa}.

Using the results of Dias et al.~\cite{Dias:2014msa}
to express $\zeta$ in terms of $\delta\rho$,
it follows that the curvature perturbation can be written in the form
\begin{equation} \label{eq:FourExpansZetaNs}
	\zeta(\vect{k}) = N_{\FouInd{a}} X^{\FouInd{a}} + \frac{1}{2} N_{\FouInd{ab}} X^\FouInd{a} X^\FouInd{b}
    .
\end{equation}
The coefficient matrices
$N_a$
and
$N_{ab}$
are given by
\begin{subequations}
\begin{equation}
	\label{eq:N_a_solution_paper}
	N_a = - \frac{\dot{\phi}_I}{2H \MPlanck^2 \epsilon}
	\begin{pmatrix}
		1 \\
      	0 \\
    \end{pmatrix},
\end{equation}
{
\renewcommand{\arraystretch}{2}
\begin{equation}
    \label{eq:N_ab_solution_paper}
    N_{ab} = \frac{1}{3H^2 \MPlanck^2 \epsilon}
    \begin{pmatrix}
        \displaystyle \frac{\dot{\phi}_I \dot{\phi}_J}{\MPlanck^2}
        \Big[
            -\frac{3}{2}
            + \frac{9}{2\epsilon}
            + \frac{3}{4\epsilon^2} \frac{V_K \pi^K}{H^3 \MPlanck^2}
        \Big]
        &
        \displaystyle \frac{3}{H\epsilon} \frac{\dot{\phi}_I \dot{\phi}_J}{\MPlanck^2}
        - \frac{3H}{k^2}
        \Big[
            \vect{k}_a \cdot \vect{k}_b
            + k_a^2
        \Big] G_{IJ} \\
        \displaystyle \frac{3}{H\epsilon} \frac{\dot{\phi}_I \dot{\phi}_J}{\MPlanck^2}
        - \frac{3H}{k^2}
        \Big[
            \vect{k}_a \cdot \vect{k}_b
            + k_b^2
        \Big] G_{IJ}
        &
        0
    \end{pmatrix}.
\end{equation}
}
\end{subequations}

\section{Numerical results}
\label{sec:NumRes}
We are now able to solve the equations obtained in
\S\ref{sec:DifferenceCanonicalCase}
and use them to compute the observable 2- and 3-point
functions of an arbitrary model with user-defined
kinetic mixing matrix.

\para{Overview}
In summary, this involves obtaining numerical solutions
to the 2- and 3-point function transport equations~\eqref{eq:2pointTransportEquation}--\eqref{eq:3pointTransportEquation},
using the $u$-tensors specified in~\eqref{eq:u1_tensor_paper}--\eqref{eq:u2_tensor_paper}.
In turn, these depend on the kinetic matrix $G_{IJ}$ and the `species tensors'
$m_{IJ}$, $A_{IJK}$, $B_{IJK}$ and $C_{IJK}$ that specify the Hamiltonian (cf. Eq.~\eqref{eq:HamiltonianSolution_paper}).
They must be determined for each model
from the general formulae~\eqref{eq:Second-Order_Kernels}
and~\eqref{eq:ATensor_paper}--\eqref{eq:CTensor_paper}.
The initial conditions are given by Eqs.~\eqref{eq:QQQinitconds_paper}--\eqref{eq:PPPinitconds_paper},
provided a suitable initial time can be found at which the massless approximation
is valid for all species.
These initial conditions also depend on $G_{IJ}$, $m_{IJ}$, $A_{IJK}$, $B_{IJK}$ and $C_{IJK}$.
Finally, Eqs.~\eqref{eq:N_a_solution_paper} and~\eqref{eq:N_ab_solution_paper}
are used to construct the correlation functions of $\zeta$.

Each of the transport tools
\href{https://transportmethod.com/mtransport}{\mtransport}, 
\href{https://transportmethod.com/cpptransport}{\cpptransport} and
\href{https://transportmethod.com/pytransport/}{\pytransport}
uses symbolic algebra
to automate the calculation of $M_{IJ}$, $A_{IJK}$, $B_{IJK}$ and $C_{IJK}$ from a specification of the kinetic
matrix $G_{IJ}$ and the potential $V$.
With explicit expressions for each tensor, it is possible to set up
the transport equations and compute suitable initial conditions.
Additionally, both {\cpptransport} and {\pytransport} automate the task of finding
a suitable initial time at which the massless approximation is valid;
in {\mtransport} this currently has to be done by hand, or a suitable initial time estimated.

\para{Notation}
When discussing concrete models we generally use the dimensionless
power spectrum $\mathcal{P}(k)$, defined in terms of the
ordinary power spectrum $P(k)$ (see Eq.~\eqref{eq:PowSpecDef})
using
\begin{equation} \label{eq:DimensionlessPowSpec}
	\mathcal{P}(k) \equiv \frac{k^3}{2\pi} P(k),
\end{equation}
The analogous quantity for the three-point function
is the `dimensionless bispectrum',
defined by
\begin{equation} \label{eq:DimensionlessBiSpec}
	\mathcal{B}(k_1, k_2, k_3) \equiv (k_1 k_2 k_3)^2 B(k_1, k_2, k_3) .
\end{equation}
We also use the reduced bispectrum,
conventionally written $\fNL(k_1, k_2, k_3)$,
which is defined to satisfy
\begin{equation} \label{eq:f_NL def}
	\frac{6}{5} \fNL (k_1, k_2, k_3) \equiv \frac{B(k_1, k_2, k_3)}{P(k_1)P(k_2) + P(k_1)P(k_3) + P(k_2)P(k_3)} .
\end{equation}
Notice that this is \emph{not} the same as the
parameter $\fNL^{\text{local}}$
measured by CMB experiments, although in models
where the bispectrum is dominantly of the `local' type it is closely
related to it.

To specify the configuration of Fourier wavenumbers that
characterize the bispectrum we use
the parametrization suggested by Fergusson \& Shellard~\cite{Fergusson:2006pr},
\begin{subequations}
	\begin{align}
		\label{eq:k_1-ShapeParam}
		k_1 &\equiv \frac{k_t}{4} \left( 1 + \alpha + \beta \right), \\
		\label{eq:k_2-ShapeParam}
		k_2 &\equiv \frac{k_t}{4} \left( 1 - \alpha + \beta \right), \\
		\label{eq:k_3-ShapeParam}
		k_3 &\equiv \frac{k_t}{2} \left( 1 - \beta \right) .
	\end{align}
\end{subequations}
The overall scale of the momentum triangle
is measured by its perimeter $k_t \equiv k_1 + k_2 + k_3$,
and its shape is measured by $\alpha$ and $\beta$.
The allowed ranges are $-1 \leq \alpha \leq 1$ and $0 \leq \beta \leq 1$.

By default, {\cpptransport} uses
its own `internal normalization' in which a distingished wavenumber $k_* = 1$
is \emph{defined} to exit the horizon at a chosen time $N_*$.
In this normalization, other wavenumbers are measured relative to $k_*$
by giving the ratios
$k/k_*$ or $k_t/k_*$ respectively. This convention means that all wavenumbers quoted in this section
are dimensionless.
In each case we quote the corresponding value of $N_*$.
Where other horizon exit times are given, these are measured relative to the initial
conditions at $N=0$.

\subsection{Obtaining the transport codes}
\label{subsec:ObtainUsingCodes}

All tools
({\mtransport}, {\cpptransport} and {\pytransport})
can be downloaded from the website \href{http://transportmethod.com}{transportmethod.com}.
At the time of writing the current version of {\pytransport}
is \href{https://zenodo.org/record/848220}{v2.0} and the current version of {\cpptransport}
is \href{https://zenodo.org/record/1183518}{2018.1}.
Alternatively, development versions of
\href{https://github.com/ds283/CppTransport}{\cpptransport}
and
\href{https://github.com/jronayne/PyTransport}{\pytransport}
can be downloaded from their respective GitHub repositories.
In this paper we focus on the new features in {\cpptransport}
that support an arbitrary metric $G_{IJ}$.

An introduction to {\cpptransport} was given in {\S}8 of Dias et al.~\cite{Dias:2016rjq}
and a comprehensive user guide is available on the arXiv~\cite{Seery:2016lko}.
When making use of the new features available in 2018.1 most steps remain the same,
with only minor variations:
\begin{itemize}
	\item To use a nontrivial metric $G_{IJ}$ it is first necessary to specify that the
	model is non-canonical by including the
	directive
	\begin{lstlisting}
		lagrangian = nontrivial_metric;
	\end{lstlisting}
	in the {\small\texttt{model}} block of the input file.
	Having done so the metric can be specified along with the potential as a list of components surrounded
	by square brackets {\small\texttt{[}} $\cdots$ {\small\texttt{]}}. For example, the metric on
	a flat two-dimensional field-space in polar coordinates would be written
	\begin{lstlisting}
		metric = [ R, R = 1; theta, theta = R^2; ];
	\end{lstlisting}
	Off-diagonal elements need be specified only for the upper or lower triangle,
	and
	entries that are not given are assumed to be zero.
	Elements can make use of subexpressions declared elsewhere in the model file.
	
	\item A suitable set of templates must be chosen for the core and implementation files
	that use correct index placement
	and employ the covariantized formulae given in~\S\ref{sec:DifferenceCanonicalCase}.
	An extra set of templates with these properties is bundled with 2018.1.
	To use them, the {\small\texttt{template}} block of the model file should read
	\begin{lstlisting}
		templates
		 { core           = "nontrivial_metric_core";
		   implementation = "nontrivial_metric_mpi";
		 };
	\end{lstlisting}
	All Riemann terms will be correctly included in
	the $u$-tensors and initial conditions, and the transport equations
	will include correct connexion components.
\end{itemize}

\subsection{Cartesian versus polar coordinates}
\label{subsec:CanoCompQsfiGel}
We begin by reproducing results for the gelaton-like scenario~\cite{Tolley:2009fg}
studied in Dias et al.~\cite{Dias:2016rjq}.
This is an `adiabatic-like' model in which a continuously-turning light field
is dressed by the fluctuations of a transverse heavy field, and has
similarities to the scenario of quasi-single field inflation~\cite{Chen:2009we}.
Because the heavy field tracks the minimum of the effective potential,
slightly displaced due to the radial motion of the light field,
the model behaves as if it has a single collective degree of freedom.

The model is most conveniently expressed in polar field-space coordinates $R$ and $\theta$,
and therefore Ref.~\cite{Dias:2016rjq} performed a coordinate transformation
to Cartesian fields $X = R \cos \theta$, $Y = R \sin \theta$
to produce a Euclidean kinetic matrix. In this section we study the model
in its original polar formulation, finding excellent agreement.
The Lagrangian is
\begin{equation} \label{eq:QSFI/Gelatonaction}
	S = \frac{1}{2} \int \d^4 x \sqrt{-g} \left[ (\partial R)^2 + R^2 (\partial \theta)^2 + 2V(R, \theta) \right],
\end{equation}
where $R$ is the heavy field and $\theta$ is the light field.
The field-space metric is
\begin{equation} \label{eq:QSFI/Gelaton_Metric}
	G_{IJ} = 
	\begin{pmatrix}
		1 & 0 \\
		0 & R^2
	\end{pmatrix}.
\end{equation}
In Ref.~\cite{Dias:2016rjq}
the potential was chosen so that it represents a circular valley
at fixed $R$. The angular velocity $\omega = \dot{\theta}/H$ was chosen so that
a rotation through $\pi$ occurred over approximately 30 e-folds.
A suitable choice is
\begin{equation} \label{eq:QSFI/GelatonPotential}
	V = V_0 \left( 1 + \frac{29\pi}{120} \theta + \frac{1}{2} \frac{\eta_R}{\MPlanck^2} (R - R_0)^2 + \frac{1}{3!} \frac{g_R}{\MPlanck^3} (R - R_0)^3 + \frac{1}{4!} \frac{\lambda_R}{\MPlanck^4} (R - R_0)^4 \right),
\end{equation}
with the parameters $V_0 = 10^{-10}\MPlanck^4$,
$\eta_R = 1 / \sqrt{3}$,
$g_R = \MPlanck^2 V_0^{-1/2}$,
$\lambda_R = 0.5 \MPlanck^3 \omega^{-1/2} V_0^{-3/4}$
and
$R_0 = (30 \MPlanck^2 / \pi^2)^{1/2}$.
\begin{figure}
    \centering
    \begin{minipage}{0.49\textwidth}
        \centering
        \includegraphics[width=1.06\textwidth]{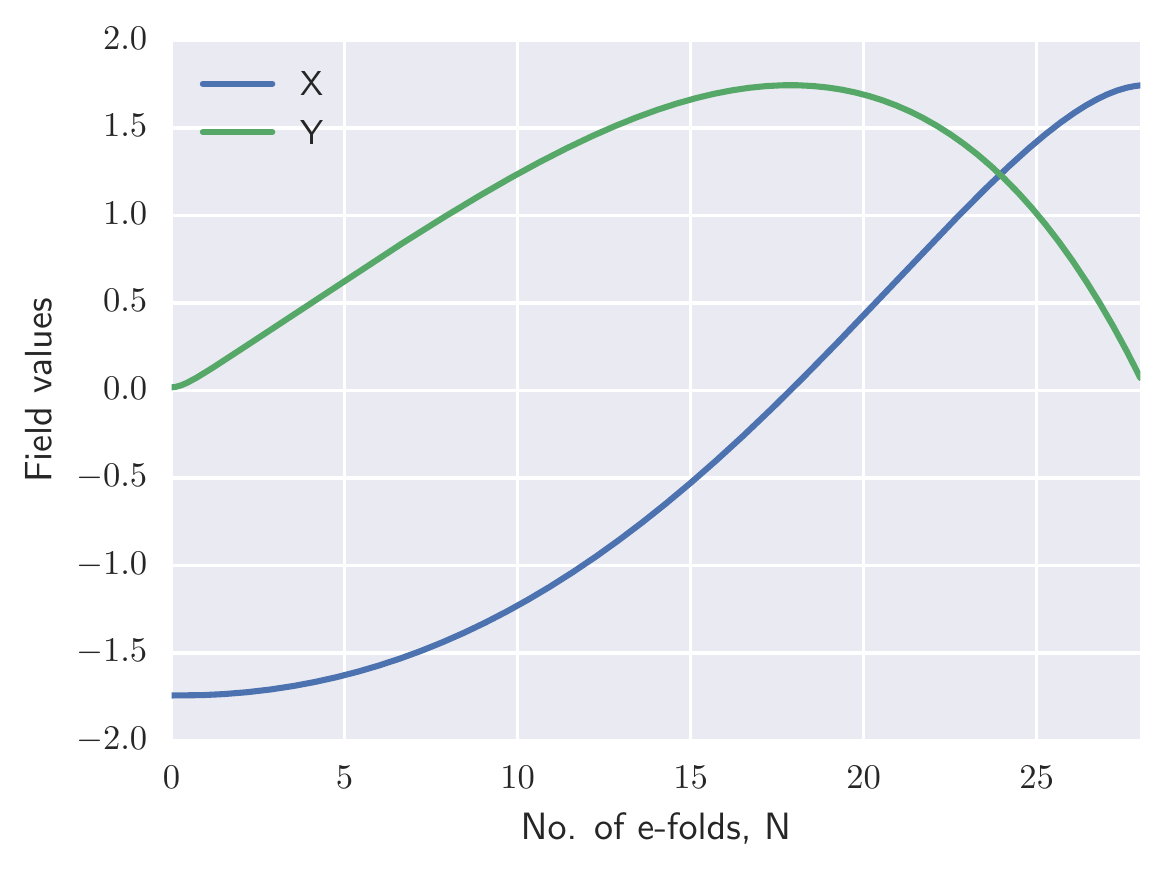} 
    \end{minipage}\hfill
    \begin{minipage}{0.49\textwidth}
        \centering
        \includegraphics[width=1.06\textwidth]{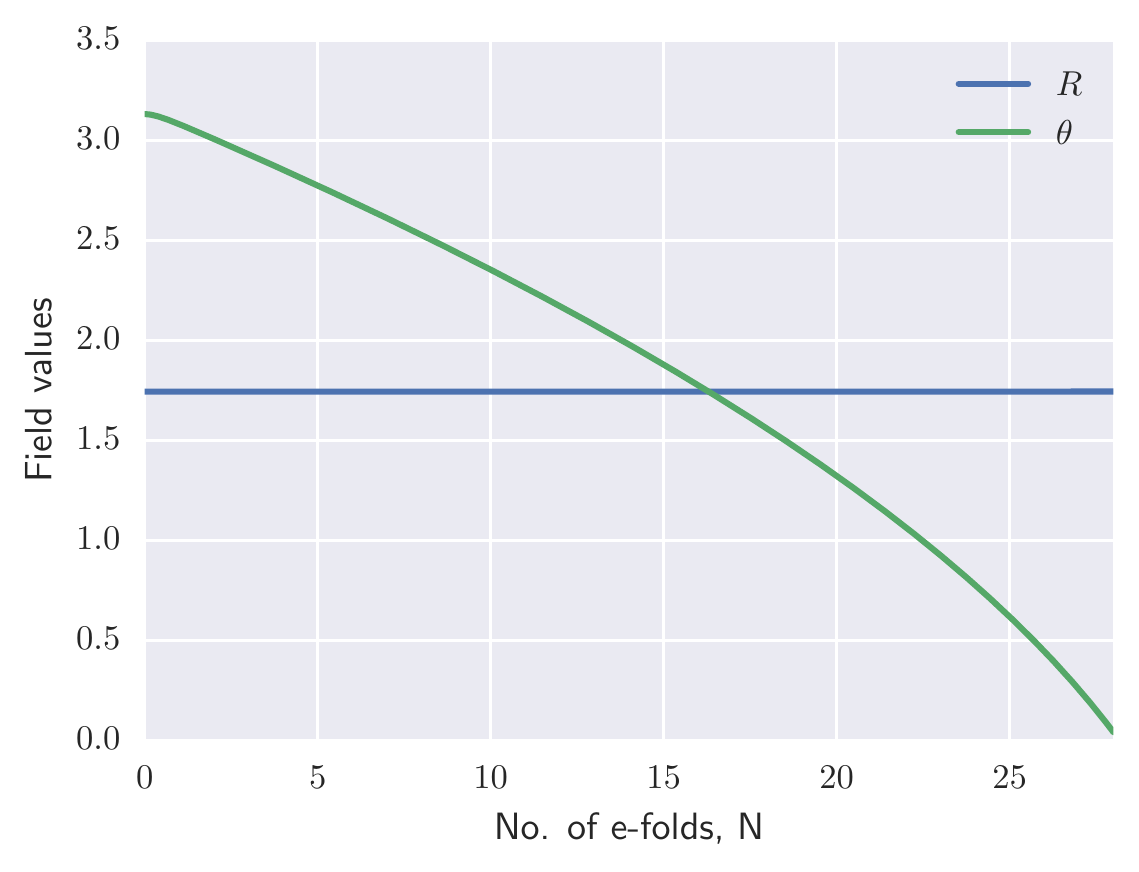} 
    \end{minipage}
    \caption{Field plots for the QSFI/Gelaton model \eqref{eq:QSFI/GelatonPotential} until end of inflation. Left: time evolution of the canonical fields $X$ and $Y$. Right: time evolution of the non-canonical fields $R$ and $\theta$.}\label{Fig:BackGsQSFIgel}
\end{figure}
\begin{figure}
    \centering
    \begin{minipage}{0.49\textwidth}
        \centering
        \includegraphics[width=1.06\textwidth]{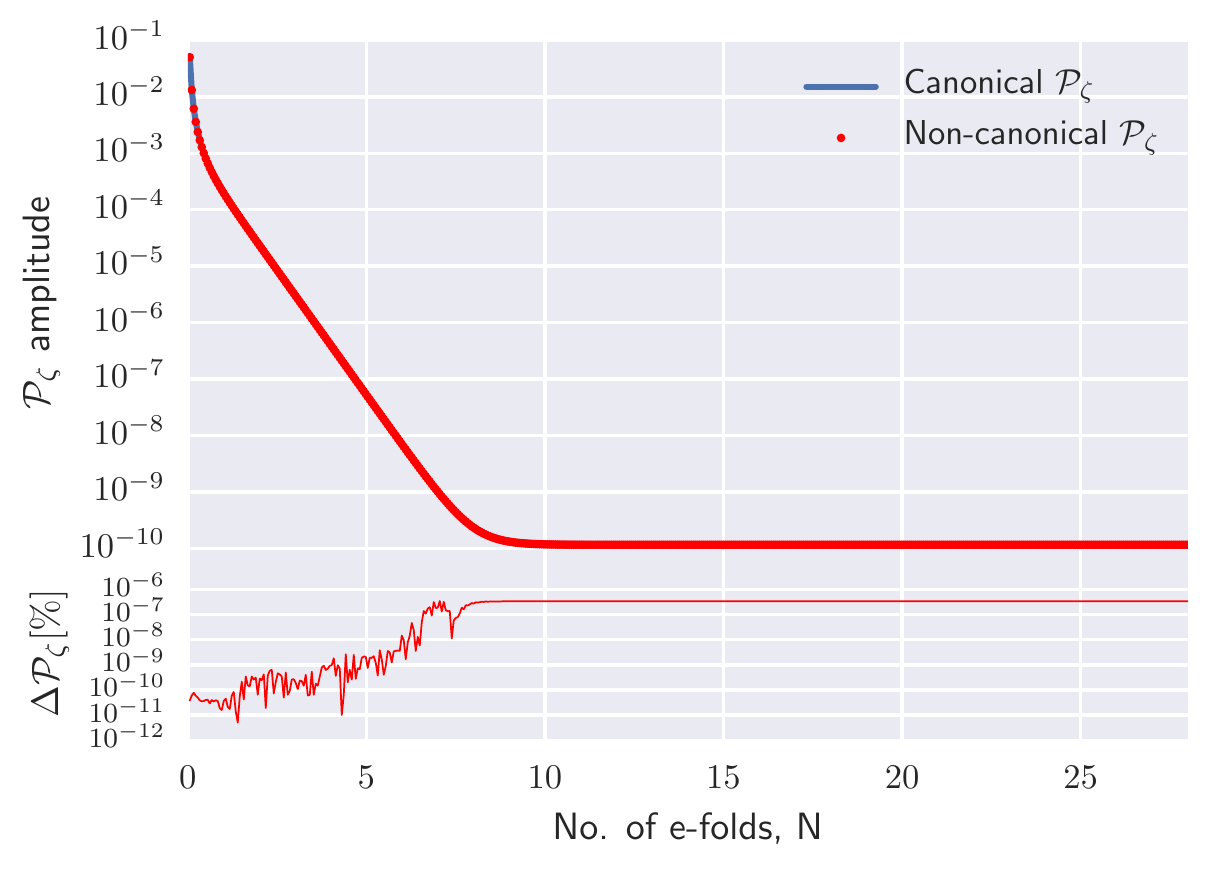} 
    \end{minipage}\hfill
    \begin{minipage}{0.49\textwidth}
        \centering
        \includegraphics[width=1.06\textwidth]{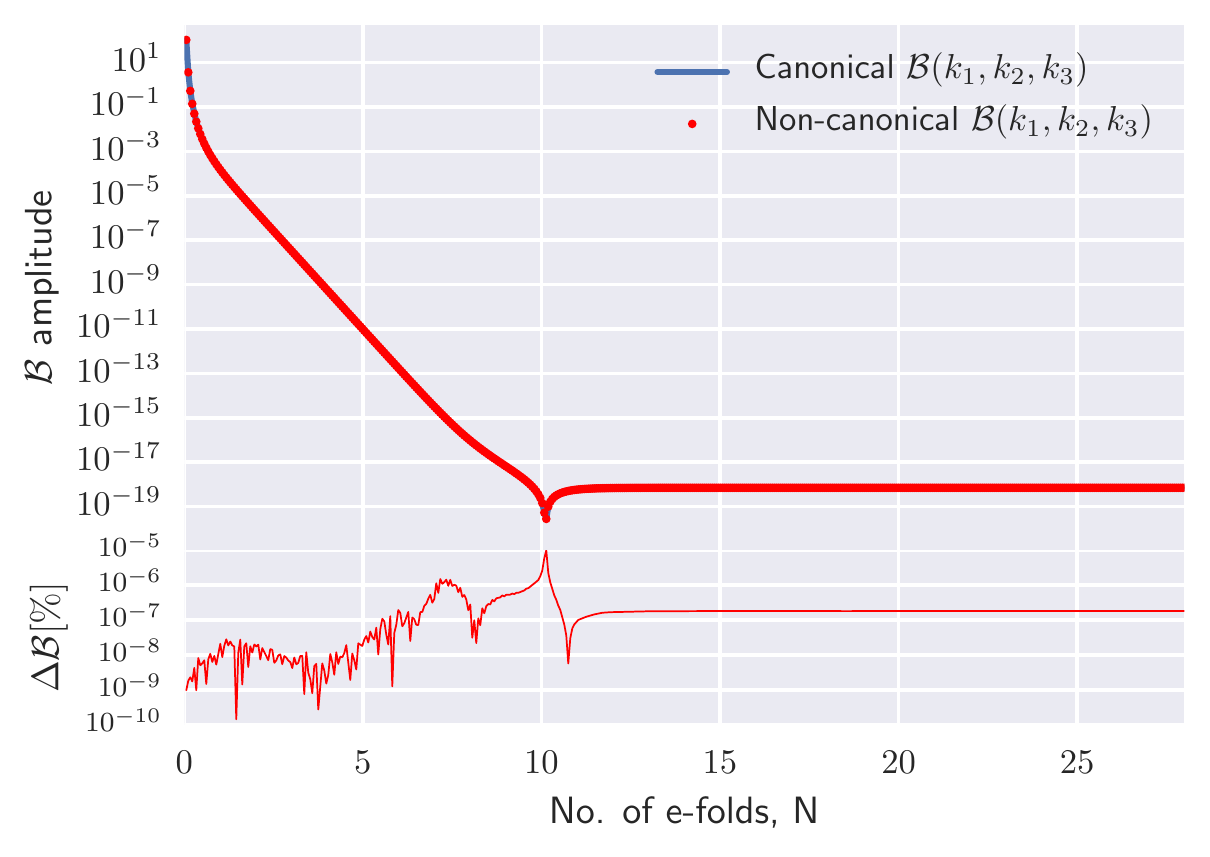} 
    \end{minipage}
    \caption{QSFI/Gelaton residual plots for the dimensionless power spectrum (left) and bispectrum (right) on equilateral configurations for a $k$ and a $k_t$ mode both leaving the horizon at $N = 8.0$ respectively.}\label{Fig:PowSpecsQSFIgel}
\end{figure}

Ref.~\cite{Dias:2016rjq} used initial conditions corresponding to
\begin{subequations}
	\begin{align}
		\label{eq:QSFI/GelatonXinit}
		X_{\mathrm{init}} &= -R_0, \\
		\label{eq:QSFI/GelatonYinit}
		Y_{\mathrm{init}} &= 10^{-2} R_0 .
	\end{align}
\end{subequations}
In polar coordinates these become
\begin{subequations}
	\begin{align}
		\label{eq:QSFI/GelatonRinit}
		R_{\mathrm{init}} &= \sqrt{X_{\mathrm{init}}^2 + Y_{\mathrm{init}}^2}, \\
		\label{eq:QSFI/GelatonThetaInit}
		\theta_{\mathrm{init}} &= \tan^{-1} \left( \frac{Y_{\mathrm{init}}}{X_{\mathrm{init}}} \right).
	\end{align}
\end{subequations}
The background evolution is plotted in Fig.~\ref{Fig:BackGsQSFIgel}.
Inflation lasts for 28 e-folds, and the field evolutions match to high accuracy.

In the left panel of
Fig.~\ref{Fig:PowSpecsQSFIgel} we plot the dimensionless power spectrum of $\zeta$
together with its residual, defined by
$\Delta \mathcal{P} = |\mathcal{P}_{\mathrm{n.can}} - \mathcal{P}_{\mathrm{can}}|/\mathcal{P}_{\mathrm{can}}$.
The results agree to better than $10^{-6} \%$.
The right panel gives a similar comparison for the dimensionless bispectrum,
showing agreement to better than $10^{-5} \%$.

\begin{figure}
    \centering
    \begin{minipage}{0.49\textwidth}
        \centering
        \includegraphics[width=1.06\textwidth]{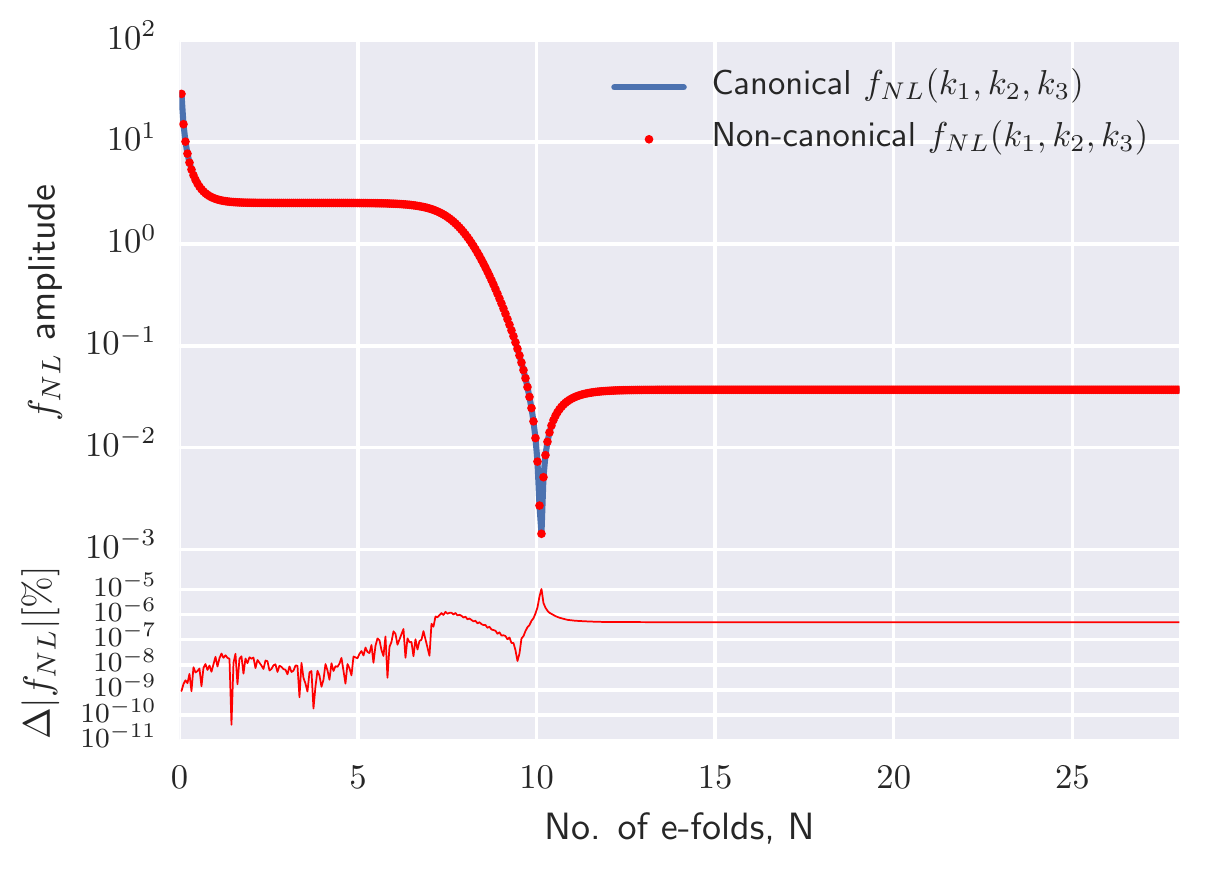} 
    \end{minipage}\hfill
    \begin{minipage}{0.49\textwidth}
        \centering
        \includegraphics[width=1.06\textwidth]{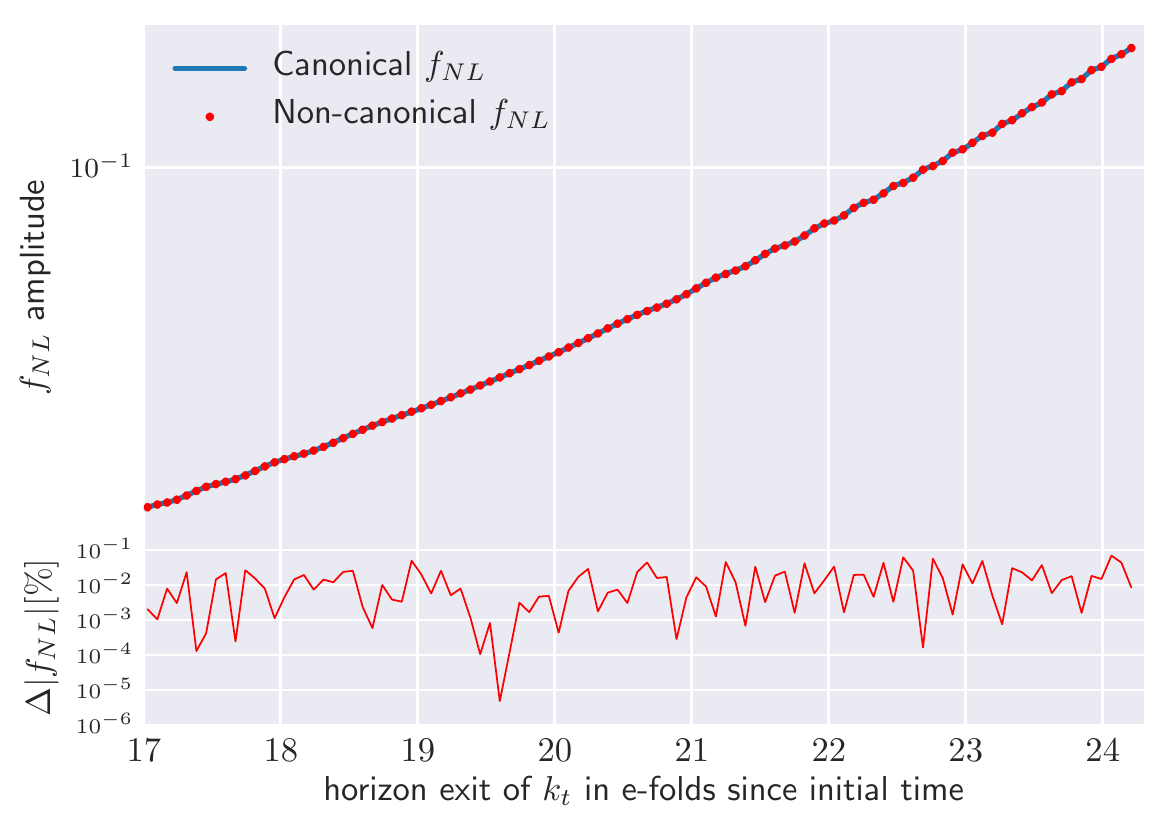} 
    \end{minipage}
    \caption{QSFI/Gelaton residual plots for the reduced bispectrum on equilateral configurations. Left: time evolution of $\fNL$ for a $k_t$ mode leaving the horizon at $N = 8.0$. Right: $k_t$-dependence of $\fNL$ for a range of $k_t$ values leaving between 17.0 and 24.2 e-folds after the initial conditions.}\label{Fig:FNLsQSFIgel}
\end{figure}

In Fig.~\ref{Fig:FNLsQSFIgel} we compare the predicted value of the reduced
bispectrum $\fNL$. The left-hand panel shows its time evolution for a single
Fourier configuration that exits the horizon at 8.0 e-folds. The results agree to
within $10^{-5} \%$, where the largest residual is given during the rapid evolution
of $f_{NL}$ during horizon crossing.

The right panel of Fig.~\ref{Fig:FNLsQSFIgel} shows the values measured
at the end of inflation as a function of wavenumbers that exit the horizon between 17.0
and 24.2 e-folds after the initial conditions are set. Here the residuals are typically 
at the $10^{-2} \%$ level with the maximum residual at 0.07\%. These are different from
the left panel due to the $k_t$ values exiting much later, at a time closer to the 
end of inflation at 28.0 e-folds where
the bispectrum has rapid small-amplitude oscillations.

Despite the $\fNL$ vs. $k_t$ plot having larger residuals,
these results indicate that the non-canonical transport formalism
agrees with its canonical counterpart to within at least $0.1\%$
when applied to this model.

\subsection{Quasi-two-field inflation}
\label{subsec:Quas2FldInflat}
Dias et al.~\cite{Dias:2015rca}
introduced a `quasi-two-field' model
in which two light scalars drive inflation.
One of these fields excites a heavy third field
via a noncanonical kinetic coupling,
giving rise to oscillatory features in the
power spectrum.
This is an extension of a simpler two-field
model suggested by Ach\'{u}carro et al.~\cite{Achucarro:2010da}.
Such oscillatory features have been well-studied
in the literature~\cite{Gao:2012uq,Achucarro:2010da,Achucarro:2013cva,Adshead:2013zfa,Flauger:2016idt}.
The power spectrum was computed using
{\mtransport} by Dias et al.~\cite{Dias:2015rca},
and the bispectrum was computed using
{\pytransport} by Ronayne et al.~\cite{Ronayne:2017qzn},
giving us an opportunity to benchmark
{\cpptransport} against the other transport tools.
Note that this is \emph{not} an empty comparison, because
although all the transport tools use the same underlying framework
they make very different numerical choices in implementation.
\begin{figure}
    \centering
    \begin{minipage}{0.49\textwidth}
        \centering
		\includegraphics[width=1.06\textwidth]{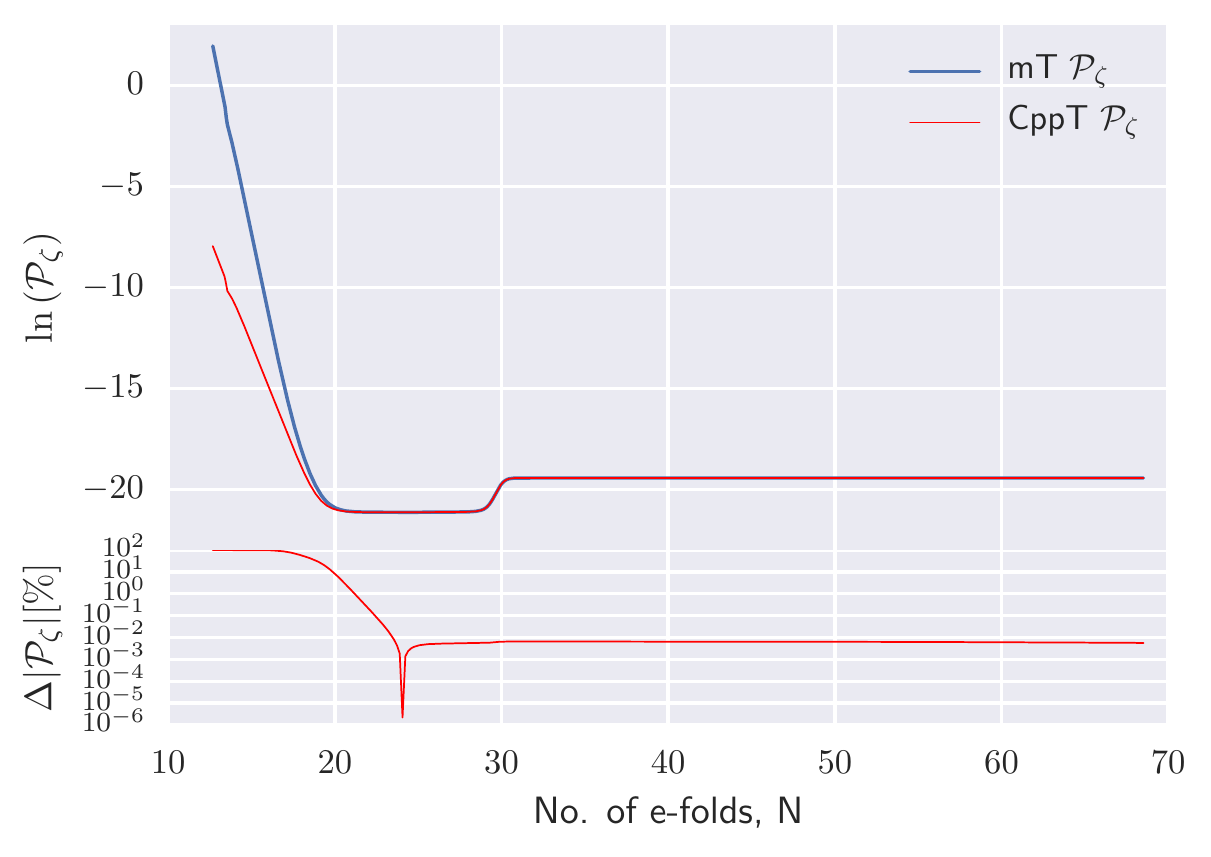} 
    \end{minipage}\hfill
    \begin{minipage}{0.49\textwidth}
        \centering
		\includegraphics[width=1.06\textwidth]{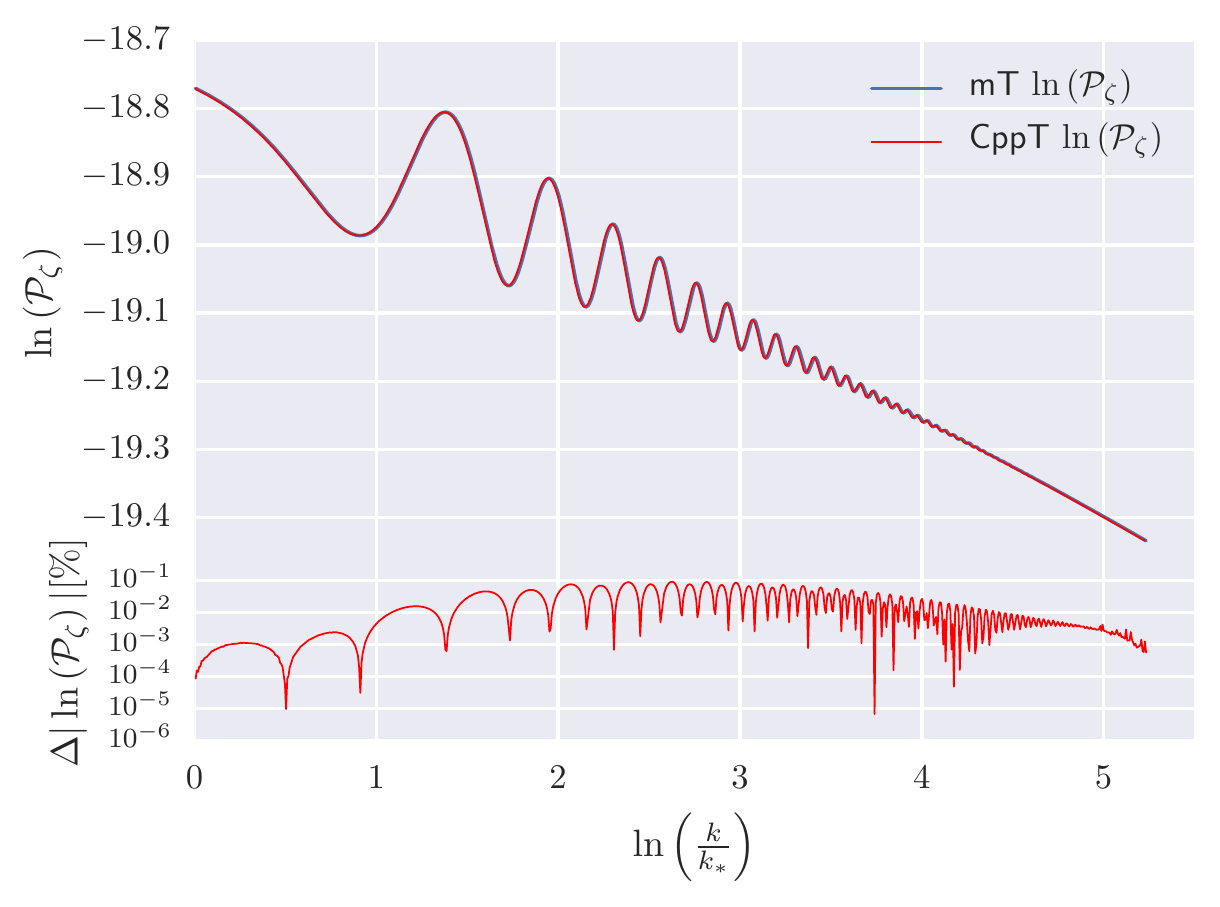} 
    \end{minipage}
    \caption{Power spectrum residuals from the quasi-two-field model. Left: residual as a function of time for the $k$-mode that exits the horizon at 19.0 e-folds from the initial time. Right: residual of $\ln (\mathcal{P}_\zeta)$ as a function of $\ln(k/k_*)$ for a range of $k$ numbers exiting the horizon between 0.0 and 5.2 e-folds after the scale, $k_*$, which exits at $N_* = 13.5$.}\label{Fig:Quas2f_mT_comps}
\end{figure}

The three fields in the model are labelled
$\phi_1$, $\phi_2$ and $\phi_3$, and the field-space metric is
\begin{equation} \label{eq:FieldMetric_Quasi2Field}
	G_{IJ} = 
	\begin{pmatrix}
	  	1 & \Gamma(\phi_1) & 0 \\
	  	\Gamma(\phi_1) & 1 & 0 \\
	  	0 & 0 & 1
	\end{pmatrix}
	,
\end{equation}
where $\Gamma(\phi_1)$ is defined to equal~\cite{Achucarro:2010da}
\begin{equation} \label{eq:GammaFunction_Quasi2Field}
	\Gamma(\phi_1) = \Gamma_0 \Big/ \cosh^2 \frac{2(\phi_1 - \phi_{1(0)})}{\Delta \phi_1}
	,
\end{equation}
where $\Gamma_0 = 0.9$ is the maximum value of $\Gamma(\phi_1)$,
$\phi_{1(0)} = 7\MPlanck$ is the value of $\phi_1$ at the apex of the turn
and $\Delta \phi_1 = 0.12 \MPlanck$ is the range of $\phi_1$ during the turn.
The potential is
\begin{equation} \label{eq:Quasi2FieldPotential}
	V = \frac{1}{2} g_1 m^2 \phi_1^2 + \frac{1}{2} g_2 m^2 \phi_2^2 + \frac{1}{2} g_3 m^2 \phi_3^2,
\end{equation}
with parameters
$g_1 = 30$,
$g_2 = 300$,
$g_3 = 30/81$,
$m = 10^{-6}$.
The initial conditions are
\begin{subequations}
	\begin{align}
		\phi_1^{\text{init}} = 10.0 \MPlanck, \\
		\phi_2^{\text{init}} = 0.01 \MPlanck, \\
		\phi_3^{\text{init}} = 13.0 \MPlanck.
	\end{align}
\end{subequations}

\para{Two-point function}
In this section, we define the residual
between the {\mtransport} and {\cpptransport} power spectrum
as
\begin{equation}
    |\Delta \mathcal{P}|
    =
    \frac{|\mathcal{P}_{\text{CppT}} - \mathcal{P}_{\text{mT}}|}
    {\mathcal{P}_{\text{mT}}} .
\end{equation}
In the left panel of Fig.~\ref{Fig:Quas2f_mT_comps}
we plot the residual as a function of time for the $k$-mode that exits
the horizon $N=19.0$ e-folds from the initial time.
During the superhorizon phase the agreement is typically at
$0.01\%$ or better, except at a small number of points
where the evolution is particularly rapid.

Note that the solutions diverge on \emph{subhorizon} scales.
As explained in Ref.~\cite{Dias:2014msa}, the curvature perturbation
$\zeta$ does not have a unique definition
on subhorizon scales, and the precise value
we assign depends which $k$-dependent terms are kept.
{\mtransport} uses the `local' form of $\zeta$
defined in Ref.~\cite{Dias:2014msa},
whereas {\cpptransport} and {\pytransport}
uses the `simple' form
(which agrees with Eqs.~\eqref{eq:N_a_solution_paper}--\eqref{eq:N_ab_solution_paper}).
At linear level these are~\cite{Dias:2014msa}
\begin{subequations}
\begin{align}
    \label{eq:zeta-local}
    \zeta_\text{local} & = \frac{1}{2H^2 \MPlanck^2 \epsilon(3-\epsilon)}
    \big( \dot{\phi}_I \dot{Q}^I + V_I Q^I \big) 
    \\
    \label{eq:zeta-simple}
    \zeta_\text{simple} & = -\frac{\dot{\phi}_I Q^I}{2H \MPlanck^2 \epsilon} .
\end{align}
\end{subequations}
The `local' form mixes $Q^I$ and $\dot{Q}^I$ whereas the `simple' form involves
only $Q^I$.
Correlation functions involving $\dot{Q}^I$ increase on subhorizon scales more
rapidly than correlation functions of $Q^I$ alone,
which accounts for the different time-dependence visible
in Fig.~\ref{Fig:Quas2f_mT_comps} on subhorizon scales.
The discrepancy is harmless.
On superhorizon scales the two forms agree to high accuracy, as they should.

Although this difference means that the $\zeta$ correlation functions cannot be
compared directly on subhorizon scales, we have verified that the
field correlation functions
(which are unambiguous)
agree to 5 significant figures.

In the right panel of Fig.~\ref{Fig:Quas2f_mT_comps}
we plot the residuals as a function of scale
for a range of $k$-modes exiting the horizon up to
5.3 e-folds from the pivot scale.
The residuals remain below $0.1\%$ over the whole range.
This shows excellent agreement between {\mtransport} and
{\cpptransport}
despite the rapid oscillations visible in the power spectrum.

\para{Three-point function}
To compare 3-point functions we use the latest
version of {\pytransport}~\cite{Ronayne:2017qzn}.
For each measure $X$ of 3-point correlations we
define the residual
$|\Delta X| = |X_{\text{CppT}} - X_{\text{PyT}}| / X_{\text{CppT}}$.

\begin{figure}
    \centering
    \begin{minipage}{0.49\textwidth}
        \centering
        \includegraphics[width=1.06\textwidth]{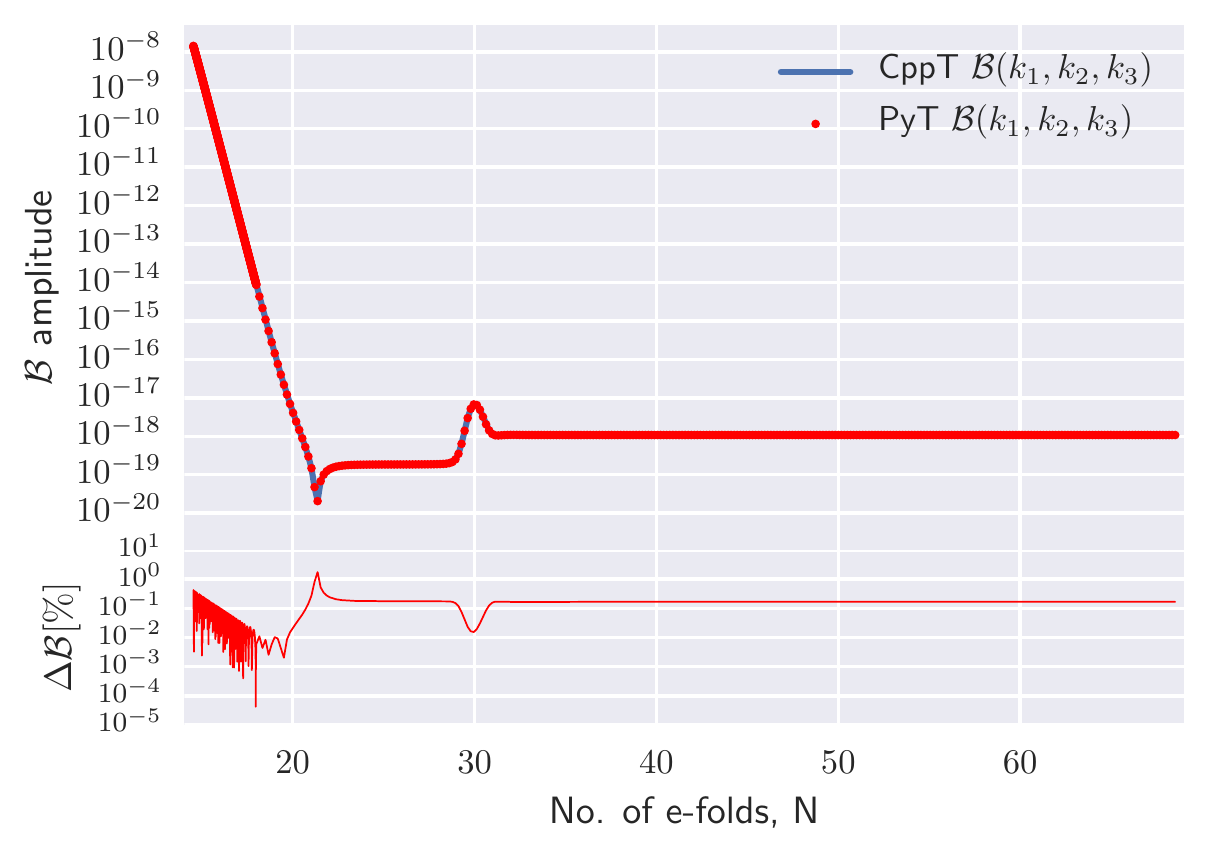} 
    \end{minipage}\hfill
    \begin{minipage}{0.49\textwidth}
        \centering
        \includegraphics[width=1.06\textwidth]{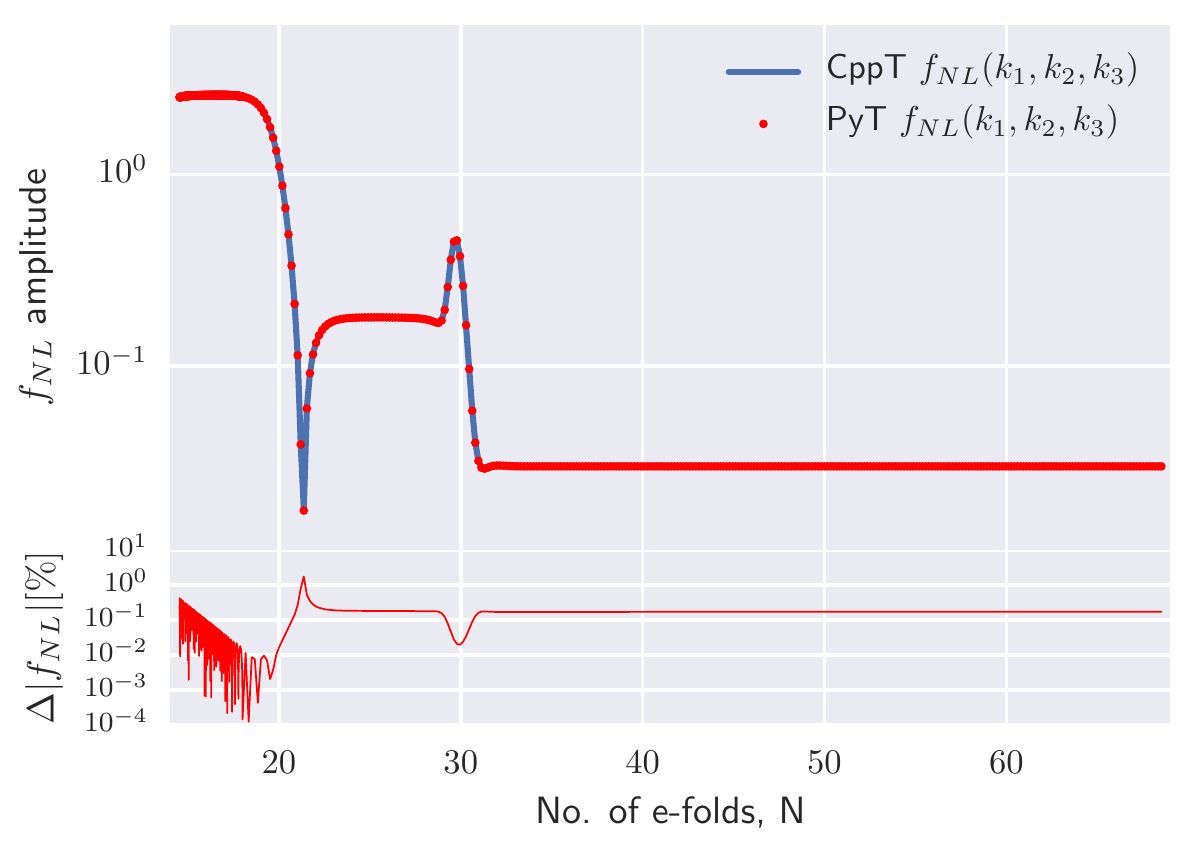} 
    \end{minipage}
    \caption{Quasi-two-field residual time-plots for three-point functions on equilateral configurations. Left: dimensionless bispectrum, $\mathcal{B}$, for a $k_t$ value that exits the horizon at $N_\text{exit} = 19.9$ plotted against time. Right: reduced bispectrum, $\fNL$, plotted against time for the same $k_t$ and $N_\text{exit}$ values.}\label{Fig:Quas2f_3pf_time}
\end{figure}

In the left-hand panel of Fig.~\ref{Fig:Quas2f_3pf_time} we plot the residual
of the dimensionless bispectrum
as a function of time
for an equilateral configuration
where $k_t$ exits the horizon roughly 20 e-folds after the
initial time.
Our results agree at roughly $0.3\%$ through most of the evolution, with short-lived
excursions to larger values at times of rapid evolution.
In the right-hand panel we give an equivalent plot for the reduced bispectrum
$\fNL$.
We conclude that the variation in numerical results between any two of the
transport tools is negligible in comparison with current experimental errors.

\begin{figure}
    \centering
    \begin{minipage}{0.49\textwidth}
        \centering
        \includegraphics[width=1.06\textwidth]{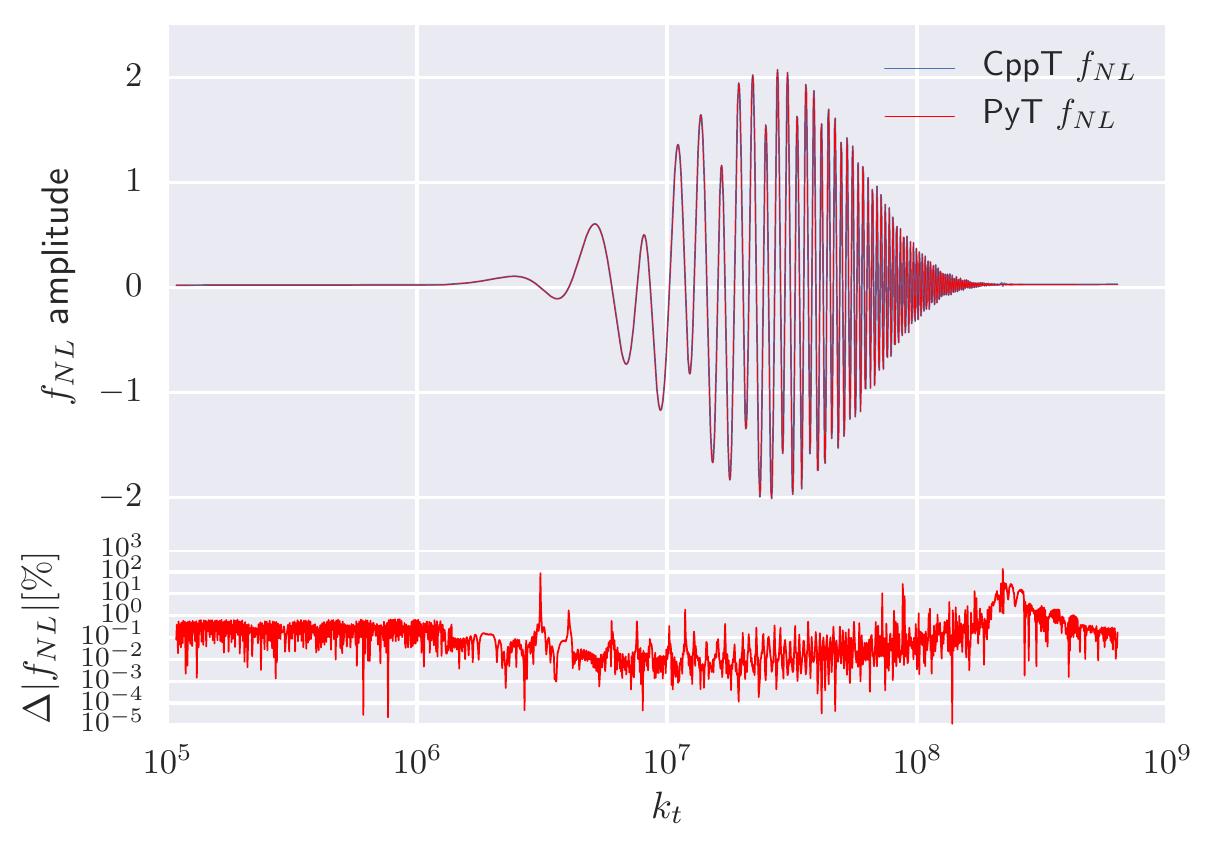} 
    \end{minipage}\hfill
    \begin{minipage}{0.49\textwidth}
        \centering
        \includegraphics[width=1.06\textwidth]{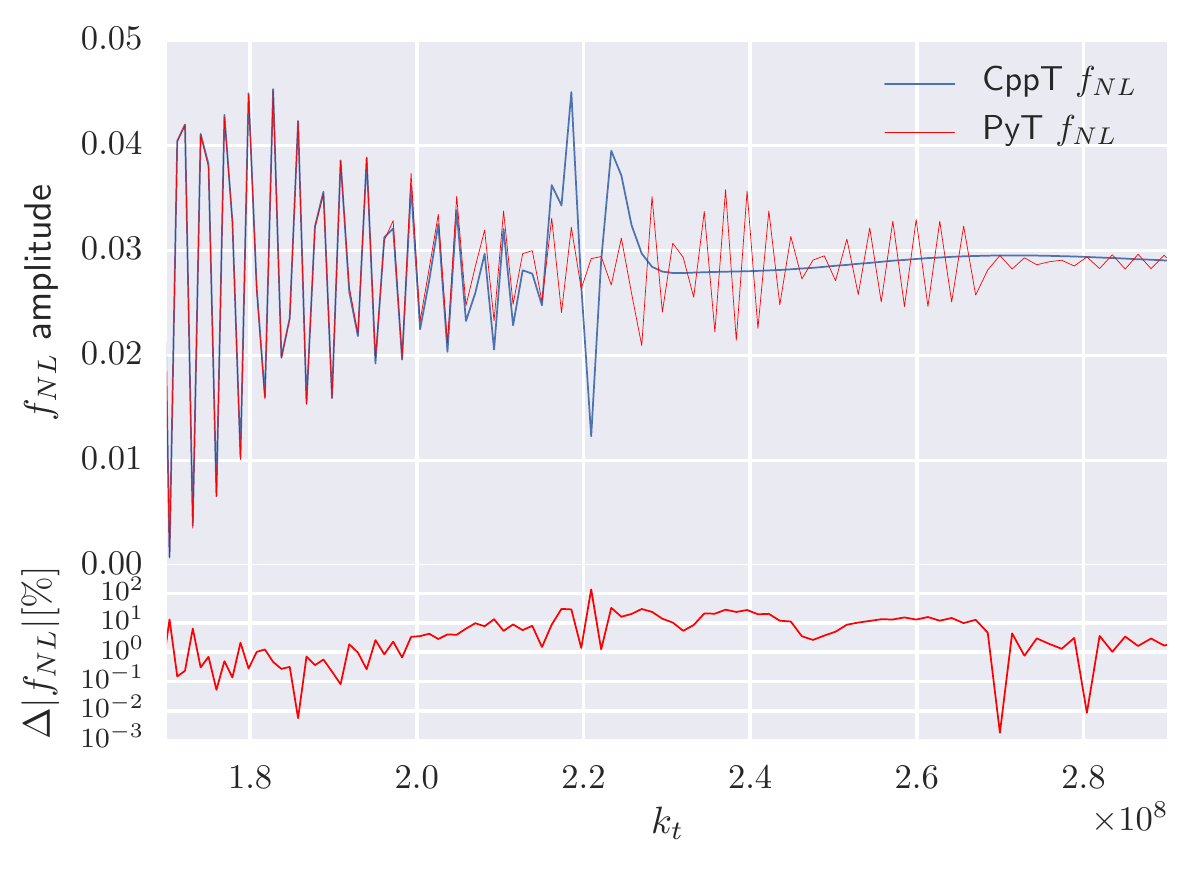} 
    \end{minipage}
    \caption{Reduced bispectrum residuals in equilateral $k_t$ space where $k_*=1$ and $N_{*\text{exit}} = 0.1$. Left: residuals for the reduced bispectrum $\fNL$ plotted against a range of $k_t$ values exiting the horizon between 10.9 \& 19.9 e-folds after inflation begins. Right: zoom-in of the largest residual at $k_t \approx 2.2 \times 10^8$.}\label{Fig:Quas2f_3pf_k}
\end{figure}

The left panel of Fig.~\ref{Fig:Quas2f_3pf_k} shows the residual of the reduced
bispectrum
as a function of $k_t$ for scales exiting the horizon between 10.9 and 19.9 e-folds
after the initial time. Agreement between {\cpptransport} and {\pytransport}
is at the level $\leq 1\%$ over almost the entire range of $k_t$, despite the extremely
rapid oscillations visible in the range $10^7 \lesssim  k_t \lesssim 10^8$.
In the right panel we show a zoomed-in section highlighting the region of
most significant disagreement.
The cause of the discrepancy is currently under investigation.
This is the only model we have encountered
in which our codes show a small disagreement of this kind.

\begin{figure}
	\centering
	\begin{minipage}{0.49\textwidth}
		\centering
		\includegraphics[width=1.06\textwidth]{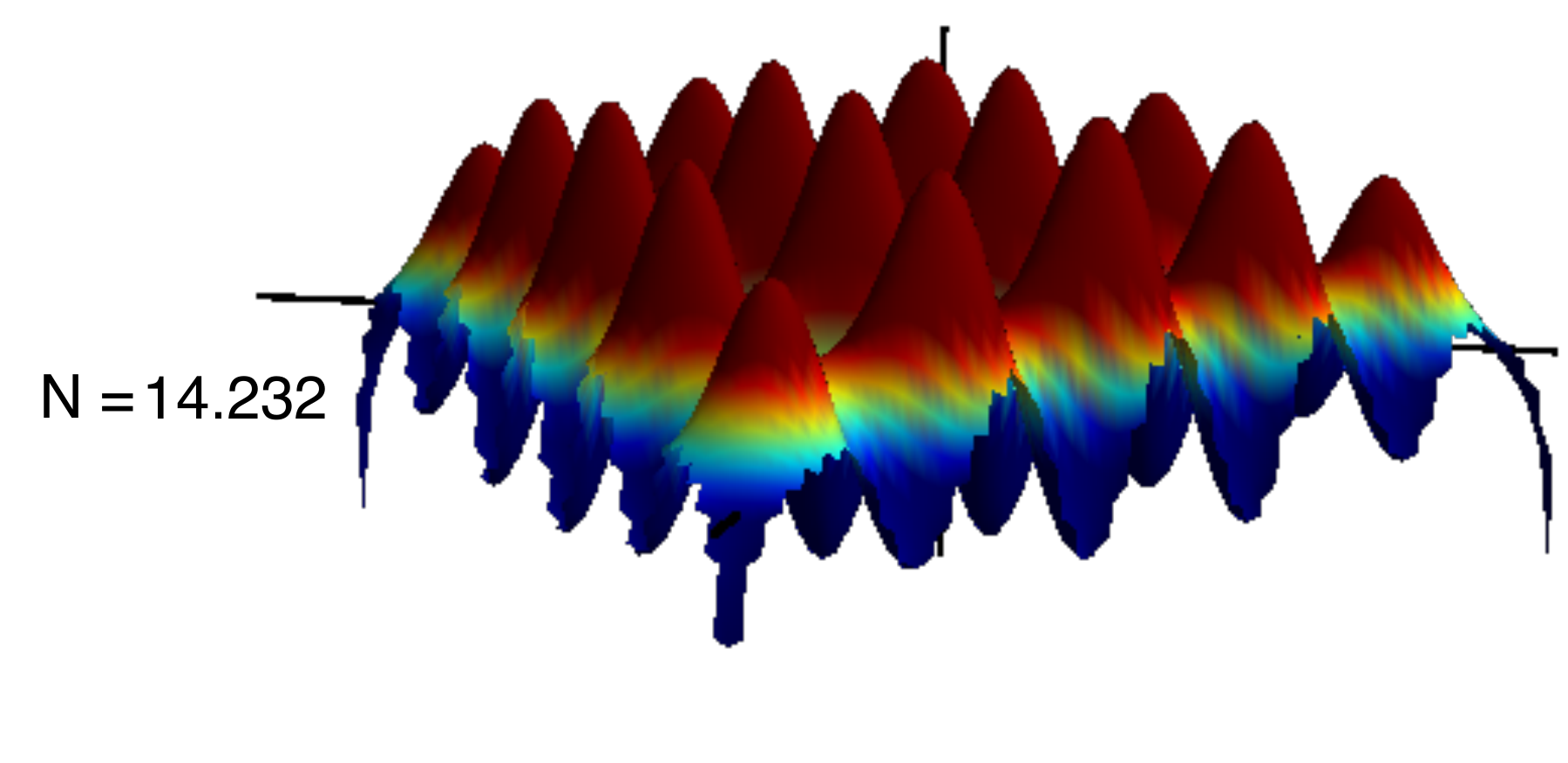}
	\end{minipage}\hfill
	\begin{minipage}{0.49\textwidth}
		\centering
		\includegraphics[width=1.06\textwidth]{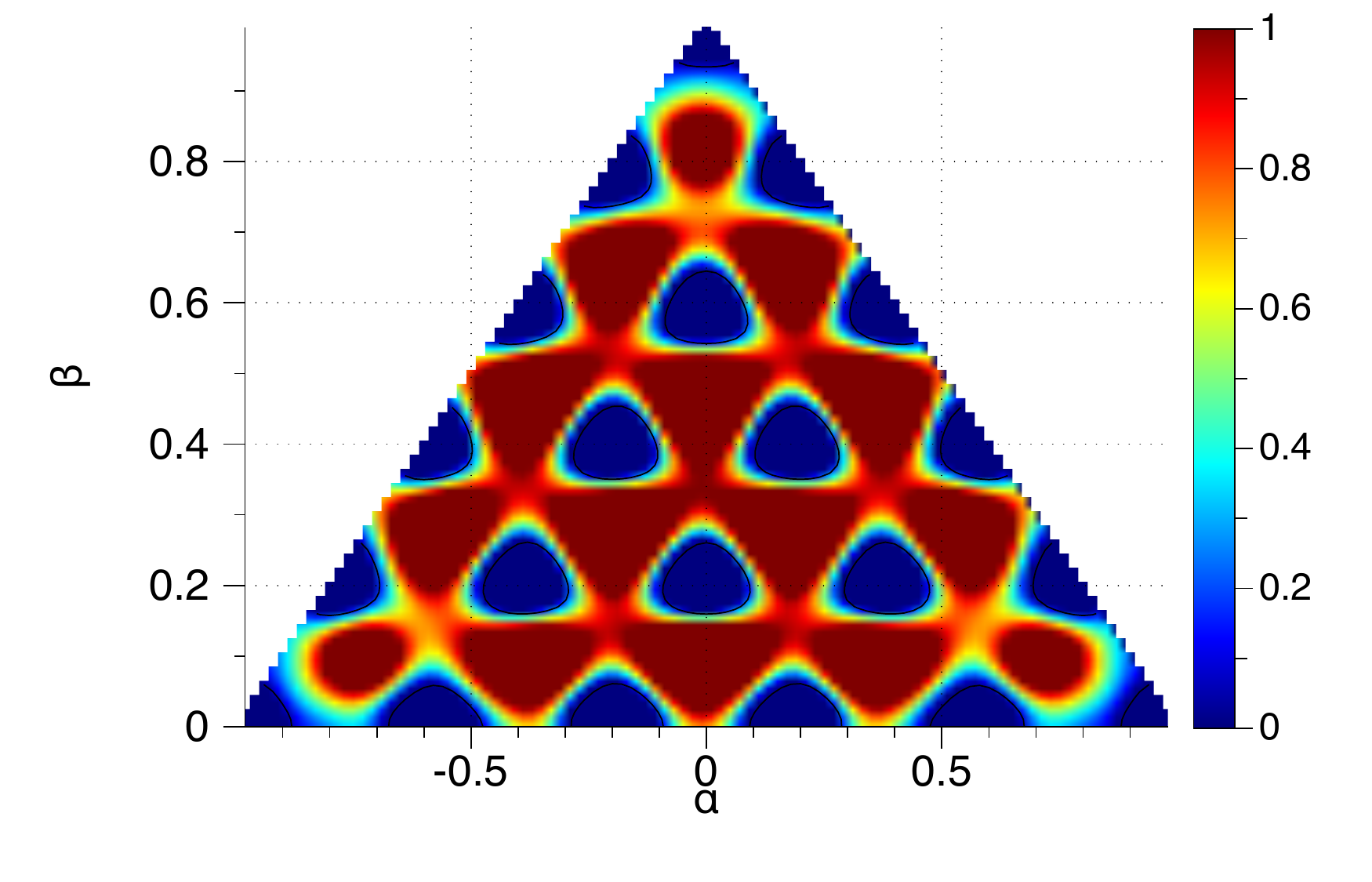}
	\end{minipage}
	\caption{Shape plots for the quasi-two-field model. Left: 3D surface plot of the dimensionless bispectrum, $\mathcal{B}(\alpha, \beta)$, taken at N = 14.232 e-folds for a range of shapes with $-0.98 \leq \alpha \leq 0.98$ and $0 \leq \beta \leq 0.99$ and a fixed $k_t$ mode with $N_\textrm{exit} = 16.6$ e-folds. Right: 2D contour plot for the same values.}\label{Fig:Quasi2F_Shapes}
\end{figure}

\para{Shape plots}
Up to this point we have focused on the bispectrum amplitude as a function of time or scale,
but important information is also encoded in the shape regarding the type of interactions
that appear in the Lagrangian.
In Fig.~\ref{Fig:Quasi2F_Shapes} we show the dimensionless bispectrum as a function of $\alpha$ and $\beta$ at
fixed $k_t$, rescaled to have unit amplitude on the equilateral configuration~\cite{Fergusson:2006pr}.
We choose $k_t$ so that the wavenumber characterizing this configuration exits the
horizon 16.6 e-folds after the initial conditions, and the plots depict the shape given 14.232 e-folds
after the initial conditions. In the left panel we show the amplitude as a surface plot with
the $z$-height representing the (rescaled) bispectrum amplitude, and in the right panel we give
a corresponding contour plot.

At the time given in Fig.~\ref{Fig:Quasi2F_Shapes}, the shape 
shows 15 separate peaks that have evolved from
an equilaterally-dominated bispectrum with a single
maximum at the equilateral configuration.
During the subhorizon phase of inflation, each region of the shape continuously subdivides,
generating further peaks.
The subdivision continues until horizon-crossing at $N \approx 17$, after which
8 peaks have formed along each side of the shape plot.
The bispectrum shape is briefly re-excited during the turn at
$N=30$ e-folds before settling to a constant
value until the end of inflation.
This behaviour is best seen in our video of the surface plot evolution
available on \href{https://vimeo.com/user81717348/quasi-2-field-inflation-shape}{Vimeo}.

\subsection{The gelaton model}
\label{subsec:Gelaton}
We now apply our tools to a new example: the gelaton model introduced by
Tolley \& Wyman~\cite{Tolley:2009fg}.
In this model a heavy gelaton field, with a mass $m \gtrsim H$, is strongly coupled to a light field
and dresses its excitations.
This causes the light field's dynamics to be modified.
Tolley \& Wyman modelled this behaviour using an action
with nontrivial kinetic mixing,
\begin{equation} \label{eq:gelaton_action}
	S =
	\frac{1}{2}
	\int \d^4 x \sqrt{-g}
	\left[
		\MPlanck^2 R
		- \partial_\mu \phi \partial^\mu \phi
		- \e{2b(\phi)} \partial_\mu \chi \partial^\mu \chi
		- V(\phi, \chi) \right].
\end{equation}
Here $\phi$ is the gelaton,
$\chi$ is the inflaton,
and we can see that the field-space metric is given by:
\begin{equation} \label{eq:GelatonModel_Metric}
	G_{IJ} = 
	\begin{pmatrix}
		1 & 0 \\
		0 & \e{2b(\phi)}
	\end{pmatrix}.
\end{equation}
The function $b(\phi)$ is chosen so that the effective mass of the
gelaton is much larger than $H$,
ensuring that it remains at the minimum of its effective potential.
This is displaced from the minimum of the bare potential
$V(\phi,\chi)$ due to the kinetic coupling.
We label the true minimum $\phi_0$, which should be determined by the
condition that the $\phi$ field is in static equilibrium,
\begin{equation} \label{eq:Gelaton_b(phi)_eq}
	V_{,\phi} (\phi_0, \chi) - 2b_{,\phi}(\phi_0) e^{2b(\phi_0)} X = 0,
\end{equation}
where $X = - \frac{1}{2} (\partial \chi)^2$ is the kinetic energy
of $\chi$.
After integrating out $\phi$ from the action~\eqref{eq:gelaton_action}
it can be shown that the resulting low-energy theory is equivalent to a $P(X,\chi)$ model~\cite{Tolley:2009fg}
in which the action is an arbitrary function of $X$ and $\chi$.
Expanding the low-energy action to second order shows that the dressed
$\chi$ fluctuations propagate with phase velocity
\begin{equation} \label{eq:GelatonSpeedSound}
	c_s^2 = \bigg( 1 + \frac{4e^{2b} (b_{,\phi})^2 \dot{\chi}^2}{\mgel^2} \bigg)^{-1},
\end{equation}
where $\mgel$ is the effective gelaton mass.
It is known that $P(X,\chi)$ models in which the speed of sound is significantly different
from unity give enhanced three-point correlations on equilateral
configurations~\cite{Seery:2005wm,Chen:2006nt,Silverstein:2003hf,Alishahiha:2004eh}.
The gelaton model will exhibit such a phenomenology if the speed of sound
can be depressed significantly below unity, $c_s \ll 1$,
while keeping the gelaton mass large, $\mgel \gtrsim H$.

\para{DBI potential}
We now specialize to the `hyperbolic manifold' scenario suggested by Tolley \& Wyman
in which $b(\phi) = g \phi / \MPlanck$.
With this choice, the dynamics of DBI inflation can be replicated by adopting the following
potential
\begin{equation} \label{eq:V_DBI_Potential}
	V_\text{DBI}(\phi, \chi) = T(\chi) \cosh \left( \frac{2g\phi}{\MPlanck} \right) - T(\chi) + W(\chi),
\end{equation}
where $g = 0.43$ is a free parameter used to adjust the gelaton mass,
$T(\chi)$ is the brane tension in the DBI interpretation, and
$W(\chi)$ is a potential representing interactions between the brane and other degrees of
freedom in the geometry.
The gelaton mass is
\begin{equation} \label{eq:GelatonFieldMass}
	\mgel^2 = 4g^2 \MPlanck^{-2} T(\chi) \exp \left( - \frac{2g \phi}{\MPlanck} \right).
\end{equation}
To fix the model we must specify $T(\chi)$ and $W(\chi)$.
We adopt
\begin{subequations}
	\begin{align} \label{eq:Gel_T_W}
		T(\chi) &= \frac{1}{2} \lambda^2 \chi^2, \\
		W(\chi) &= \Lambda^4 - \frac{1}{2} m^2 \chi^2,
	\end{align}
\end{subequations}
where $\lambda = 0.001 \MPlanck$,
$\Lambda = 0.005 \MPlanck$,
and $m = 10^{-5} \MPlanck$.
The potential $W(\chi)$ is chosen to keep the expectation value of $\chi$ sub-Planckian.
It can be assumed to be representative of any hilltop potential provided $\chi$
does not become too large.

The initial conditions for the two fields are
$\phi_{\mathrm{init}} = 1 \times 10 ^{-3} \MPlanck$
and
$\chi_{\mathrm{init}} = 1 \times 10^{-4} \MPlanck$ respectively.
\begin{figure}
    \centering
    \begin{minipage}{0.49\textwidth}
        \centering
        \includegraphics[width=1.06\textwidth]{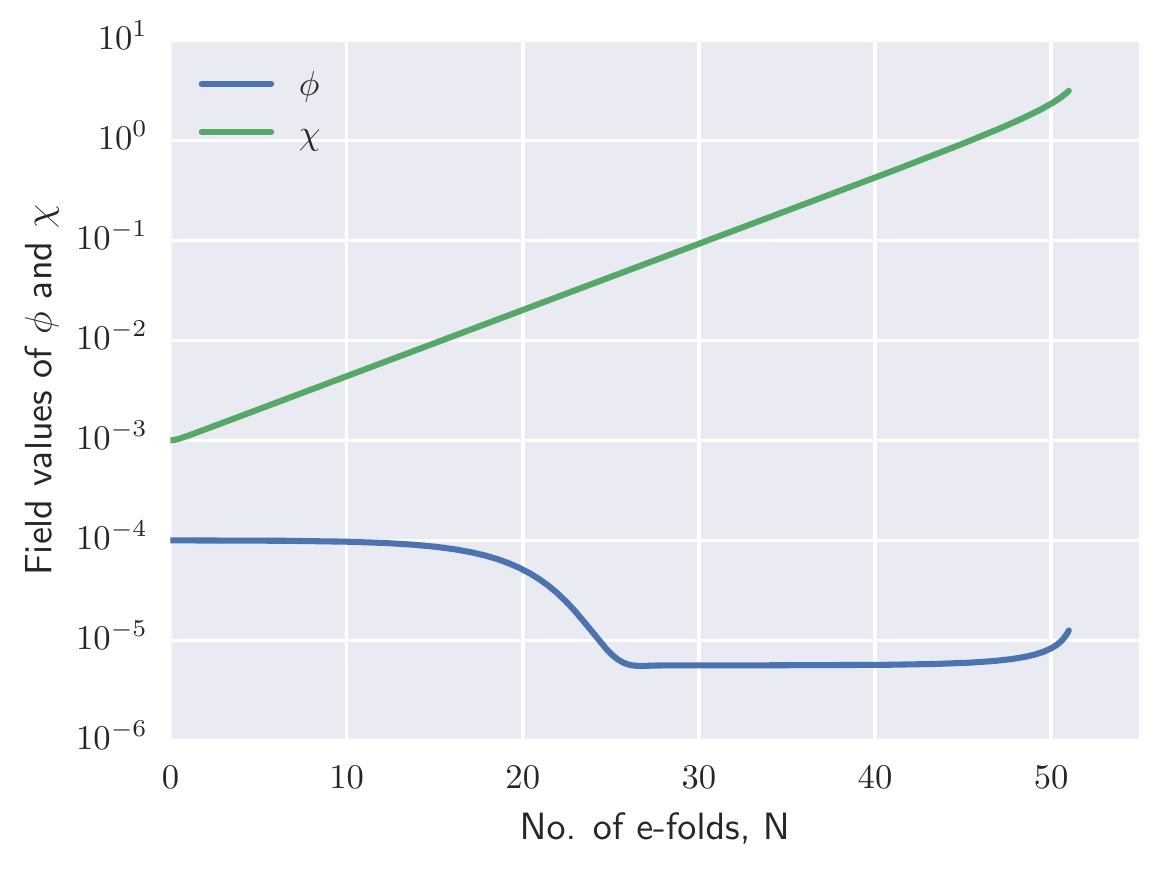} 
    \end{minipage}\hfill
    \begin{minipage}{0.49\textwidth}
        \centering
        \includegraphics[width=1.06\textwidth]{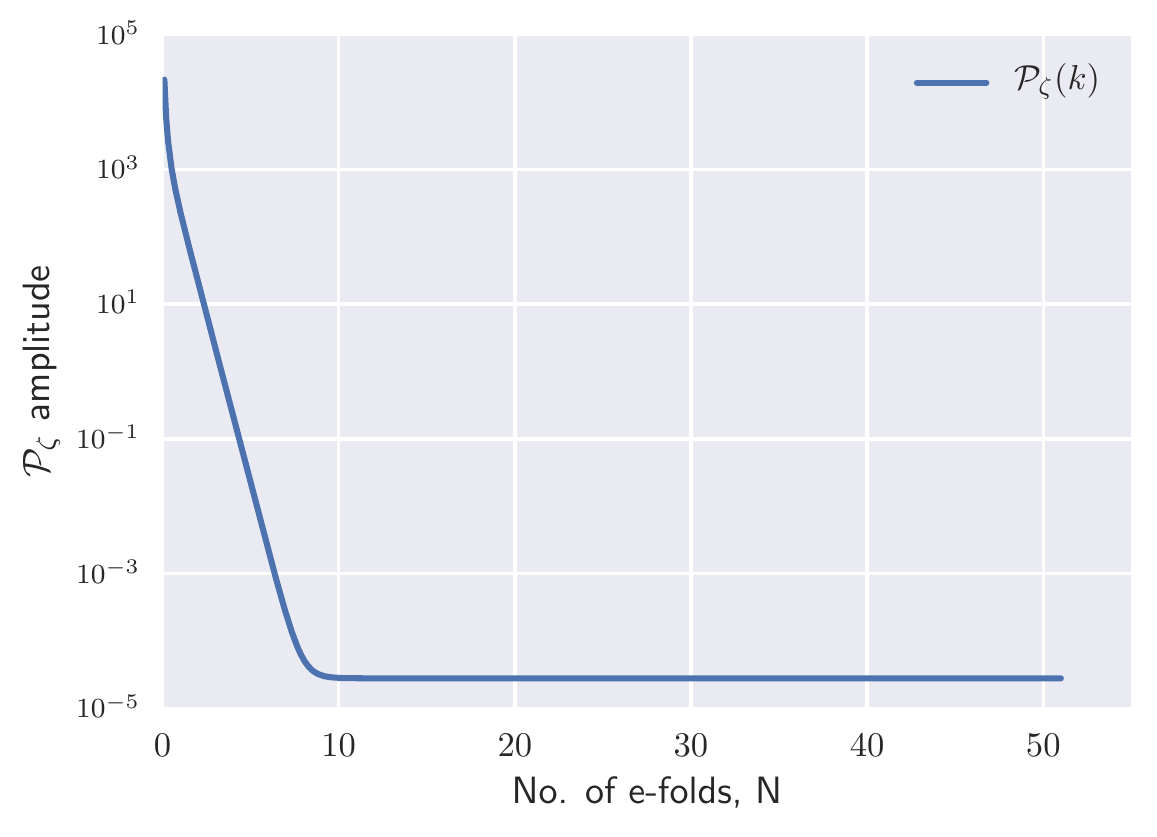} 
    \end{minipage}
    \caption{Background fields and power spectrum for the gelaton model. Left: e-fold evolution of fields $\phi$ and $\chi$ with inflation ending at $N=51$. Right: dimensionless power-spectrum $\mathcal{P_{\zeta}}$ for a $k$ mode exiting the horizon 8.0 e-folds after the initial conditions.}\label{Fig:Gelaton_backg_2pf}
\end{figure}

\begin{figure}
    \centering
    \begin{minipage}{0.49\textwidth}
        \centering
        \includegraphics[width=1.06\textwidth]{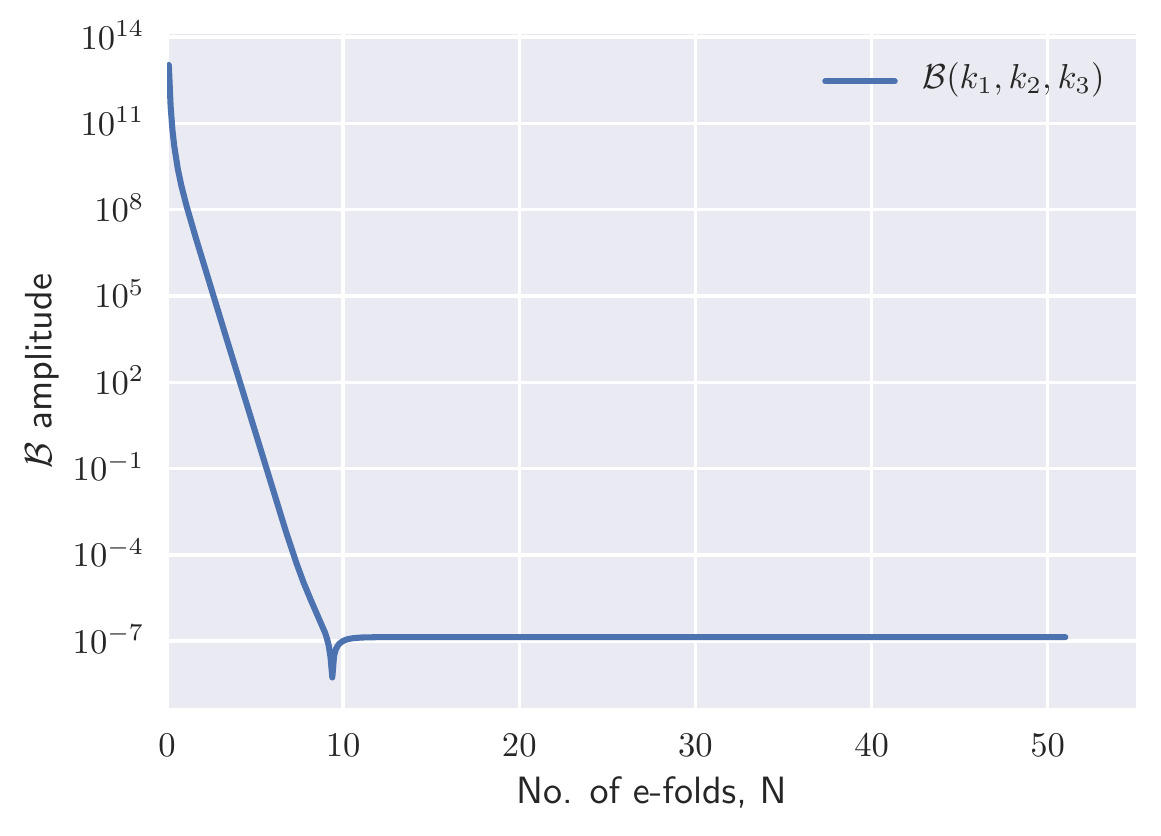} 
    \end{minipage}\hfill
    \begin{minipage}{0.49\textwidth}
        \centering
        \includegraphics[width=1.06\textwidth]{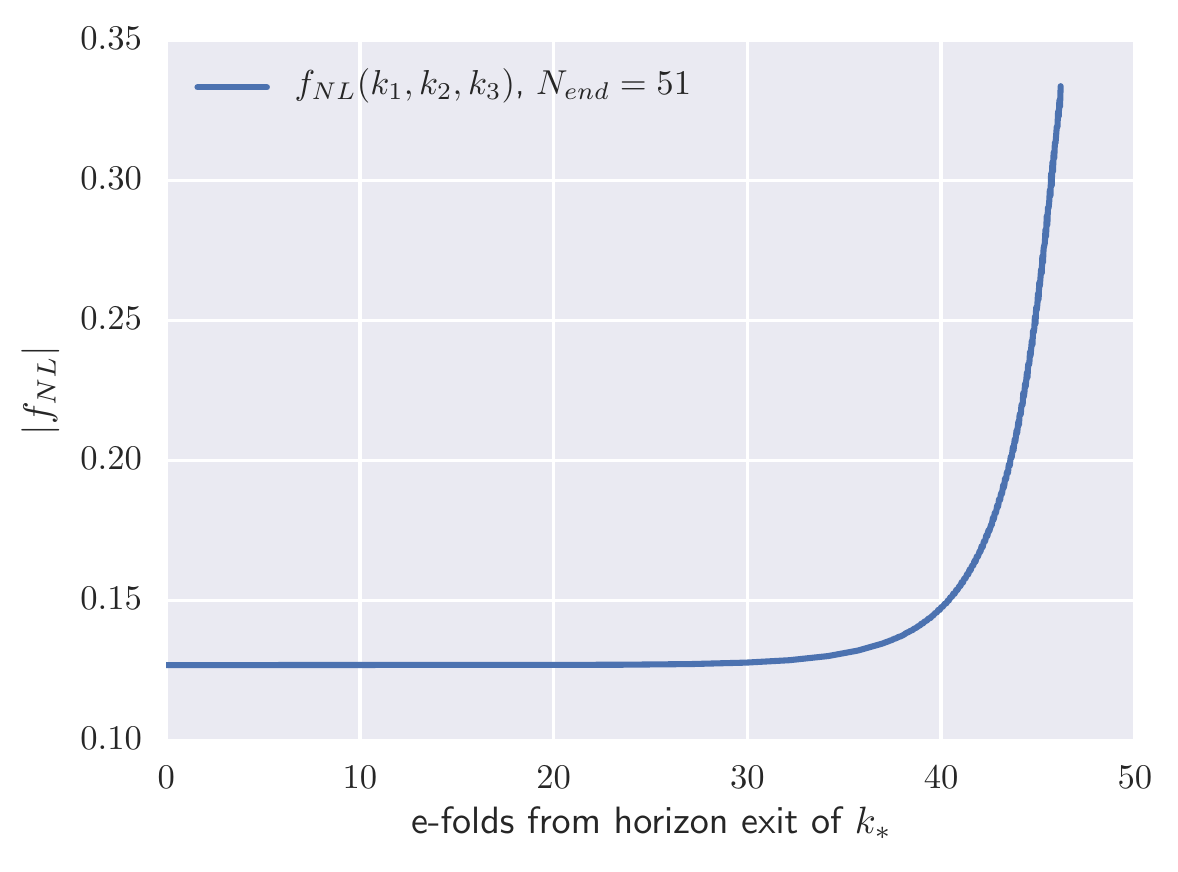} 
    \end{minipage}
    \caption{Left: dimensionless bispectrum $\mathcal{B}$ for an equilateral configuration where $k_t = 3$ with each individual $k$ mode exiting the horizon at $N = 8.0$. Right: reduced bispectrum $\fNL$ plotted against a range of $k$ values exiting the horizon between 0-46 e-folds after the scale $k_*$ with $N_{*\textrm{exit}} = 3.0$.}\label{Fig:Gelaton_3pfs}
\end{figure}

\para{Results}
We perform numerical computations with the full two-field model, to determine
whether the low-energy effective description containing only the
dressed light fluctuation is an accurate representation of the dynamics.
We find very good agreement between our numerical results and the
predictions of the low-energy effective theory.

In the left panel of Fig.~\ref{Fig:Gelaton_backg_2pf}
we plot the evolution of the background fields from their initial values at 
$N = 0$ until the end of inflation at $N_{\textrm{end}} = 51$.
At early times the evolution of $\chi$ is dominated by its kinetic coupling.
The $\phi$ field is driven by the $\cosh$ term in $V_{\text{DBI}}$.
In the right panel we show the evolution of the power spectrum for a single $k$-mode leaving the
horizon at $N=8.0$.
It exhibits smooth decay inside the horizon and asymptotes to a constant value on superhorizon
scales, as it should for an effectively single-field model.

In the left panel of Fig.~\ref{Fig:Gelaton_3pfs}
we plot the evolution of the dimensionless bispectrum for an equilateral configuration
with fixed $k_t$
corresponding to horizon exit at a time $N_{\text{exit}} = 8$.
In the right panel we show the reduced bispectrum $\fNL$
evaluated on equilateral configurations
as a function of scale, for a range of $k_t$ exiting the horizon
between $N=0$ and $N=46$ e-folds after the scale $k_*$ which exits 3.0 e-folds after the initial time.
We see that, with this choice of parameters, the enhancement of equilateral configurations
is only modest, yielding $|\fNL| \approx 0.13$ for a large range of $k$ before the
end of inflation causes $|\fNL|$ to grow slightly as $\epsilon$ increases.
\begin{figure}
    \centering
    \begin{minipage}{0.49\textwidth}
        \centering
        \includegraphics[width=1.06\textwidth]{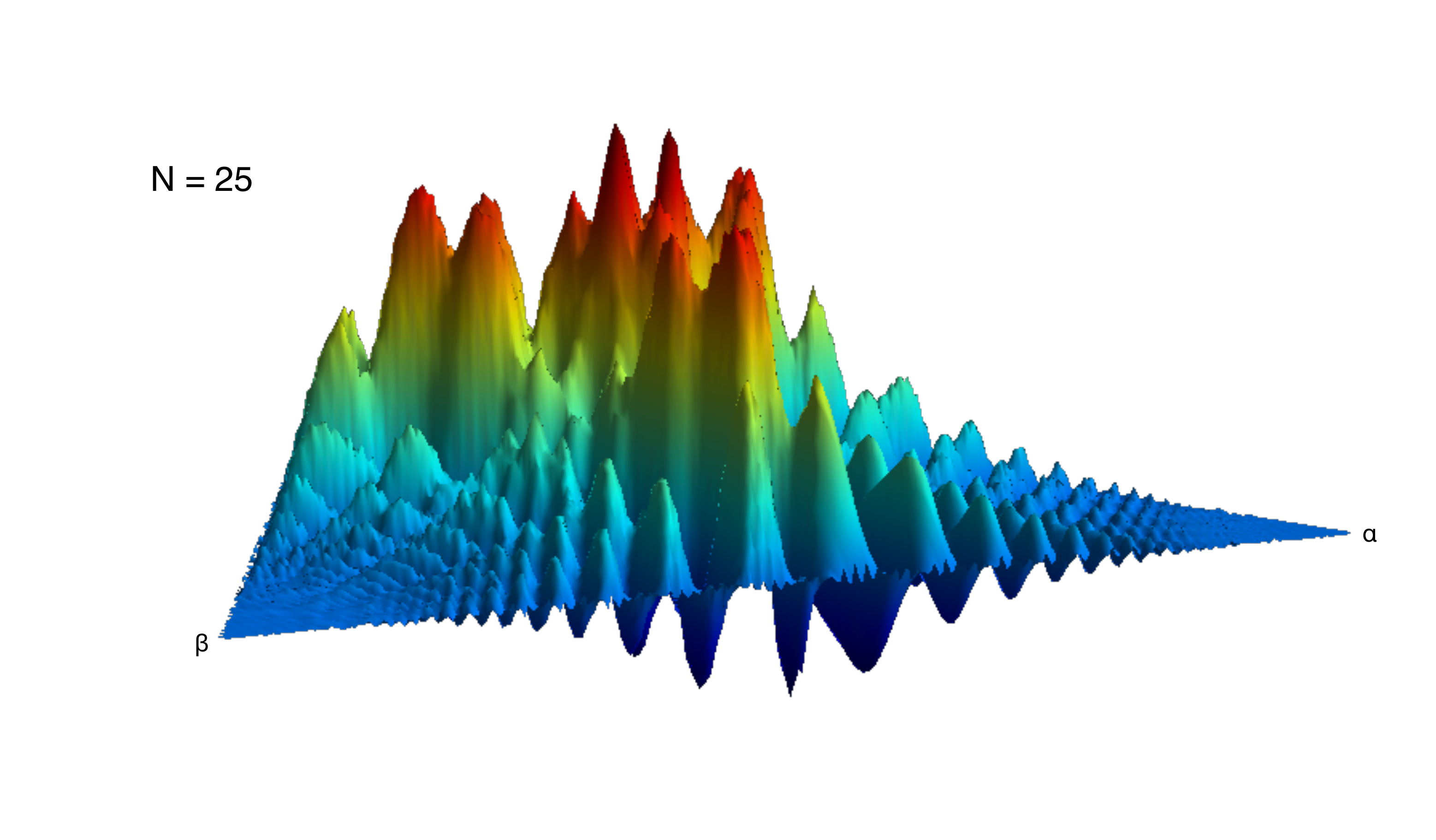} 
    \end{minipage}\hfill
    \begin{minipage}{0.49\textwidth}
        \centering
        \includegraphics[width=1.06\textwidth]{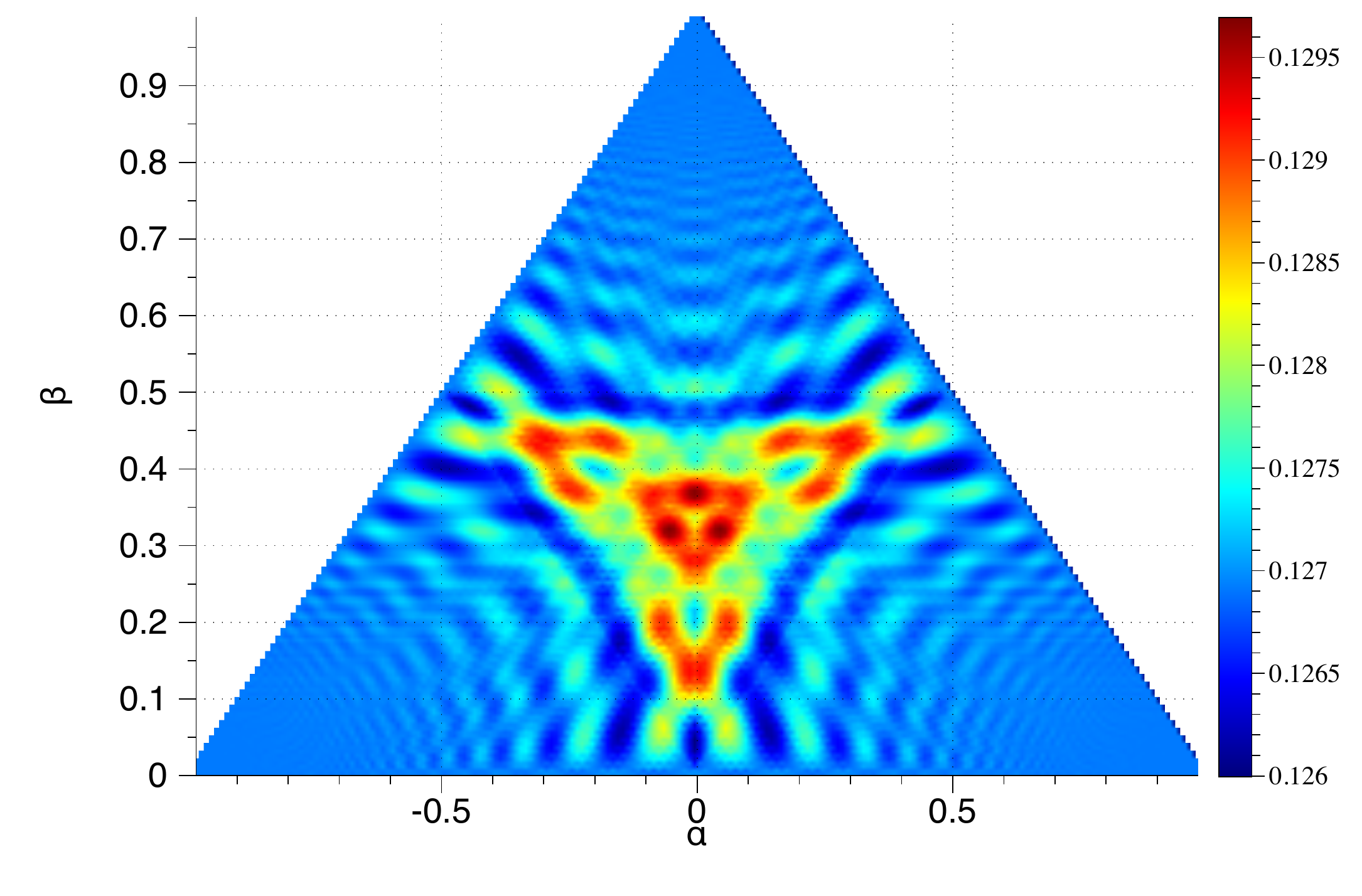} 
    \end{minipage}
    \caption{Left: gelaton model surface plot showing the shape of the reduced bispectrum, $\fNL(\alpha, \beta)$, taken at a time $N=25.0$ e-folds for a $k_t$ mode that leaves the horizon 18.9 e-folds after the initial time. Right: contour plot of the data in the left panel.}\label{Fig:Gelaton_shapes}
\end{figure}

In Fig.~\ref{Fig:Gelaton_shapes} we plot the shape of the
reduced bispectrum $\fNL(\alpha, \beta)$ for a single $k_t$-value
that exits the horizon 18.9 e-folds after the initial conditions.
As before, the left panel shows a
three-dimensional surface plot and the right panel
shows the corresponding contour plot.
Both are evaluated at time $N=25.0$, when the time
dependence has settled down to become constant.
At peak, $|\fNL| \approx 0.1297$ in agreement with
the values plotted in Fig.~\ref{Fig:Gelaton_3pfs}
(for a different value of $k_t$),
which is still some way from the smallest observable
value $|\fNL| \approx 10$.
The shape plot shows that the detailed structure of the bispectrum
is somewhat complicated, although it resembles the
equilateral template in its overall structure.

In the next section we will show that an observationally-relevant
amplification of the bispectrum is difficult to achieve for
a gelaton model of this type,
because consistency constraints give very little parameter space
to depress the speed of sound.

\subsubsection{Gelaton model parameter constraints} \label{subsubsec:GelParaConstraints}

The effective single-field description of the gelaton model is applicable
only when the gelaton mass $\mgel$ is significantly larger than $H$.
For smaller masses we must revert to the full two-field description.
We now argue that the modest amplitude of $\fNL$ seen in Figs.~\ref{Fig:Gelaton_3pfs}
and~\ref{Fig:Gelaton_shapes}
is a consequence of simultaneously satisfying this and other
consistency conditions.

\para{Gelaton mass}
First, we require $\mgel^2 \gg H^2$. With our choice of tension $T(\chi)$, Eq.~\eqref{eq:GelatonFieldMass}
shows that
\begin{equation} \label{eq:GelMasswithT}
	\mgel^2 = 2g^2 \MPlanck^{-2} \lambda^2 \chi^2 \exp \left( - \frac{2g \phi}{\MPlanck} \right).
\end{equation}
Evidently, if the argument of the exponential term is large then the
gelaton mass will be exponentially suppressed.
Therefore we suppose $|2g\phi /\MPlanck| \lesssim 1$,
allowing a Taylor expansion of $\mgel$.
The leading term is
\begin{equation} \label{eq:GelMassTaylor}
	\mgel^2 \approx 2g^2 \MPlanck^{-2} \lambda^2 \chi^2 + \cdots .
\end{equation}
To estimate the Hubble parameter we assume that the slow-roll approximation
applies, making the kinetic terms are sub-dominant to the potential.
Under these circumstances a reasonable approximation to
$H^2$ will be
\begin{equation} \label{eq:HsqGelEstimate}
	H^2 \approx
	\frac{V_{\text{DBI}}}{3\MPlanck^2}
	=
	\frac{1}{6} \frac{\lambda^2}{\MPlanck^2} \chi^2
	\left(
		\cosh \frac{2g \phi}{\MPlanck} - 1
	\right)
	+
	\frac{1}{3\MPlanck^2}
	\left(
		\Lambda^4 - \frac{1}{2} m^2 \chi^2 
	\right),
\end{equation}
where $V_\text{DBI}$ from Eq.~\eqref{eq:V_DBI_Potential} has been inserted
assuming our choices for
$T(\chi)$ and $W(\chi)$.

Our assumption that the exponential in Eq.~\eqref{eq:GelMasswithT}
is not significantly suppressed makes the $\cosh$ term in~\eqref{eq:HsqGelEstimate}
negligible.
Therefore the most significant contribution to $H$ will come from the
potential $W(\chi)$.
Meanwhile,
to prevent higher order terms become relevant we must constrain
the negative term $m^2 \chi^2 / 2$ to be significantly smaller than the
hilltop amplitude $\Lambda^4$.
This yields
\begin{equation} \label{eq:Gel_Wterm_constraint}
	\chi^2 \ll \frac{2\Lambda^4}{m^2}.
\end{equation}
In this regime the dominant contribution to the Hubble rate will come from the hilltop,
\begin{equation} \label{eq:HsqFinalGelEstimate}
	H^2 \approx \frac{\Lambda^4}{3}.
\end{equation}
Eqs.~\eqref{eq:GelMassTaylor} and~\eqref{eq:HsqFinalGelEstimate}
can be used together with the consistency condition $\mgel^2 \gg H^2$
to yield a \emph{minimum} value of the $\chi$ expectation value,
\begin{equation} \label{eq:m_gel_chi_constraint}
	\chi^2 \gg \frac{\Lambda^4}{6g^2 \lambda^2}.
\end{equation}
Consistency of
Eqs.~\eqref{eq:Gel_Wterm_constraint} and~\eqref{eq:m_gel_chi_constraint} yields a constraint on the
mass $m^2$,
\begin{equation} \label{eq:Gel_msq_upperbound}
	m^2 \ll 12 g^2 \lambda^2.
\end{equation}

\para{Speed of sound}
Second, to give at least modest suppression of the sound
speed we suppose $c_s^2 \ll 10/11 \approx 0.9$.
Eq.~\eqref{eq:GelatonSpeedSound} then requires
\begin{equation} \label{eq:c_s_approx_init}
	\frac{1}{c_s^2} = 1 + \frac{2}{\lambda^2} \left( \frac{\dot{\chi}}{\chi} \right)^2 \gg \frac{11}{10} ,
\end{equation}
where, as before, we have performed a Taylor expansion in exponentials of $\phi$.
The slow-roll approximation can be used to estimate $\dot{\chi}$,
\begin{equation} \label{eq:gel_chidot_est}
	\dot{\chi}^2 = \frac{m^4 \chi^2}{9H^2} = \frac{m^4 \chi^2}{3 \Lambda^4} .
\end{equation}
Combining Eq.~\eqref{eq:gel_chidot_est} and~\eqref{eq:c_s_approx_init} now yields a lower bound for $m^2$,
\begin{equation} \label{eq:Gel_msq_lowerbound}
	m^2 \gg \sqrt{ \frac{3 \lambda^2 \Lambda^4}{20} }.
\end{equation}

\begin{figure}
    \centering
    \begin{minipage}{0.49\textwidth}
        \centering
        \includegraphics[width=1.06\textwidth]{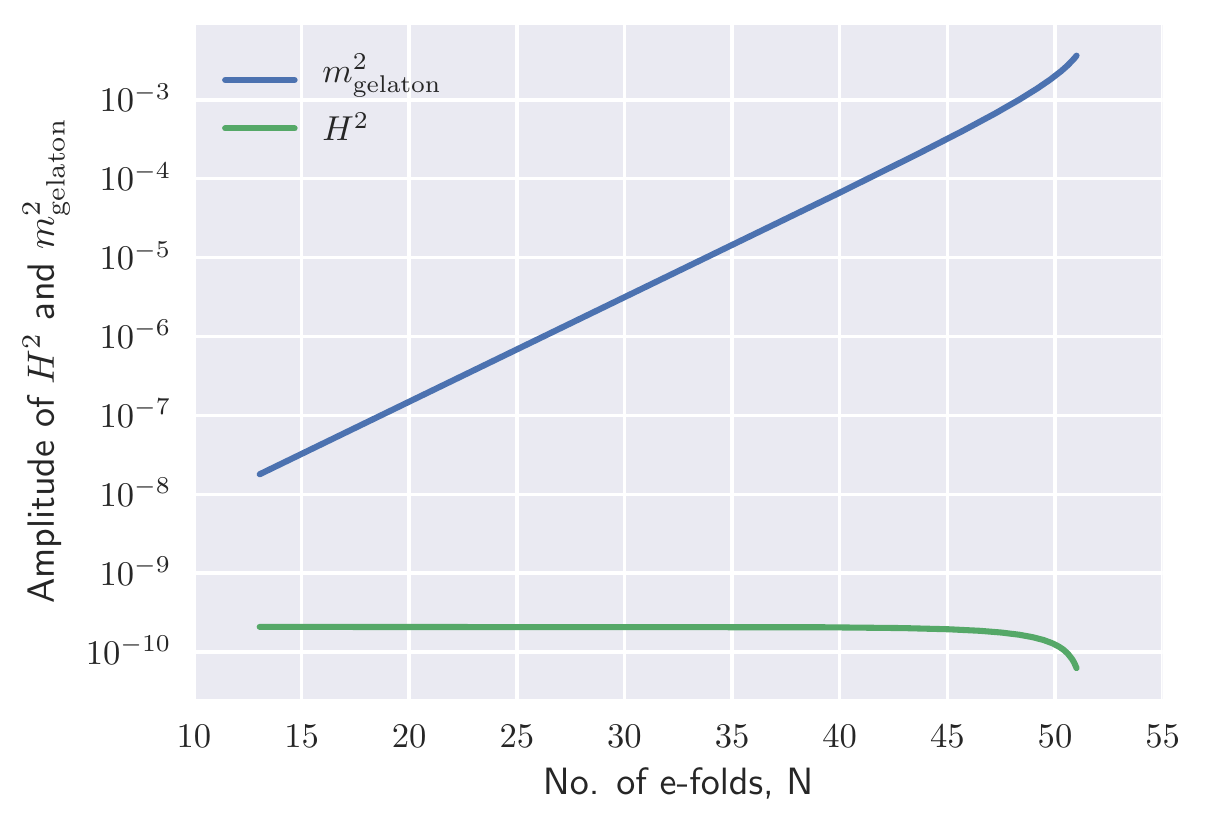} 
    \end{minipage}\hfill
    \begin{minipage}{0.49\textwidth}
        \centering
        \includegraphics[width=1.06\textwidth]{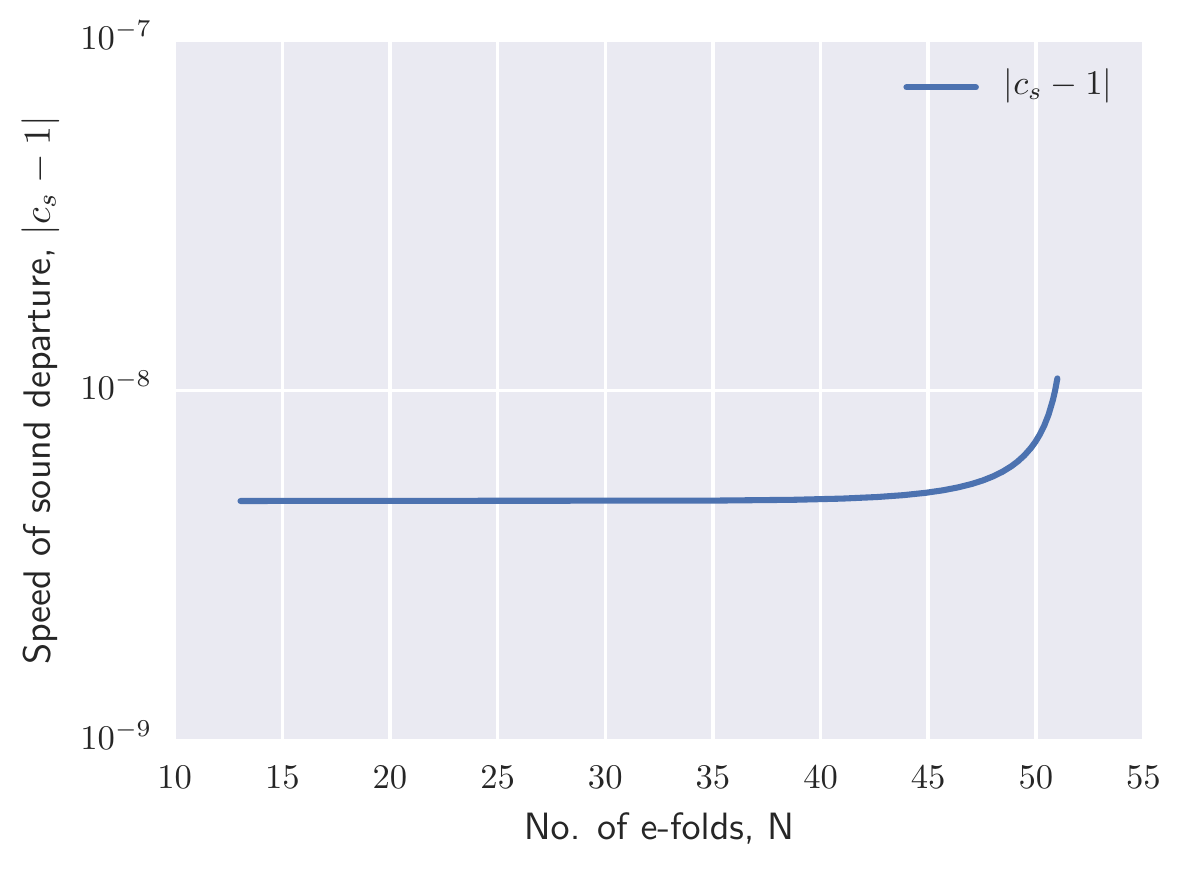}
    \end{minipage}
    \caption{Left: a plot of the gelaton mass $\mgel^2$ and $H^2$ demonstrating that the constraint $\mgel^2 \gg H^2$ is satisfied. Right: a plot showing the departure in the speed of sound, $|c_s - 1|$ is very small due to the constraints described in \S\ref{subsubsec:GelParaConstraints}.}\label{Fig:Gelaton_Constraints}
\end{figure}

The constraint is the principal obstruction to finding parameter combinations
that would yield significant amplification of the equilateral correlations.
Most obviously, Eq.~\eqref{eq:Gel_msq_upperbound}
creates a tension with the upper bound~\eqref{eq:Gel_msq_upperbound}
causing the available parameter window for $m$ to be rather narrow.
The lower limit scales parametrically with $\lambda$ whereas the upper limit scales with $\lambda^2$,
and therefore one strategy to increase the size of the window is to increase $\lambda$.
Unfortunately, Eq.~\eqref{eq:c_s_approx_init}
shows that increasing $\lambda$ will typically force the speed of sound towards unity
unless $\dot{\chi}/\chi$ can be changed to compensate.
This cannot happen in the slow-roll regime because~\eqref{eq:gel_chidot_est} shows that $\dot{\chi}/\chi$ is independent
of $\lambda$.

For example, with the above choices of $g$, $\lambda$ and $\Lambda$, the window for $m$ is
$0.00139 \MPlanck \ll m \ll 0.00145 \MPlanck$. This is so narrow that
it is not really possible to have
the strong `$\ll$' inequality satisfied on either side.
As we will see below, our choice $m = 10^{-5} \MPlanck$ amply satisfies
the upper bound~\eqref{eq:Gel_msq_upperbound}
and is sufficient to guarantee
$\mgel \gg H^2$, but it violates the lower bound and therefore does not
yield a suppressed speed of sound.

The limits on the $\chi$ expectation value~\eqref{eq:Gel_Wterm_constraint}
and~\eqref{eq:m_gel_chi_constraint} give another constraint.
Both limits scale with $\Lambda^4$ and therefore the relative size of the
window does not change with scaling $\Lambda$.
Instead, we must rely on changing the parameters $m$ or $g\lambda$
that appear in the denominators of~\eqref{eq:Gel_Wterm_constraint}
and~\eqref{eq:m_gel_chi_constraint} respectively.
We have already seen that $m$ is tightly constrained, making the upper limit
practically fixed once $\Lambda$ is prescribed.
Also, if $m$ is not too close to its lower limit then it will also scale roughly
with $\lambda$.
Therefore parametrically widening the available window for the $\chi$ expectation
value depends on increasing $g$ to decrease the lower limit relative to the upper one.
Unfortunately $g$ must be fairly small in order to keep
$\e{2g\phi/\MPlanck}$ reasonable small.
If the exponential becomes too large then $\epsilon$ typically grows also, causing
inflation to end exponentially quickly.
Therefore, in addition to the small range of $m$, there is a very small range of $\chi$
values that satisfy the constraints for a suppressed speed of sound.
In our example the range is roughly
$0.0237 \MPlanck \ll \chi \ll 0.0254 \MPlanck$.
This means that it is typically not possible to sustain enhanced three-point correlations
for a significant number of e-folds.

We have not succeeded in finding parameter combinations that give a significant
enhancement to equilateral correlations
while respecting the consistency conditions of the theory.
This does not rule out the possibility that the gelaton model
can do so, but it would require a different functional form for the potential
or the brane tension.
We have verified that similar constraints operate for the simplest monomial
chaotic models $W(\chi) \propto \chi^n$
for integer $n$, and that these constraints likewise lead to very narrow
windows for $m$ and $\chi$.
A modification to the brane tension is possible, but any exotic form would
need careful microphysical justification.

\para{Numerical comparison}
In the left panel of Fig.~\ref{Fig:Gelaton_Constraints} we plot the
gelaton mass, $\mgel^2$, together with the Hubble rate, $H^2$.
We demonstrate that $\mgel^2 \gg H^2$ so that it is consistent to integrate out
the gelaton.
Comparison of our numerical results and the analytic estimates
given in this section shows that our approximations for
$\mgel$, $c_s$, $H$ and $\dot{\chi}$
are each accurate to within an order of magnitude.
In the right panel we plot the departure of the speed of sound from unity,
$|c_s - 1|$.
This is very small, with approximate value $|c_s-1| \approx 10^{-8}$
using our parameter choices.

Together,
the constraints on the gelaton model with a hilltop potential mean
that it is possible to get an inflating solution lasting for approximately 50 e-folds,
but only without significant amplification of equilateral non-Gaussianities.
A similar conclusion applies if we replace the hilltop potential by a monomial
large-field model.
This does not rule out the possibility that a different gelaton model
could achieve significant enhancement, but the resulting model is likely to
more complex than the one considered here.

\section{Conclusions}
\label{sec:Concs}

We have applied the transport method to calculate the primordial bispectrum
produced by inflationary models containing non-canonical kinetic terms.
To do so we leverage the formalism of covariant perturbations
suggested by Gong \& Tanaka~\cite{Gong:2011uw}
to obtain a covariant Hamiltonian up to third order
(\S\S\ref{subsec:FieldCovarFormalism}--\ref{subsec:Hamiltonian}).
In agreement with other analyses, we show that up to a small number of Riemann terms
appearing in $M_{IJ}$, $A_{IJK}$ and $B_{IJK}$, the formalism covariantizes na\"{\i}vely.
Moreover, the initial conditions and gauge transformation to $\zeta$
also covariantize na\"{\i}vely provided index positioning is respected
(\S\S\ref{subsec:NewInitialConds}--\ref{subsec:GaugeTrans_Short}).
In \S\ref{subsec:CovarTransEqs} we demonstrate how to obtain a covariant hierarchy
of transport equations.

We have implemented these equations in a new version of the {\cpptransport}
tool, which is now capable of handling models with an arbitrary kinetic mixing
matrix. At this time, all three transport tools
({\mtransport}, {\cpptransport} and {\pytransport}) support models of this kind
and we can perform a meaningful comparison between them.
We find excellent agreement between the different codes (\S\ref{subsec:Quas2FldInflat}),
with differences typically less than $1\%$.
We also find excellent agreement for the same model written in different
field-space coordinates (\S\ref{subsec:CanoCompQsfiGel})
for which differences typically manifest at less than $0.1\%$.

In~\S\ref{subsec:Gelaton} we used {\cpptransport} to obtain numerical
predictions for a concrete implementation of the gelaton model.
We find good agreement between our numerical results
(which capture the full dynamics of the two-field model)
and the
predictions of the single-field effective description in which the
gelaton dresses the light fluctuations, giving them a suppressed speed of sound.
We find a small boost in the equilateral bispectrum
at the level $\left| \fNL(k_1, k_2, k_3) \right| \approx 0.1$
on equilateral configurations.
We give an analytic argument that it is not possible to achieve
more dramatic enhancements,
at least with the potential $V_\text{DBI}$ designed to reproduce the
dynamic of the Dirac--Born--Infeld model,
without considering more exotic forms
for the potential or brane tension.

To summarise, we have extended the automated numerical framework presented by Dias et al.~\cite{Dias:2016rjq} to include more complex models with a non-trivial kinetic term $G_{IJ}(\phi) \partial_\mu \phi^I \partial^\mu \phi^J$. As before, this allows numerical calculation of all tree-level contributions to the bispectrum and includes physical effects both before and after horizon-crossing. Practically, this means that observable statistics can be found for inflationary models containing a non-trivial kinetic sector, which can include supergravity theories (eg. Refs.~\cite{Kallosh:2013hoa, Kallosh:2013yoa}) or models motivated by string-theory (eg. Refs.~\cite{Ibanez:2014swa, Bielleman:2015lka}). In future we plan to extend {\cpptransport} to allow sampling over the prior probabilities for initial conditions or Lagrangian parameters, enabling estimates of the important observable parameters such as $n_s$ or $r$ in multiple-field models~\cite{Mortonson:2010er, Easther:2011yq, Norena:2012rs, Price:2014xpa}.

\acknowledgments

The work reported in this paper has been supported by the European Research Council under the European
Union's Seventh Framework Agreement (FP7/2007--2013) and ERC Grant Agreement No. 308082.
We would like to thank David Mulryne and John Ronayne for helpful conversations.

\appendix
\section{Appendix: Detailed calculations}
\label{app:DetailedCalcs}

\subsection{Perturbed action in curved field space}
\label{subsec:PertAction}

We begin with an action coupled to $\Nfields$ scalar fields $\phi^I$, minimally coupled to gravity with a self-interaction potential $V$,
\begin{equation} \label{eq:OriginalAction}
S \supseteq \frac{1}{2} \int \d^4x \sqrt{-g} \left[R - G_{IJ} g^{\mu \nu} \partial_{\mu} \phi^{I} \partial_{\nu} \phi^{J} -2V \right],
\end{equation}
where $R$ is the Ricci scalar, $g \equiv \det g$ and we use Greek indices and upper case Roman indices for the space-time and field-space coordinates respectively.
The kinetic mixing matrix $G_{IJ}$ is symmetric and positive definite, and can be
regarded as a metric on field-space.
The case of canonical kinetic terms $G_{IJ} = \delta_{IJ}$ corresponds
to a flat, Euclidean metric.

In this section we simplify calculations by setting the Planck mass to unity,
$\MPlanck = 1$.

\subsubsection{Field-covariant perturbations}
\label{subsubsec:FieldPerts}

For a bispectrum calculation we require an expansion in the field perturbations up to third order,
where each fluctuation is given by a coordinate displacement $\delta \phi^I = \phi^{I} (\vect{x}, t) - \phi^I (t)$.
Here, $\phi^I (t)$ is the background field and $\phi^I (\vect{x}, t)$ is the perturbed field.
Unfortunately this expression is not covariant under a change of field coordinates.
To obtain a covariant formulation
we follow the treatment of Gong \& Tanaka \cite{Gong:2011uw}, who focused on
the unique geodesic connected the field-space coordinates of the perturbed and unperturbed fields.
We take this geodesic to be labelled by an affine parameter $\lambda$,
with normalization adjusted so that $\lambda=0$ at the unperturbed coordinate at
$\lambda=1$ at the perturbed coordinate.
The initial tangent vector to the geodesic is then defined by
\begin{equation} \label{eq:InitVelocity}
	Q^I \equiv \left. \frac{\d\phi^I}{\d\lambda} \right|_{\lambda=0}.
\end{equation}
We can then assume parallel transport for our affine parameter $\lambda$ and write the geodesic equation as
\begin{equation} \label{eq:GeodesicEquation}
	D_\lambda^2 \phi^I = \frac{\d^2 \phi^I}{\d\lambda^2} + \Gamma^I_{JK} Q^J Q^K = 0,
\end{equation}
where $D_\lambda$ denotes a covariant derivative $Q^I \nabla_I$ and $\Gamma^I_{JK}$ is a field-space Christoffel symbol. We can then introduce a covariant Taylor expansion of the perturbation, $\delta \phi^I$,
\begin{equation} \label{eq:TaylorPerts}
	\delta \phi^I = \frac{\d \phi^I}{\d\lambda} + \frac{1}{2!} \frac{\d^2\phi^I}{\d\lambda^2} + \cdots.
\end{equation}
(Note that the appearance of this equation depends on our normalization convention
for $\lambda$, but its physical content is independent of it.)
Equations \eqref{eq:InitVelocity} \& \eqref{eq:GeodesicEquation} can then be inserted into Eq.~\eqref{eq:TaylorPerts} to obtain
\begin{equation} \label{eq:FieldPerturbation}
	\delta \phi^I = Q^I - \frac{1}{2!} \Gamma^I_{JK} Q^J Q^K + \cdots ,
\end{equation}
where `$\cdots$' denotes terms cubic and higher in $Q^I$ that we have neglected.
When applying these perturbations to the action in Eq.~\eqref{eq:OriginalAction}, we will only need to use this formalism for the kinetic part in the second and third terms as the Ricci scalar is zero in the spatially flat gauge. Before doing this however, we will need the field-covariant background equations which can be found \cite{Lee:2005bb} similarly
\begin{equation} \label{eq:CovarFieldEq}
	D_t \dot{\phi}^I + 3H \dot{\phi}^I + G^{IJ}V_{,J} = 0,
\end{equation}
\begin{equation} \label{eq:CovarFriedmannEq}
	3H^2 = \rho = \frac{1}{2} G_{IJ} \partial_\mu \phi^I \partial^\mu \phi^J + V(\phi),
\end{equation}
\begin{equation} \label{eq:CovarEpsilon}
	\epsilon \equiv - \frac{\dot{H}}{H^2} = \frac{G_{IJ} \dot{\phi}^I \dot{\phi}^J}{2H^2},
\end{equation}
which are the field-covariant Klein-Gordon equation, Friedman equation and the inflation condition respectively. The covariant time-derivative in Eq.~\eqref{eq:CovarFieldEq} appears frequently in expressions and is defined by
\begin{equation} \label{eq:CovarTimeDerivative}
	D_t Q^I = \dot{Q}^I + \dot{\phi}^J \Gamma^I_{JK} Q^K.
\end{equation}
Next we define $X = - G_{IJ} g^{\mu \nu} \partial_{\mu} \phi^{I} \partial_{\nu} \phi^{J} -2V$ and apply $\lambda$ derivatives up to third order which will add new terms to our perturbed action. However we first need the field-covariant derivative of $\partial_\mu \phi^I$ which is given by
\begin{equation} \label{eq:CovarDeriv d_mu phi}
	D_\lambda \partial_\mu \phi^I = \partial_\mu Q^I + \Gamma^I_{JK} \partial_\mu \phi^J Q^K \equiv D_\mu Q^I.
\end{equation}
Then Eq.~\eqref{eq:CovarDeriv d_mu phi} is used to give the first $\lambda$ derivative on $X$,
\begin{equation} \label{eq:Xepans1}
	\left. D_\lambda X \right\rvert_{\lambda=0} = -g^{\mu \nu} \partial_\mu \phi_I D_\nu Q^I - V_{;I}Q^I,
\end{equation}
which gives no new terms. The second derivative yields
\begin{equation} \label{eq:Xepans2}
	\left. D_\lambda^2 X \right\rvert_{\lambda=0} = -g^{\mu\nu} \left\{ R_{IJKL}\partial_\mu\phi^I \partial_\nu\phi^L Q^J Q^K + D_\mu Q_I D_\nu Q^I \right\} - V_{;IJ} Q^I Q^J,
\end{equation}
where we see that working with a non-canonical field metric has introduced a curvature term over the field coordinates. Finally the third derivative gives
\begin{align} \label{eq:Xepans3}
\begin{split}
	\left. D_\lambda^3 X \right\rvert_{\lambda=0} = &-g^{\mu \nu} \left\{ R_{MJKL;I} \partial_\mu\phi^M \partial_\nu \phi^L Q^I Q^J Q^K + 4R_{IJKL} \partial_\nu \phi^L D_\mu Q^I Q^J Q^K \right\} \\
	&- V_{;IJK} Q^I Q^J Q^K,
\end{split}		
\end{align}
where we have a Riemann tensor term as before as well as a covariant derivative of the curvature term. Equations \eqref{eq:Xepans1}--\eqref{eq:Xepans3} will be inserted into the action in Eq.~\eqref{eq:OriginalAction} along with the metric perturbations to find the perturbed action later.

\subsubsection{ADM decomposition and metric perturbations}
\label{subsubsec:ADMexpansMetPert}

We will follow the treatment of Maldacena \cite{Maldacena:2002vr,Seery:2005gb,Seery:2005wm} and use the (3+1) ADM decomposition \cite{Arnowitt:1962hi} of the metric which is given by
\begin{equation} \label{eq:ADMdecomp}
\d s^2 = -N^2 \, \d t^2 + h_{ij} \left( \d x^i + N^i \d t \right) \left( \d x^j + N^j \d t \right),
\end{equation}
where $N$ is the lapse function, $N^i$ is the shift vector, and $h_{ij}$ is the spatial metric. With this decomposition, the action in \eqref{eq:OriginalAction} can now be rewritten using the Gauss--Codazzi relation to remove the Gibbons--Hawking--York boundary term \cite{York:1972sj,Gibbons:1976ue}
\begin{equation} \label{eq:ADMaction}
	S = \frac{1}{2} \sqrt{h} \int \d^4 x \left\{ N \left( R^{(3)} - G_{IJ} h^{ij} \partial_i \phi^I \partial_j \phi^J - 2V \right) + \frac{1}{N} \left( \pi^I \pi_I + E^{ij}E_{ij} - E^2 \right) \right\},
\end{equation}
where $R^{(3)}$ is the Ricci scalar built from the spatial metric and $E_{ij}$ is related to the extrinsic curvature of constant slices,
\begin{equation} \label{eq:ExtrinsCurvature}
	E_{ij} = \frac{1}{2} \left( \dot{h}_{ij} - N_{i|j} - N_{j|i} \right),
\end{equation}
where the vertical bar index denotes a covariant derivative compatible with $h_{ij}$ and we have made the definition
\begin{equation} \label{eq:ADMpi}
	\pi^I \equiv \dot{\phi}^I - N^j \phi^I_{|j}.
\end{equation}
We will later expand the lapse function and shift vector in terms of scalar perturbations in our field perturbations for the spatially flat gauge with $h_{ij} = a^2 \delta_{ij}$,
\begin{subequations} \label{eq:LapseShiftExpans}
\begin{align}
	N &= 1 + \alpha_1 + \alpha_2 + \cdots \\
	N^i &= \theta_1^{,i} + \theta_2^{,i} +\cdots,
\end{align}
\end{subequations}
where the subscripts $1, 2, ...$ indicate the order of expansion and $\alpha$ is a perturbation in the lapse function with $\theta$ being an expansion in the shift vector.

\subsubsection{Perturbing the action}
\label{subsubsec:pertAction}

We must now insert the expressions for the kinetic term, $X = - G_{IJ} g^{\mu \nu} \partial_{\mu} \phi^{I} \partial_{\nu} \phi^{J} -2V$, found in equations \eqref{eq:Xepans1}--\eqref{eq:Xepans3} as well as the metric perturbations found in equations \eqref{eq:ADMaction} and \eqref{eq:LapseShiftExpans} into our action found in Eq.~\eqref{eq:OriginalAction} which gives the results found by Elliston et al. \cite{Elliston:2012ab}
\begin{align} \label{eq:Order2Action}
\begin{split}
	S_2 = \frac{1}{2} \int \d^4x \ a^3 \bigg\{ &\alpha_1 \left[ -6H^2 \alpha_1 + \dot{\phi}_I\dot{\phi}^I \alpha_1 - 2\dot{\phi} D_t Q^I - 2V_{,I} Q^I \right] \\
	& -\frac{2}{a^2} \partial^2 \theta_1 \left[ 2H\alpha_1 - \dot{\phi}_I Q^I \right] \\
	& +R_{KIJL} \dot{\phi}^K \dot{\phi}^L Q^I Q^J + D_t Q_I D_t Q^I - \partial_i Q_I \partial^i Q^I - V_{;IJ}Q^I Q^J
\end{split}
\end{align}
and
\begin{align}
\begin{split} \label{eq:Order3Action}
	S_{3} = &\frac{1}{2} \int \d^4x \ a^3 \bigg\{ 6H^2\alpha_{1}^3 + \frac{4H}{a^2} \alpha_1^2 \partial^2\theta_{1} - \frac{\alpha_1}{a^4} \left( \partial_i \partial_j \theta_{1} \partial_i \partial_j \theta_{1} - \partial^2 \theta_{1} \partial^2 \theta_{1} \right) \\ 
	 &-\alpha_1^3 \dot{\phi}_I \dot{\phi}^I + 2\alpha_1^2 \dot{\phi}_I D_t Q^I + \frac{2}{a^2} \alpha_1 \dot{\phi}_I \partial_i \theta_{1} \partial_i Q^I - \alpha_1 R_{K(IJ)L} \dot{\phi}^K \dot{\phi}^L Q^I Q^J \\
	 & -\alpha_1 \left( D_t Q_I D_t Q^I + \frac{1}{a^2} D_i Q_I D_i Q^I \right) - \frac{2}{a^2} \partial_i \theta_{1}D_t Q_I \partial_i Q^I +\frac{4}{3} R_{I(JK)L} \dot{\phi}^L D_t Q^I Q^J Q^K \\
	 & \left. +\frac{1}{3} R_{(I|LM|J;K)} \dot{\phi}^M \dot{\phi}^L Q^I Q^J Q^K - \alpha_1 V_{;(IJ)} Q^I Q^J - \frac{1}{3} V_{;(IJK)} Q^I Q^J Q^K \right\},
\end{split}
\end{align}	
at second and third order respectively where brackets around indices indicate that they can be cyclically permuted and vertical bars exclude indices from that symmetrisation. It should be noted that neither of these actions contain second order terms in the lapse and shift just like the canonical case but we will need them regardless because they are used in the gauge transformation calculation. 

\subsubsection{Applying constraints for Fourier-space action}
\label{subsubsec:ConstaintFourierAction}

We may now vary the second-order action in Eq.~\eqref{eq:Order2Action} with respect to the lapse and shift to find expressions for $\alpha_1$ and $\theta_1$ in terms of the perturbed fields $Q^I$ which are
\begin{equation} \label{eq:Alpha_1}
	\alpha_1 = \frac{\dot{\phi}_I Q^I}{2H} ,
\end{equation}
and
\begin{equation} \label{eq:Theta_1}
	\theta_1 = \frac{a^2}{2H} \partial^{-2} \left( -V_{,I}Q^I -\dot{\phi}_I D_t Q^I + 2\alpha_1 \left[ -3H^2 +\frac{1}{2} \dot{\phi}_I \dot{\phi}^I \right] \right).
\end{equation}
Here $\partial^{-2}$ denotes the inverse Laplacian operator over spatial coordinates and Eq.~\eqref{eq:Theta_1} may be further simplified using the background Eq.~\eqref{eq:CovarFieldEq} and Eq.~\eqref{eq:Alpha_1} for an expression in terms of fields only. As mentioned previously, we also need second-order expressions for the lapse and shift which are found by varying Eq.~\eqref{eq:ADMaction} with respect to $N$ and $N^i$ and then expanding perturbatively to find
\begin{equation} \label{eq:Alpha_2}
	\alpha_2 = \frac{\alpha_1^2}{2} + \frac{1}{2H} \partial^{-2} \left\{ \partial_i(D_t Q^I) \partial^i Q_I + D_t Q^I D_t Q_I + \frac{1}{a^2} \left( \partial^2 \alpha_1 \partial^2 \theta_1 - \partial_i \partial_j \alpha_1 \partial_i \partial_j \theta_1 \right) \right\},
\end{equation}
and
\begin{align} \label{eq:Theta_2}
	\begin{split}
		\theta_2 = \frac{a^2}{4H} \partial^{-2} &\bigg\{ 2\alpha_1 \left( \frac{4H}{a^2} \partial^2 \theta_1 + 2 \dot{\phi}^I D_t Q_I \right) - V_{IJ} Q^I Q^J - D_t Q^I D_t Q_I \\
		&+ \frac{1}{a^2} \left( 2 \dot{\phi}^I \partial_i \theta_1 \partial_i Q_I - \partial_i Q^I \partial_i Q_I + \frac{1}{a^2} \left( \partial^2 \theta_1 \partial^2 \theta_1 - \partial_i \partial_j \theta_1 \partial_i \partial_j \theta_1 \right) \right) \\
		&+ 2H^2 (2\alpha_2 - 3\alpha_1^2) ( \epsilon - 3 ) - R_{IJKL} Q^I \dot{\phi}^J \dot{\phi}^K Q^L \bigg\}.
	\end{split}
\end{align}
Equations \eqref{eq:Theta_1} and \eqref{eq:Theta_2} can then be used to give the Hamiltonian constraint for a non-trivial metric on super-horizon scales,
\begin{align} \label{eq:HamiltonianConstraint}
	\begin{split}
		0 = \mbox{} &
		V_I Q^I
		+ \frac{1}{2} V_{IJ} Q^I Q^J
		+ \dot{\phi}^I D_t Q_I
		+ \frac{1}{2} D_t Q_I D_t Q^I
		+ \frac{1}{2} R_{IJKL} Q^I \dot{\phi}^J \dot{\phi}^K Q^L
		\\
		&
		+ H^2 
		\left(
			2\alpha_1
			+ 2\alpha_2
			- 3\alpha_1^2
		\right) (3 - \epsilon)
		- 2\alpha_1 \dot{\phi}^I D_t Q_I,
	\end{split}
\end{align}
where the spatial derivatives have been omitted due to them decaying on super-horizon scales. Equations \eqref{eq:Alpha_1} and \eqref{eq:Theta_1} can be used to rewrite the second- and third-order actions in terms of only background fields and their perturbations. We would also like to write these in terms of the Fourier modes instead of spatial coordinates so we must adopt a convention and notation to express this. Therefore we use bold sans-serif indices to indicate an integration over Fourier modes for an index contraction such as
\begin{equation} \label{eq:FourierConvention1}
	A_{\FouInd{I}}B^{\FouInd{I}} = \sum_I \int \frac{\d^3 k_I}{(2 \pi)^3} A_{I} (\vect{k}_I) B^{I} (\vect{k}_I),
\end{equation}
where the indices on the right-hand side represent phase-space coordinate labels and indices may be changed to be co- or contravariant using the field-space metric $G_{IJ}$. This can be a problem if the $\delta$-function $G_{\FouInd{IJ}} = (2\pi)^3 G_{IJ} \delta(\vect{k}_I + \vect{k}_J)$ is produced, because this reverses the sign of a momentum label; we use a prime on the index to indicate this,
\begin{equation} \label{eq:FourierConvention2}
	A_{\FouInd{I}}B^{\FouInd{\FlipInd{I}}} = \sum_I \int \frac{\d^3 k_I}{(2 \pi)^3} A_{I} (\vect{k}_I) B^{I} (- \vect{k}_I).
\end{equation}
Using these conventions and by substituting equations \eqref{eq:Alpha_1} and \eqref{eq:Theta_1} into the second- and third-order actions in equations \eqref{eq:Order2Action} and \eqref{eq:Order3Action}, we find
\begin{align}
\begin{split} \label{eq:PerturbedAction}
	S_\phi = \frac{1}{2} \int \d t \ a^3 \Big\{ &G_{\FouInd{IJ}} D_t Q^\FouInd{I} D_t Q^\FouInd{J} + M_{\FouInd{IJ}} Q^\FouInd{I} Q^\FouInd{J} + \\
	&A_{\FouInd{IJK}}Q^\FouInd{I} Q^\FouInd{J} Q^\FouInd{K} + B_{\FouInd{IJK}} Q^\FouInd{I} Q^\FouInd{J} D_t Q^\FouInd{K} + C_{\FouInd{IJK}} D_t Q^\FouInd{I} D_t Q^\FouInd{J} Q^\FouInd{K} \Big\},
\end{split}
\end{align} 
where the second-order and third-order parts of the action are written on the first and second lines respectively. The second order kernels are given by
\begin{align}
	\begin{split} \label{eq:G_IJ}
	G_{\FouInd{IJ}} &= (2\pi)^3 G_{IJ} \delta(\vect{k}_1 + \vect{k}_2), \\
	M_{\FouInd{IJ}} &= (2\pi)^3 \delta(\vect{k}_1 + \vect{k}_2) \left( \frac{\vect{k}_1 \cdot \vect{k}_2}{a^2} G_{IJ} - m_{IJ} \right),
	\end{split}
\end{align}
where $m_{IJ}$ satisfies
\begin{equation} \label{eq:m_IJ}
	m_{IJ} = V_{;IJ} - R_{KIJL}\dot{\phi}^K \dot{\phi}^L - \frac{1}{a^3} D_t \left( \frac{a^3 \dot{\phi}_I \dot{\phi}_J}{H} \right).
\end{equation}
Then the third-order kernels are given by
\begin{subequations}
\begin{align}
\begin{split} \label{eq:Akernel}
	A_{\FouInd{IJK}} &= (2\pi)^3 \delta(\vect{k}_1 + \vect{k}_2 + \vect{k}_3) A_{IJK},
\end{split}\\
\begin{split} \label{eq:Bkernel}
	B_{\FouInd{IJK}} &= (2\pi)^3 \delta(\vect{k}_1 + \vect{k}_2 + \vect{k}_3) B_{IJK},
\end{split}\\
\begin{split} \label{eq:Ckernel}
	C_{\FouInd{IJK}} &= (2\pi)^3 \delta(\vect{k}_1 + \vect{k}_2 + \vect{k}_3) C_{IJK},
\end{split}
\end{align}
\end{subequations}
with
\begin{subequations}
\begin{align}
\begin{split} \label{eq:ATensor}
	A_{IJK} & = -\frac{1}{3} V_{;IJK} - \frac{\dot{\phi}_I V_{;JK}}{2H} + \frac{\dot{\phi}_I \dot{\phi}_J Z_K}{4H^2} + \frac{\dot{\phi}_I Z_J Z_K}{8H^3} \left( 1 - \frac{(\vect{k}_2 \cdot \vect{k}_3)^2}{k_2^2 k_3^2} \right) \\ 
	& \quad +\frac{\dot{\phi}_I \dot{\phi}_J \dot{\phi}_K}{8H^3} (6H^2 - \dot{\phi}^2 )-\frac{\dot{\phi}_K \dot{\phi}^L \dot{\phi}^M}{2H} R_{L(IJ)M} + \frac{1}{3}R_{(I|LM|J;K)}\dot{\phi}^L \dot{\phi}^M \\
	& \quad +\frac{\dot{\phi}_I G_{JK}}{2H} \frac{\vect{k}_2 \cdot \vect{k}_3}{a^2},
\end{split}\\
\begin{split} \label{eq:BTensor}
	B_{IJK} & = \frac{4}{3} R_{I(JK)L} - \frac{\dot{\phi}_I Z_J \dot{\phi}_K}{4H^3} \left( 1 - \frac{(\vect{k}_2 \cdot \vect{k}_3)^2}{k_2^2 k_3^2} \right) +\frac{\dot{\phi}_I \dot{\phi}_J \dot{\phi}_K}{4H^2} - \frac{Z_I G_{JK}}{H} \frac{\vect{k}_1 \cdot \vect{k}_2}{k_1^2},
\end{split}\\
\begin{split} \label{eq:CTensor}
	C_{IJK} & = \frac{G_{IJ} \dot{\phi}_K}{2H} +\frac{\dot{\phi}_I \dot{\phi}_J \dot{\phi}_K}{8H^3} \left( 1 - \frac{(\vect{k}_1 \cdot \vect{k}_2)^2}{k_1^2k_2^2} \right) + \frac{\dot{\phi}_I G_{JK}}{H} \frac{\vect{k}_1 \cdot \vect{k}_3}{k_1^2},
\end{split}
\end{align}
\end{subequations}
and where $Z_I$ is given by
\begin{equation} \label{eq:Z_I}
	Z_I = D_t \dot{\phi}^I + \frac{\dot{\phi}_I \dot{\phi}_J \dot{\phi}^J}{2H}.
\end{equation}
Expression \eqref{eq:ATensor} should be symmetrised over all three indices and expressions \eqref{eq:BTensor} and \eqref{eq:CTensor} should be symmetrised over the first two indices where an exchange of indices corresponds with a matching change of $\vect{k}$ vectors. The results for these kernels are identical to those found for the canonical case in \cite{Dias:2016rjq} apart from the addition of Riemann terms appearing on the second line of $A_{IJK}$ above and in the first term of $B_{IJK}$. We also note that the last term in $A_{IJK}$ is proportional to $(k/a)^2$ so will grow exponentially on sub-horizon scales which we will later need to treat separately when computing initial conditions.

\subsection{Transport method}
\label{subsec:TransportMethod}

We want to use the action we have found in the previous section to find evolution equations for our correlation functions and therefore compute the 2- and 3-point functions on sub- and super-horizon scales. For this we can use the transport method as first detailed in \cite{Mulryne:2009kh,Mulryne:2010rp,Seery:2012vj,Mulryne:2013uka}, which relates correlation functions of Heisenberg picture operators to those in the interaction picture where the Heisenberg equations of motion can be used to give evolution equations of interaction-picture fields.

\subsubsection{Correlation functions}
\label{subsubsec:CorrFuncs}

We begin by defining Heisenberg fields and their momenta as $Q^I$ and $P^I$ respectively, which we then can use to write a Hamiltonian split into free and interacting parts,
\begin{equation} \label{eq:Ham_Decomp}
	H(Q, P) = H_0(Q, P) + \Hint(Q, P),
\end{equation}
where the index $0$ denotes the free part and $int$ gives the interacting part.	Next we must define our new interaction-picture operators using some unitary operator, $F$, as
\begin{align}
	\begin{split} \label{eq:IntPictureFields}
		q^I &= F^{\dag} Q^I F, \\
		p_J &= F^{\dag} P_J F,
	\end{split}
\end{align}
where $q^I$ and $p_J$ are in the interaction picture. From these relations, it is simple to rewrite a vacuum expectation value of Heisenberg picture operators, $\mathcal{O}(Q,P)$, in terms of interaction picture operators,
\begin{align}
	\begin{split} \label{eq:HeisVev_IntVev}
		\braket{\vac|\mathcal{O}(Q,P)|\vac} = \braket{ \vac| F \mathcal{O}(q,p) F^{\dag} | \vac },
	\end{split}
\end{align}
where $\braket{\vac|\cdots|\vac}$ denotes an expectation value in the Minkowski vacuum.
We can use the Heisenberg equation of motion, $\d Q / \d t = - \im [Q, H(Q,P)]$, to show that the differential equation needed to find the unitary operator $F$ is
\begin{equation} \label{eq:IntOperODE}
	\frac{\d F}{\d t} = \im F \Hint(q,p),
\end{equation}
where the equation for $F^\dag$ is found by taking the complex conjugate. These differential equations can be solved using a power-series method to give the solution
\begin{equation} \label{eq:Fsolution}
	F = \AntiTimeOrder \exp \left( \im \int^t \Hint(t') \d t' \right)
\end{equation}
where $\AntiTimeOrder$ is the anti-time ordering operator which orders its argument in terms of increasing time. We can set the lower limits of these integrals by using a theory by Gell-Mann and Low \cite{GellMann:1951rw} which states that the vacuum state of an interacting theory can be related to the ground state of a non-interacting theory with an adiabatic `switch on' of the interacting theory. Then the integrals are performed over contours deformed into the complex plane in the distant past with analytic continuation used to define the fields for each ladder operator. These results are used in Eq.~\eqref{eq:HeisVev_IntVev}, yielding
\begin{equation} \label{eq:VEVsolution}
	\braket{\vac| \mathcal{O}(X) |\vac} = \left\langle 0 \left| \AntiTimeOrder \exp \left( \im \int^t_{-\infty^{+}} \Hint(t') \d t' \right) \mathcal{O}(x) \TimeOrder \exp \left( -\im \int^t_{-\infty^{-}} \Hint (t'') \d t'' \right) \right| 0 \right\rangle,
\end{equation}
where $-\infty^+$ and $-\infty^-$ show that the integration contour should be deformed into the positive and negative imaginary half-planes respectively with $X^a = (Q^I,P^J)$ and $x^a = (q^I,p^J)$ defined as phase-space vectors containing fields and momenta in the Heisenberg and interaction picture respectively. This is known as the `in--in' formalism \cite{Adshead:2009cb} used for computing correlation functions and is a sum over all possible `out' states for the theory.

\subsubsection{Evolution equations}
\label{subsubsec:EvoEqs}

We can now use these relations between Heisenberg and interaction picture fields along with our Fourier convention found in equations \eqref{eq:FourierConvention1} and \eqref{eq:FourierConvention2} to write the Hamiltonian as
\begin{equation} \label{eq:HamiltFourierExpans}
	H = \frac{1}{2!} H_{\FouInd{ab}} X^\FouInd{a} X^\FouInd{b} + \frac{1}{3!} H_{\FouInd{abc}} X^\FouInd{a} X^\FouInd{b} X^\FouInd{c} + \cdots ,
\end{equation}
where all fields are in the Heisenberg picture and `$\cdots$' denotes higher-order terms. This allows the Heisenberg equations of motion to be written
\begin{equation} \label{eq:HeisenbergEOMs}
	D_t X^\FouInd{a} = u\indices{ ^{\FouInd{a}} _{\FouInd{b}} } X^\FouInd{b} + \frac{1}{2!} u\indices{ ^{\FouInd{a}} _{\FouInd{bc}}} X^\FouInd{b} X^\FouInd{c} + \cdots ,
\end{equation}
which gives definitions for the `$u$-tensors', $u\indices{^{\FouInd{a}}_{\FouInd{b}}}$ and $ u\indices{ ^{\FouInd{a}} _{\FouInd{bc}} } $. There is also a Christoffel symbol appearing on the left hand side of \eqref{eq:HeisenbergEOMs} because of the field-covariant time derivative defined in Eq.~\eqref{eq:CovarTimeDerivative}. For our action in Eq.~\eqref{eq:PerturbedAction}, we choose the free part of the Hamiltonian to be the quadratic terms in perturbations and the interacting part is given by the cubic terms. The time evolution of an interaction-picture field is
\begin{equation} \label{eq:IntPicTimeEvo}
	D_t x^\FouInd{a} = u\indices{ ^{ \FouInd{a} } _{ \FouInd{b} } } x^\FouInd{b}.
\end{equation}
This allows us to use equations \eqref{eq:VEVsolution} and \eqref{eq:HamiltFourierExpans} to give tree-level two- and three-point correlation functions
\begin{subequations}
	\begin{align}
		\label{eq:2pntCorrFn} \left\langle X^{\FouInd{a}} X^{\FouInd{b}} \right\rangle &= \left\langle 0 \left| x^\FouInd{a} x^\FouInd{b} \right| 0 \right\rangle, \\
		\label{eq:3pntCorrFn} \left\langle X^\FouInd{a} X^\FouInd{b} X^\FouInd{c} \right\rangle &= \left\langle 0 \left| \left[ \frac{\im}{3!} \int^t H_{\FouInd{def}} x^\FouInd{d} x^\FouInd{e} x^\FouInd{f} \ \d t', x^\FouInd{a} x^\FouInd{b} x^\FouInd{c} \right] \right| 0 \right\rangle.
	\end{align}
\end{subequations}
Evolution equations can now be found for the two-point function first by differentiating Eq.~\eqref{eq:2pntCorrFn} with respect to time and using Eq.~\eqref{eq:IntPicTimeEvo} to simplify the result.
We find
\begin{equation} \label{eq:2PointCorrFnEvolution}
	D_t \left\langle X^\FouInd{a} X^\FouInd{b} \right\rangle = u\indices{^{\FouInd{a}}_{\FouInd{c}}} \left\langle X^\FouInd{c} X^\FouInd{b} \right\rangle + u\indices{^{\FouInd{b}}_{\FouInd{c}}} \Big\langle X^\FouInd{a} X^\FouInd{c} \Big\rangle,
\end{equation}
where $u\indices{^{\FouInd{a}}_{\FouInd{b}}}$ can be found by finding the Hamiltonian from our action and then using the Heisenberg equations from it to compare with Eq.~\eqref{eq:HeisenbergEOMs} above. The evolution equation for the 3-point correlation function is slightly harder to calculate than the 2-point function because it requires rewriting some of the commutation relations found after differentiating Eq.~\eqref{eq:3pntCorrFn} as seen in Ref.~\cite{Dias:2016rjq}. The result is
\begin{equation} \label{eq:3PointCorrFnEvolution}
	D_t \left\langle X^\FouInd{a} X^\FouInd{b} X^\FouInd{c} \right\rangle = u\indices{^{\FouInd{a}}_{\FouInd{d}}} \left\langle X^\FouInd{d} X^\FouInd{b} X^\FouInd{c} \right\rangle + u\indices{^{\FouInd{a}}_{\FouInd{de}}} \Big\langle X^\FouInd{d} X^\FouInd{b} \Big\rangle \Big\langle X^\FouInd{e} X^\FouInd{c} \Big\rangle + 2 \ \mathrm{perms},
\end{equation}
where there are contributions from both of the $u$-tensors defined in Eq.~\eqref{eq:HamiltFourierExpans} above and the permutations preserve the ordering of indices. Equations \eqref{eq:2PointCorrFnEvolution} and \eqref{eq:3PointCorrFnEvolution} both contain a Christoffel symbol term for each of the phase-space indices appearing on the left hand side of each equation. These equations may be further simplified by defining $\braket{X^{\FouInd{a}} X^{\FouInd{b}}} \equiv (2\pi)^3 \delta(\vect{k}_a + \vect{k}_b) \Sigma^{ab}$ and $\braket{X^{\FouInd{a}} X^{\FouInd{b}} X^{\FouInd{c}}} \equiv (2\pi)^3 \delta(\vect{k}_a + \vect{k}_b + \vect{k}_c) \alpha^{abc}$ as the two- and three-point functions to obtain
\begin{subequations}
	\begin{align}
		\label{eq:2pointTransportEquationApp}
		D_t \Sigma^{ab} &= u\indices{^a_c} \Sigma^{cb} + u\indices{^b_c} \Sigma^{ac} , \\
		\label{eq:3pointTransportEquationApp}
		D_t \alpha^{abc} &= u\indices{^a_d} \alpha^{dbc} + u\indices{^a_{de}} \Sigma^{db} \Sigma^{ec} + 2 \ \mathrm{cyclic} \ (a\rightarrow b \rightarrow c),
	\end{align}
\end{subequations}
The two differential equations found in equations \eqref{eq:2pointTransportEquationApp} and \eqref{eq:3pointTransportEquationApp} can both be solved numerically to find a power spectrum or bispectrum for an inflation theory and only require calculation of the $u$-tensors and initial conditions.

\subsubsection{Calculating the $u$-tensors}
\label{subsubsec:CalcUtensors}

As mentioned previously, we must find the Hamiltonian from our action in Eq.~\eqref{eq:PerturbedAction} so we begin by defining the momentum canonically conjugate to the field perturbations $Q^I$,
\begin{equation} \label{eq:ConjugateMomenta}
	P_\FouInd{I} (t) = \frac{\delta S_\phi}{\delta (D_t Q^\FouInd{I})},
\end{equation}
with a variational derivative defined by
\begin{equation} \label{eq:VarDerivDefinition}
	\frac{\delta [ Q^\FouInd{I} (\vect{k}_I, t) ]}{\delta [ Q^\FouInd{J} (\vect{k}_J, t') ]} = \delta_J^I (2\pi)^3 \delta(t - t') \delta(\vect{k}_I + \vect{k}_J) = \delta_\FouInd{J}^\FouInd{I} \delta(t - t') .
\end{equation}
Equations \eqref{eq:ConjugateMomenta} and \eqref{eq:VarDerivDefinition} can then be used on Eq.~\eqref{eq:PerturbedAction} to obtain the momentum,
\begin{equation} \label{eq:P_I solution}
	P_\FouInd{I} = a^3 \bigg\{ D_t Q_\FouInd{I} + \frac{1}{2} B_{\FouInd{JK\FlipInd{I}}} Q^\FouInd{J} Q^\FouInd{K} + C_{\FouInd{\FlipInd{I}JK}} P^\FouInd{J} Q^\FouInd{K} \bigg\},
\end{equation}
where a prime on an index indicates a sign reversal of momentum.
From Eq.~\eqref{eq:P_I solution}, it is simple to rearrange for $D_t Q_\FouInd{I}$,
\begin{equation} \label{eq:D_t Q_I}
	D_t Q_\FouInd{I} = \frac{P_\FouInd{I}}{a^3} - \frac{1}{2} B_{\FouInd{JK\FlipInd{I}}} Q^\FouInd{J} Q^\FouInd{K} - C_{\FouInd{\FlipInd{I}JK}} P^\FouInd{J} Q^\FouInd{K}.
\end{equation}
Then we may use the relation $H = \int \d t \, [ P^\FouInd{I} (D_t Q_{\FouInd{\FlipInd{I}}}) - L]$ and rescale the momentum by a factor of $a^3$ as $P_\FouInd{I} \rightarrow a^3 P_\FouInd{I}$ to obtain the Hamiltonian
\begin{align}
	\begin{split} \label{eq:HamiltonianSolution}
		H = \frac{1}{2} \int \d t \ a^3 \Big( &G_{\FouInd{IJ}} P^\FouInd{I} P^\FouInd{J} - M_{\FouInd{IJ}} Q^\FouInd{I} Q^\FouInd{J} - \\
		&A_{\FouInd{IJK}} Q^\FouInd{I} Q^\FouInd{J} Q^\FouInd{K} - B_{\FouInd{IJK}} Q^\FouInd{I} Q^\FouInd{J} P^\FouInd{K} - C_{\FouInd{IJK}} P^\FouInd{I} P^\FouInd{J} Q^\FouInd{K} \Big),
	\end{split}
\end{align}
where the terms on the first line are quadratic in perturbations and the terms on the second line are cubic in perturbations which represent $H_0$ and $\Hint$ in Eq.~\eqref{eq:Ham_Decomp} respectively. Next we must find the Heisenberg equations for the fields $Q^I$ and $P^I$, which are given by
\begin{subequations}
	\begin{align}
		\label{eq:HeisenbergEqQ^I}
		D_t Q^\FouInd{I} &= -\im [Q^\FouInd{I}, H], \\
		\label{eq:HeisenbergEqP^I}
		D_t P^\FouInd{I} &= -\im [P^\FouInd{I}, H] - 3H P^\FouInd{I},
	\end{align}
\end{subequations}
where Eq.~\eqref{eq:HeisenbergEqP^I} is slightly different from the typical canonical relation because of the rescaled momentum. If the Hamiltonian in Eq.~\eqref{eq:HamiltonianSolution} is inserted into equations \eqref{eq:HeisenbergEqQ^I} and \eqref{eq:HeisenbergEqP^I}, then we find
\begin{equation} \label{eq:D_tQ^I Solution}
	D_t Q^\FouInd{I} = \delta^{\FouInd{\FlipInd{I}}}_\FouInd{J} P^\FouInd{J} - \frac{1}{2} B\indices{_{\FouInd{JK}}^{\FouInd{\FlipInd{I}}}} Q^\FouInd{J} Q^\FouInd{K} - C\indices{^{\FouInd{\FlipInd{I}}}_{\FouInd{JK}}} P^\FouInd{J} Q^\FouInd{K}, 
\end{equation}
and
\begin{equation} \label{eq:D_tP^I Solution}
	D_t P^\FouInd{I} = -3H\delta_\FouInd{J}^{\FouInd{\FlipInd{I}}} P^\FouInd{J} + M\indices{ ^{\FouInd{\FlipInd{I}}} _{\FouInd{J}} } Q^\FouInd{J} + \frac{3}{2} A\indices{^{\FouInd{\FlipInd{I}}}_{\FouInd{JK}}} Q^\FouInd{J} Q^\FouInd{K} + B\indices{^{\FouInd{\FlipInd{I}}}_{\FouInd{JK}}} Q^\FouInd{J} P^\FouInd{K} + \frac{1}{2} C\indices{_{\FouInd{JK}}^{\FouInd{\FlipInd{I}}}} P^\FouInd{J} P^\FouInd{K}.
\end{equation}
By comparing the linear terms in equations \eqref{eq:D_tQ^I Solution} and \eqref{eq:D_tP^I Solution} with Eq.~\eqref{eq:HeisenbergEOMs}, we first find the $u\indices{^{\FouInd{a}}_{\FouInd{b}}}$ tensor to be
\begin{equation} \label{eq:u_ab tensor}
u\indices{^{\FouInd{a}}_{\FouInd{b}}} = 
 \begin{pmatrix}
  0 & \delta^{\FouInd{\FlipInd{I}}}_\FouInd{J} \\
  M\indices{^{\FouInd{\FlipInd{I}}}_{\FouInd{J}}} & -3H\delta_\FouInd{J}^{\FouInd{\FlipInd{I}}}
 \end{pmatrix},
\end{equation}
where we identify each row with terms coming from the evolution equation for $Q$ and $P$ respectively and each column as having terms proportional to $Q$ and $P$ respectively. Similarly, we find the $u\indices{^{\FouInd{a}}_{\FouInd{bc}}}$ tensor to be
\begin{equation} \label{eq:u_abc tensor}
	u\indices{^{\FouInd{a}}_{\FouInd{bc}}} =
	\begin{Bmatrix}
		\begin{pmatrix}
			- B\indices{_{\FouInd{JK}}^{\FouInd{\FlipInd{I}}}} & - C\indices{^{\FouInd{\FlipInd{I}}}_{\FouInd{JK}}} \\
			3 A\indices{^{\FouInd{\FlipInd{I}}}_{\FouInd{JK}}} & B\indices{^{\FouInd{\FlipInd{I}}}_{\FouInd{KJ}}}
		\end{pmatrix} \\
		\\
		\begin{pmatrix}
			-C\indices{^{\FouInd{\FlipInd{I}}}_{\FouInd{KJ}}} & 0 \\
			B\indices{^{\FouInd{\FlipInd{I}}}_{\FouInd{JK}}} & C\indices{_{\FouInd{KJ}}^{\FouInd{\FlipInd{I}}}}
		\end{pmatrix}
	\end{Bmatrix},
\end{equation}
where the rules are the same as before for each 2-by-2 matrix and the extra index $\FouInd{c}$ identifies which 2-by-2 matrix is being referred to. There are also further simplifications to be made regarding the primed indices in \eqref{eq:u_ab tensor} and \eqref{eq:u_abc tensor}. For both of the equations above, it is the index $\FouInd{I}$ that has a prime which corresponds with a sign reversal of all momenta in equations \eqref{eq:G_IJ} and \eqref{eq:ATensor}--\eqref{eq:CTensor}. However because the $k$ terms in these equations always appear as an inner product of pairs of momenta, then all of the sign reversal will be cancelled out. This means our $u$-tensors can be written with plain phase-space indices,
\begin{subequations}
\begin{align}
	\label{eq:u1_tensor_app}
	u\indices{^a_b} =&
 	\begin{pmatrix}
  		0 & \delta^{I}_J \\
  		M\indices{^I_J} & -3H\delta_J^I
 	\end{pmatrix}, \\
 	\label{eq:u2_tensor_app}
 	u\indices{^a_{bc}} =&
	\begin{Bmatrix}
		\begin{pmatrix}[1.1]
			- B\indices{_{JK}^I} & - C\indices{^I_{JK}} \\
			3 A\indices{^I_{JK}} & B\indices{^I_{KJ}}
		\end{pmatrix} \\
		\\
		\begin{pmatrix}
			-C\indices{^I_{KJ}} & 0 \\
			B\indices{^I_{JK}} & C\indices{_{KJ}^I}
		\end{pmatrix}
	\end{Bmatrix}.
\end{align}
\end{subequations}
It should be noted that all of the extra terms added by the non-canonical field metric here are contained within the kernels introduced earlier and the only other differences are caused by Christoffel symbols in field coordinate space coming from covariant derivatives.

\subsection{Initial conditions}
\label{subsec:InitConds}

Having found the differential equations needed to be solved for a numerical implementation, the next task is to use the formalism developed in section \ref{subsec:TransportMethod} to find appropriate initial conditions for the equations giving both the 2- and 3-point correlation functions. We will again need to be careful to ensure that our expressions are kept field-covariant and to find any new field curvature times arising from the inclusion of a non-canonical field metric.

\subsubsection{2-point correlation functions}
\label{subsubsec:2pntCorrFns}

We begin by writing the second-order action in terms of our perturbed fields,
\begin{equation} \label{eq:S_2 pert fields}
	S_{(2)} = \frac{1}{2} \int \d t \ a^3 \Big\{ - G_{IJ} \partial_\mu Q^I \partial^\mu Q^J - \mathcal{M}_{IJ} Q^I Q^J \Big\},
\end{equation}
where $\mathcal{M}_{IJ}$ is a mass-term encompassing all terms involving potentials and other non-kinetic terms. This calculation is done using the path-integral formalism so we integrate by parts whilst assuming boundary terms vanish at infinity and change the time variable to conformal time, defined by $\d t = a \, \d\eta$. We find
\begin{align}
	\begin{split} \label{eq:S_2 conformal time}
		S_{(2)} &= -\frac{1}{2} \int \d\eta \, \d^3 x \ a^2 Q^I \left[ G_{IJ} \left( \mathcal{D}_\eta^2 + 2 \frac{a'}{a} \mathcal{D}_\eta - \partial_i \partial_i \right) + a^2 \mathcal{M}_{IJ} \right] Q^J \\
		&= -\frac{1}{2} \int \d\eta \, \d^3 x \left\{ a^2 Q^I \Delta_{IJ} Q^J \right\},
	\end{split}
\end{align}
where we have written a covariant derivative over conformal time as $\mathcal{D}_\eta$ and use a prime ($'$) to indicate a derivative $\d / \d\eta$ and defined the quantity $\Delta_{IJ}$ as the differential operator in brackets $(\cdots)$ above. We now seek to use Eq.~\eqref{eq:VEVsolution} to find the two-point correlation function but we must distinguish between fields on the left anti-time-ordered product and the right time-ordered product which we do using a $Q_+$ and $Q_-$ field respectively. Therefore, there are four separate two-point functions for the correlations between `$++$', `$+-$', `$-+$' and `$--$' fields which need to be calculated with the `in--in' formalism.

It can be shown \cite{Weinberg:2005vy} that Eq.~\eqref{eq:VEVsolution} is written in the path integral formalism with the action above as
\begin{equation} \label{eq:VEV PI formalism}
	Z = \int[DQ_+^I DQ_-^I ] \exp \left\{ -\frac{\im}{2} \int_{\tau_0}^\tau \d\eta \d^3 x \ a^2 \bar{Q}^I
	\begin{pmatrix}
		\Delta & \  \\
		\ & -\Delta 
	\end{pmatrix}_{IJ} Q^J \right\},
\end{equation}
where $Q^I = (Q_+^I, Q_-^I)$ and $\bar{Q}^I$ denotes the transpose matrix with $\tau_0$ being a time well before horizon-crossing and $\tau$ being the time we're seeking initial conditions for. We define the two point function with a time-ordered product of fields to be
\begin{equation} \label{eq:Time-orderedFieldCorr}
	D_{++}^{JK'}(\eta, \vect{x} ; \sigma, \vect{y}) = \braket{T Q_+^J(\eta, \vect{x}) Q_+^{K'}(\sigma, \vect{y})},
\end{equation}
with similar definitions for the other products of fields and unprimed indices label tangent spaces at $\eta$ and primed ones label tangent spaces at $\sigma$. Using the rules of Gaussian integration for a matrix with vectors that are transpose to one another and by making a Fourier transform on $D_{\pm \pm}$ to diagonalise the dependence on $\vect{x}$ and $\vect{y}$, we can calculate $D_{++}$ using the following differential equation
\begin{equation} \label{eq:ODEforD_++}
	G_{IJ} \left( \mathcal{D}_\eta^2 + 2\mathcal{H} \mathcal{D}_\eta + k^2 \right) D_{++}^{JK'} (\vect{k}) = - \frac{\im}{a^2} G_{I}^{K'} \delta ( \eta - \sigma),
\end{equation}
where we have set $\mathcal{H} \equiv a'/a$ as the conformal Hubble constant and we're now ignoring the $M_{IJ}$ term but \textit{only} for the initial conditions in the early, sub-horizon times where they will make a small contribution. We would now like to factorise the tensor structure so we introduce a bi-tensor $\Pi^{JK'}$ which must solve $\mathcal{D}_\eta \Pi^{JK'} = 0$ and a bi-scalar $\Delta_{\pm \pm} (\eta, \sigma, \vect{k})$ that contains all of the dimensionful quantities. This means we can now write the 2-point function as
\begin{equation} \label{eq:D_++withPiRequirement}
	D_{++}^{JK'} (\eta, \sigma, \vect{k}) = \Pi^{JK'} \Delta_{++} (\eta, \sigma, \vect{k}) \quad \textrm{with} \quad \frac{D}{\d\eta} \Pi^{JK'} = 0.
\end{equation}
We are now able to make this substitution into Eq.~\eqref{eq:ODEforD_++} where the bi-tensor can now be factorised out,
\begin{equation} \label{eq:FactorisedODE_D++}
	G_{IJ} \Pi^{JK'} \left( \Delta_{++}'' + 2\mathcal{H} \Delta_{++}' + k^2 \Delta_{++} \right) = - \frac{\im}{a^2} G_{I}^{K'} \delta ( \eta - \sigma).
\end{equation}
The evolution equation for $\Pi^{JK'}$ can be solved using
\begin{align}
	\begin{split} \label{eq:TrajectoryPropagator}
		\mathcal{D}_\eta \Pi^{JK'} =& \frac{d \Pi^{JK'}}{\d\eta} + \Gamma^J_{LM} \frac{d\phi^L}{\d\eta} \Pi^{MK'} = 0 \\
		\implies &\Pi^{JK'} = \hat{P} \exp \left( - \int_\sigma^\eta \d\tau \ \Gamma^{J''}_{L''M''} \frac{d\phi^{L''}}{\d\tau} \right) G^{M'K'},
	\end{split}
\end{align}
where $\hat{P}$ indicates the exponential is path-ordered and double primed indices label tangent spaces evaluated at $\tau$. This bi-tensor is known as the `trajectory propagator' which is the parallel propagator evaluated along the inflationary trajectory with the boundary condition chosen so that when $\sigma \rightarrow \eta$, we have $\Pi^{JK'} \rightarrow G^{JK'}$. This means that field metric dependence is removed from Eq.~\eqref{eq:FactorisedODE_D++} and $\Delta_{++}$ satisfies
\begin{equation} \label{eq:Delta++_ODE}
	\left( \mathcal{D}_\eta^2 + 2\mathcal{H} \mathcal{D}_\eta + k^2 \right) \Delta_{++} = - \frac{\im}{a^2} \delta ( \eta - \sigma).
\end{equation}
This equation is identical to the canonical field-space solution and we see that the complexity introduced by the field-space metric is captured by the trajectory propagator and the use of the in--in formalism. Now we only need to identify each of the different field combinations mentioned earlier. From the boundary conditions in Eq.~\eqref{eq:VEV PI formalism} it can be seen \cite{Weinberg:2005vy,Elliston:2012ab} that `$++$' and `$--$' as well as `$+-$' and `$-+$' field combinations are Hermitian conjugates of one another. This yields the following solutions for the 2-point correlation function,
\begin{subequations}
	\begin{align}
		\label{eq:D++_sol}
		D_{++}^{IJ'} &= (2\pi)^3 \delta(\vect{k}_1 + \vect{k}_2) \Pi^{IJ'} \frac{H_*^2}{2k^3} (1 + \im k\eta)(1 - \im k\sigma) e^{\im k(\sigma - \eta)}, \\
		\label{eq:D-+_sol}
		D_{-+}^{IJ'} &= (2\pi)^3 \delta(\vect{k}_1 + \vect{k}_2) \Pi^{IJ'} \frac{H_*^2}{2k^3} (1 + \im k\eta)(1 - \im k\sigma) e^{\im k(\sigma - \eta)}
	\end{align}
\end{subequations}
with $D_{--}^{IJ'}$ and $D_{+-}^{IJ'}$ being given by the complex conjugates of equations \eqref{eq:D++_sol} and \eqref{eq:D-+_sol} respectively with $H_*$ denoting the Hubble parameter taken at horizon crossing. At equal-time with $\sigma = \eta$, these all give the same solution so that the field-field initial condition is
\begin{equation} \label{eq:QQinit_conds}
	\braket{Q^I(\vect{k}_1) Q^{J}(\vect{k}_2)}_{\mathrm{init}} = (2\pi)^3 \delta(\vect{k}_1 + \vect{k}_2) G^{IJ} \left( \frac{1}{2ka^2} + \frac{H^2}{2k^3} \right) ,
\end{equation}
where we have used $\eta = -1/aH$ to remove time dependence. Similarly for field-momentum, momentum-field and momentum-momentum correlations, we have
\begin{subequations}
	\begin{align}
		\label{eq:QPinit_conds}
		\braket{Q^I(\vect{k}_1) P^{J}(\vect{k}_2)}_{\mathrm{init}} &= (2\pi)^3 \delta(\vect{k}_1 + \vect{k}_2) G^{IJ} \left( - \frac{H}{2 k a^2} + \frac{\im}{2 a^3} \right), \\
		\label{eq:PQinit_conds}
		\braket{P^I(\vect{k}_1) Q^{J}(\vect{k}_2)}_{\mathrm{init}} &= (2\pi)^3 \delta(\vect{k}_1 + \vect{k}_2) G^{IJ} \left( - \frac{H}{2 k a^2} - \frac{\im}{2 a^3} \right), \\
		\label{eq:PPinit_conds}
		\braket{P^I(\vect{k}_1) P^{J}(\vect{k}_2)}_{\mathrm{init}} &= (2\pi)^3 \delta(\vect{k}_1 + \vect{k}_2) G^{IJ} \left( \frac{k}{2 a^4} \right).
	\end{align}
\end{subequations}
In summary, the introduction of the trajectory propagator, $\Pi^{IJ'}$, has ensured we are tracking all of the fields correctly on sub-horizon scales before becoming $G^{IJ}$ on equal time correlations.

\subsubsection{3-point correlation functions}
\label{subsubsec:3pntCorrFns}

For calculation of the 3-point correlation function initial conditions, it is more convenient to use the operator formalism as used in Eq.~\eqref{eq:VEVsolution}. Each of the exponential functions are expanded using the in--in formalism and the leading-order, non-vanishing terms are given by
\begin{equation} \label{eq:3pntCorrFn_Leading-Order}
	\braket{X^I X^J X^K} \subseteq \bigg\langle 0 \bigg| i \int_{-\infty}^\eta \d\tau \left[ H_{\mathrm{int}} , X^I (\eta, \vect{k}_1) X^J(\eta, \vect{k}_2) X^K(\eta, \vect{k}_3) \right] \bigg| 0 \bigg\rangle,
\end{equation}
where $H_{\mathrm{int}} \equiv H_{\FouInd{LMN}} X^\FouInd{L} X^\FouInd{M} X^\FouInd{N}$ which comes from the cubic terms in the action in Eq.~\eqref{eq:PerturbedAction} along with the kernels defined in equations \eqref{eq:Akernel}--\eqref{eq:CTensor}. If we perform a Fourier transform on the $X$ terms in $H_{\mathrm{int}}$, the first term from the commutator is
\begin{equation} \label{eq:3pntCorrFn_FourierTransform}
	\braket{X^I X^J X^K} \subseteq i \int_{-\infty}^\eta \d\tau \ H_{\FouInd{LMN}} \int \frac{\Pi_n \d^3 q_n}{(2\pi)^9} (2\pi)^3 \delta ( \Sigma \vect{q}_i ) \left\langle 0 \left| X^\FouInd{L}_{q_1} X^\FouInd{M}_{q_2} X^\FouInd{N}_{q_3} X_{k_1}^I X_{k_2}^J X_{k_3}^K \right| 0 \right\rangle,
\end{equation}
where we have compacted our notation for each $Q$'s dependence on wave number by placing it as a subscript. Now we can Wick-contract between different fields to rewrite this in terms of 2-point functions as
\begin{equation} \label{eq:3pntCorrFn_WickRotate}
	\braket{X^I X^J X^K} \subseteq i \int_{-\infty}^\eta \d\tau \ H_{\FouInd{LMN}} \int \frac{\Pi_n \d^3 q_n}{(2\pi)^9} (2\pi)^3 \delta ( \Sigma \vect{q}_i ) \left\{ \braket{X^{\FouInd{L}}_{q_1} X_{k_1}^I} \braket{X^\FouInd{M}_{q_2} X_{k_2}^J} \braket{X^\FouInd{N}_{q_3} X_{k_3}^K} + \textrm{cyclic} \right\},
\end{equation}
where `cyclic' indicates there are extra terms omitted that are permutations of the field labels on the inner products. Now we can use this to find $\braket{Q^I Q^J Q^K}$ whilst just using the $A_{LMN}$ term from $H_{\mathrm{int}}$ as an example to evaluate the Fourier integral using Eq.~\eqref{eq:QQinit_conds} as
\begin{align}
	\begin{split} \label{eq:Q_IQ_JQ_K-unsimplified}
		\braket{Q^I Q^J Q^K} \subseteq i (2\pi)^3 \delta ( \Sigma \vect{k}_i ) \frac{H_*^6}{8(\prod_i k_i^3)} \Pi^{IL} \Pi^{JM} \Pi^{KN} \left( 1 + \im k_1 \eta \right) \left( 1 + \im k_2 \eta \right) \left( 1 + \im k_3 \eta \right) e^{-k_t \eta} \\
		\times \int_{-\infty}^\eta \d\tau \left\{ \left( 1 - \im k_1 \tau \right) \left( 1 - \im k_2 \tau \right) \left( 1 - \im k_3 \tau \right) e^{i k_t \tau} A_{LMN}(\tau) \right\},
	\end{split}
\end{align}
where we have defined $k_t = k_1 + k_2 + k_3$ and $\prod_i k_i^3$ indicates a product of $k^3$ terms. We would like to remove the $A_{LMN}$ term from the $\tau$ integral which can be done using a Taylor series at time $N_*$ and using some more trajectory propagators between times $\eta$ and $N_*$ as
\begin{equation} \label{eq:KernelExpansion}
	A_{LMN} \approx \Pi_L^i \Pi_M^j \Pi_N^k \left\{ A_{ijk}|_* + (N - N_*) \frac{\d}{\d N} A^{ijk} \bigg|_* + \cdots \right\},
\end{equation}
where we use lower-case indices here to indicate that we're in the $N_*$ tangent space. Indices between trajectory propagators contract in the normal way (ie. $\Pi^i_L \Pi^{IL} = \Pi^{Ii}$) so that when we insert this approximation into Eq.~\eqref{eq:Q_IQ_JQ_K-unsimplified} whilst keeping the lowest order terms and multiplying by 2 for the complex conjugate of a \textit{real} observable, we find that the 3-point function is given by
\begin{align}
	\begin{split} \label{eq:Q_IQ_JQ_K solution}
		\braket{Q^I Q^J Q^K} &\subseteq \Pi^{Ii} \Pi^{Jj} \Pi^{Kk} \cdot \im (2\pi)^3 \delta ( \Sigma \vect{k}_i ) \frac{H_*^6 A^*_{ijk}}{4(\prod_i k_i^3)} \\
		&\times \left( 1 + \im k_1 \eta \right) \left( 1 + \im k_2 \eta \right) \left( 1 + \im k_3 \eta \right) e^{-k_t \eta} \\
		&\times \int_{-\infty}^\eta \d\tau \left\{ \left( 1 - \im k_1 \tau \right) \left( 1 - \im k_2 \tau \right) \left( 1 - \im k_3 \tau \right) e^{i k_t \tau} \right\},
	\end{split}
\end{align}
where we have placed `constant' terms on the first line, the second line is the `external polynomial' and the third line is the `internal polynomial'. While all the `constants' on the first line are not constant, the same terms do appear for every 3-point correlation possible. The external polynomial is determined by the particular interaction chosen on the left hand side of Eq.~\eqref{eq:3pntCorrFn_WickRotate} so could be $Q^I Q^J Q^K$, $P^I Q^J Q^K$, $P^I P^J Q^K$ or $P^I P^J P^K$ where $P^I \equiv dQ^I/dt$ which are easy to handle because they are effective constants in the calculation. The internal polynomials however come from the particular cubic Hamiltonian term chosen in Eq.~\eqref{eq:HamiltonianSolution} and must be carefully integrated to keep leading-order, real and imaginary terms to ensure the result of Eq.~\eqref{eq:Q_IQ_JQ_K solution} is \textit{real} with its factor of $\im$.

\para{External polynomials}
\label{para:ExternalPolynomials}
There are 4 different types of external polynomials from each of the possible 3-point interactions that need to be computed. From Eq.~\eqref{eq:Q_IQ_JQ_K solution} above, we can see that $Q^I$ contributes the following polynomial,
\begin{equation} \label{eq:Q^I_approxform}
	Q^I(\eta) \approx \left( 1+\im k\eta \right) e^{-\im k\eta}.
\end{equation}
It is then simple to take a derivative with respect to $\eta$ to find
\begin{equation} \label{eq:P^I_approxform}
	P^I(\eta) \approx -\im k \left( 1 + \im k\eta \right) e^{-\im k\eta} + \im k e^{-\im k\eta} = k^2 \eta e^{-\im k\eta}.
\end{equation}
Using these relations, we can find each of the possible polynomials as
\begin{subequations}
	\begin{align}
		\label{eq:ExtPolynom-QQQ} \braket{Q^I Q^J Q^K} &= \left( 1 + \im k_t \eta - K^2 \eta^2 - \im k_1 k_2 k_3 \eta^3 \right) e^{-\im k_t \eta}, \\
		\label{eq:ExtPolynom-PQQ} \braket{P^I Q^J Q^K} &= \left( k_1^2 \eta + \im k_1^2 (k_2 + k_3) \eta^2 - k_1^2 k_2 k_3 \eta^3 \right) e^{-\im k_t \eta}, \\
		\label{eq:ExtPolynom-PPQ} \braket{P^I P^J Q^K} &= \left( k_1^2 k_2^2 \eta^2 + \im k_1^2 k_2^2 k_3 \eta^3 \right) e^{-\im k_t \eta}, \\
		\label{eq:ExtPolynom-PPP} \braket{P^I P^J P^K} &= \left( k_1^2 k_2^2 k_3^2 \right) e^{-\im k_t \eta},
	\end{align}
\end{subequations}
where we have defined $K^2 \equiv k_1 k_2 + k_1 k_3 + k_2 k_3$ in \eqref{eq:ExtPolynom-QQQ}.

\para{Internal polynomials}
\label{para:InternalPolynomials}
From Eq.~\eqref{eq:ActionWithSlow&FastAtensors} above, we have 4 different vertex integrals to perform where we keep the highest-order terms in $\eta$ to ensure we have the correct initial conditions on sub-horizon scales. From Eq.~\eqref{eq:Q_IQ_JQ_K solution}, we see that $Q(\tau)$ is given by
\begin{equation} \label{eq:Q(tau)}
	Q(\tau) = (1 - \im k \tau) e^{\im k \tau},
\end{equation}
which we can differentiate to obtain $P(\tau)$ as:

\begin{equation} \label{eq:P(tau)}
	P(\tau) = \frac{dQ(\tau)}{d t} = \frac{k^2 \tau}{a} e^{\im k \tau}.
\end{equation}
As mentioned at the end of section \ref{subsubsec:ConstaintFourierAction}, the $A_{IJK}$ kernel contains a `fast' term that grows exponentially on sub-horizon scales whereas all of the other terms are `slow' and do not grow quickly. In order to numerically model inflationary paradigms that exhibit one or both of these behaviours, we split up the third-order action as follows
\begin{equation} \label{eq:ActionWithSlow&FastAtensors}
	S_{\phi}^{(3)} = \int \d\tau \frac{a^4}{2} \left\{ \left( \frac{\dot{\phi}^I G^{JK}}{2H} \frac{\vect{k}_2 \cdot \vect{k}_3}{a^2} + \Aslow^{IJK} \right) Q_I Q_J Q_K + \frac{1}{a} B_{IJK} Q^I Q^J P^K + \frac{1}{a^2} C_{IJK} P^I P^J Q^K \right\},
\end{equation} 
where $\Aslow^{IJK}$ denotes the `slow' term which is $A^{IJK}$ with the first term above removed. We can then insert equations \eqref{eq:Q(tau)} and \eqref{eq:P(tau)} into Eq.~\eqref{eq:ActionWithSlow&FastAtensors} and use $\tau = -1/aH$ to obtain the internal polynomials,
\begin{subequations}
	\begin{align}
			\label{eq:IntPolynom-Afast} \Afast^{IJK} &= \frac{\dot{\phi}^I G^{JK} (\vect{k}_2 \cdot \vect{k}_3)}{4H^3} \left\{ \frac{k_1 k_2 k_3}{k_t} \eta + \frac{\im}{k_t} \left( K^2 + \frac{k_1 k_2 k_3}{k_t} \right) + \mathcal{O}(\eta^{-1}) \right\} e^{\im k_t \eta} + \text{perms.}, \\
			\label{eq:IntPolynom-A} \Aslow^{IJK} &= \frac{\Aslow^{IJK}}{2H^3} \left\{ \frac{k_1 k_2 k_3}{k_t} \frac{1}{\eta} + \frac{\im}{k_t \eta^2} \left( K^2 - \frac{k_1 k_2 k_3}{k_t} \right) +\mathcal{O}(\eta^{-3}) \right\} e^{\im k_t \eta} + \mathrm{perms.}, \\
			\label{eq:IntPolynom-B} B^{IJK} &= -\frac{B^{IJK}}{2H^3} \left\{ \im \frac{k_1 k_2 k_3^2}{k_t} - \frac{(k_1 + k_2) k_3^2}{k_t} \frac{1}{\eta} + \mathcal{O}(\eta^{-2}) \right\} e^{\im k_t \eta} + \text{perms.}, \\
			\label{eq:IntPolynom-C} C^{IJK} &= \frac{C^{IJK}}{2H^2} \left\{ -\frac{k_1^2 k_2^2 k_3}{k_t} - \im \frac{k_1^2 k_2^2}{k_t} \left( 1 + \frac{k_3}{k_t} \right) \right\} e^{\im k_t \eta} + \text{perms.}
	\end{align}
\end{subequations}

\para{3-point initial conditions}
\label{para:3pntinitconds}
Now we use the external polynomials in equations \eqref{eq:ExtPolynom-QQQ}--\eqref{eq:ExtPolynom-PPP} with the internal polynomials in equations \eqref{eq:IntPolynom-Afast}--\eqref{eq:IntPolynom-C} with the `constant' terms found in Eq.~\eqref{eq:Q_IQ_JQ_K solution} to obtain the initial conditions for a correlation of 3 fields,
\begin{align}
	\begin{split} \label{eq:QQQinitconds}
		\braket{Q^I Q^J Q^K}_{\mathrm{init}}
		=
		\frac{(2\pi)^3 \delta (\vectktot)}{4a^4 k_1 k_2 k_3 k_t}
		\bigg\{
		  &
		  \frac{\dot{\phi}^I G^{JK}}{4H \MPlanck^2} \vect{k}_2 \cdot \vect{k}_3
		  + \frac{a^2}{2} \Aslow^{IJK}
		  - C^{IJK} \frac{k_1 k_2}{2}
		  \\
    	  & \mbox{}
		  + \frac{a^2 H}{2} B^{IJK}
		      \bigg[
		          \frac{(k_1 + k_2)k_3}{k_1 k_2}
		          - \frac{K^2}{k_1 k_2}
		      \bigg]
		  + \text{5 perms}
		\bigg\}
		,		
	\end{split}
\end{align}
with a correlation of 1 momentum and 2 fields,
\begin{align}
	\begin{split} \label{eq:PQQinitconds}
		\braket{P^I Q^J Q^K}_{\mathrm{init}}
		& =
		\frac{(2\pi)^3 \delta (\vectktot)}{4 a^4 (k_1 k_2 k_3)^2 k_t}
		\\
		& \hspace{-1.7cm}\times
		\bigg\{
		  k_1^2 (k_2 + k_3)
		  \bigg[
		      \frac{\dot{\phi}^I G^{JK}}{4H \MPlanck^2} \vect{k}_2 \cdot \vect{k}_3
		      + \frac{a^2}{2} \Aslow^{IJK}
		      - C^{IJK} \frac{k_1 k_2}{2}
		      + \text{5 perms}
		  \bigg]
		\\
		& \hspace{-1cm} +
		  k_1
		  \bigg[
		      {-\frac{\dot{\phi}^I G^{JK}}{4H \MPlanck^2}} \vect{k}_2 \cdot \vect{k}_3 
		      \Big(
		          K^2 + \frac{k_1 k_2 k_3}{k_t}
		      \Big)
		      - \frac{a^2}{2} \Aslow^{IJK}
		      \Big(
		          K^2 - \frac{k_1 k_2 k_3}{k_t}
		      \Big)
		\\
		& +
		      B^{IJK} \frac{k_1 k_2 k_3^2}{2H}
		      + C^{IJK} \frac{k_1^2 k_2^2}{2}
		      \Big(
		          1 + \frac{k_3}{k_t}
		      \Big)
		      + \text{5 perms}
		  \bigg]
        \bigg\}
        ,
	\end{split}
\end{align}
with a correlation of 2 momenta and a field,
\begin{align}
	\begin{split} \label{eq:PPQinitconds}
		\braket{P^I P^J Q^K}_{\mathrm{init}} 
		& =
		\frac{(2\pi)^3 \delta (\vectktot)}{4 a^6 H^2 (k_1 k_2 k_3)^2 k_t}
		\\
		& \hspace{-1.7cm}\times
		\bigg\{
		  k_1^2 k_2^2 k_3 
		  \bigg[
		      {-\frac{\dot{\phi}^I G^{JK}}{4H \MPlanck^2}} \vect{k}_2 \cdot \vect{k}_3
		      - \frac{a^2}{2} \Aslow^{IJK}
		      + C^{IJK} \frac{k_1 k_2}{2}
		      - \frac{a^2 H}{2} B^{IJK} \frac{(k_1 + k_2) k_3}{k_1 k_2}
		\\
		      & \hspace{3mm} + \text{5 perms}
		  \bigg]
		  +
		  k_1^2 k_2^2
		  \bigg[
		      \frac{a^2 H}{2} B^{IJK} k_3
		      + \text{5 perms}
		  \bigg]
		\bigg\}
		,
	\end{split}
\end{align}
and a correlation of 3 momenta,
\begin{align}
	\begin{split} \label{eq:PPPinitconds}
		\braket{P^I P^J P^K}_{\mathrm{init}}
		=
		\frac{(2\pi)^3 \delta (\vectktot)}{4 a^6 H^2 k_1 k_2 k_3 k_t}
		\bigg\{
		  & 
		  \frac{\dot{\phi}^I G^{JK}}{4H \MPlanck^2} \vect{k}_2 \cdot \vect{k}_3
		  \Big(
		      K^2 + \frac{k_1 k_2 k_3}{k_t}
		  \Big)
		  + \frac{a^2}{2} \Aslow^{IJK}
		  \Big(
		      K^2 - \frac{k_1 k_2 k_3}{k_t}
		  \Big) \\
		  & \mbox{}
		  - B^{IJK} \frac{k_1 k_2 k_3^2}{2H}
		  - C^{IJK} \frac{k_1^2 k_2^2}{2}
		  \Big(
		      1 + \frac{k_3}{k_t}
		  \Big)
		  + \text{5 perms}
		\bigg\}
		.
	\end{split}
\end{align}
where `perms.' indicates there are terms omitted which are cyclic permutations of the indices but \emph{only within the surrounding brackets} of where the permutation instruction is given.

\subsection{Gauge transformation to curvature perturbations}
\label{subsec:GaugeTransCurvs}

The final calculation needed before finding numerical results for inflationary models that use a non-trivial metric is a gauge transformation that translates our correlations functions in phase-space (eg. $\braket{Q^I Q^J P^K}$) into correlation functions of the curvature perturbation, $\zeta$. We follow much of the same treatment as in \cite{Dias:2014msa} and use some of their results that still apply with a non-trivial field metric in order to find the gauge transformations used in our code. 

\subsubsection{Calculating $\zeta$}
\label{subsubsec:CalculatingZeta}

We would like to switch from the spatially-flat gauge used in our calculations so far to the uniform density gauge mainly because $\zeta$ is a quantity that is conserved to all orders in perturbation theory \cite{Lyth:2004gb,Malik:2003mv} and can then be used to calculate the power spectrum and bispectrum for an inflation model. As in \cite{Dias:2014msa}, we use an exponential mapping of the Lie derivative that is used to change gauges,
\begin{equation} \label{eq:LieDerivExpMap}
	x^\mu(p) \rightarrow x^\mu(p') = \exp \left( \mathcal{L_{\vect{\xi}}} \right) x^\mu (p) .
\end{equation}
The Lie derivative is performed along a vector, $\vect{\xi}$, which is given by
\begin{equation} \label{eq:LieDerivVector}
	\mathcal{L}_{\vect{\xi}} \implies \vect{\xi} = \xi^0 \frac{\partial}{\partial f} + \xi^i \frac{\partial}{\partial x^i},
\end{equation}
where $f$ here is a label for a time on the flat hypersurface. This exponential mapping is then used with a Taylor expansion on fields and their derivatives to find equations that translate fields in one gauge to another. These expressions can then be applied to the $h_{ij} \d x^i \d x^j$ part of the ADM decomposition found in Eq.~\eqref{eq:ADMdecomp} to rewrite it in terms of uniform density quantities. Finally, the ADM expression for the curvature perturbation, $\zeta = \det(h_{ij})/a^6$, is used to find
\begin{equation} \label{eq:ADMexpressionZeta}
	\zeta = H \xi^0 + \frac{H}{4} \frac{\partial(\xi^0)^2}{\partial f} + \frac{\dot{H}}{2} (\xi^0)^2,
\end{equation}
where we have chosen to write the gauge transformation only in terms of $\xi^0$ and we have neglected spatial gradients due to them vanishing on the super-horizon scales we are interested in. We can also use the above expression to find the density perturbation in the uniform-density gauge, $\delta \rho(u)$, by employing the $\delta N$ formula \cite{Lyth:2004gb} to identify $\zeta$ with $\delta \rho(u)$ and substitute $\dot{\rho} \rightarrow \dot{N} = H$ with $\ddot{\rho} \rightarrow \ddot{N} = \dot{H}$ to give
\begin{equation} \label{eq:delta_rho_expansion}
	\delta \rho(u) = \delta \rho + \dot{\rho} \xi^0 + \dot{\delta \rho} \xi^0 + \frac{\dot{\rho}}{2} \xi^0 \dot{\xi}^0 + \frac{\ddot{\rho}}{2} \left( \xi^0 \right)^2,
\end{equation}
where spatial gradients have been dropped. Equations \eqref{eq:LieDerivExpMap}--\eqref{eq:delta_rho_expansion} were first found in \cite{Dias:2014msa} and still apply in the non-trivial field space used in our calculations. Eq.~\eqref{eq:delta_rho_expansion} can be used with $\delta \rho(u) = 0$ to find first- and second-order expressions for $\xi^0$ which are then substituted into Eq.~\eqref{eq:ADMexpressionZeta},
\begin{equation} \label{eq:zeta(rho,delta_rho)}
	\zeta = -H \frac{\delta \rho}{\dot{\rho}} + H \frac{\delta \rho}{\dot{\rho}} \frac{\delta \dot{\rho}}{\dot{\rho}} - \frac{H}{2} \frac{\ddot{\rho}}{\dot{\rho}} \left( \frac{\delta \rho}{\dot{\rho}} \right)^2 + \frac{\dot{H}}{2} \left( \frac{\delta \rho}{\dot{\rho}} \right)^2.
\end{equation}
An expression for $\rho$ is now needed specifically for our matter theory. We may assume that the perfect fluid equations apply, in which
case the stress-energy tensor satisfies
\begin{equation} \label{eq:Stress-EnergyTensor}
	T\indices{^a_b} = \partial^a \phi^I \partial_b \phi_I - \delta^a_b \left( \frac{1}{2} \partial_c \phi^I \partial^c \phi_I + V \right).
\end{equation}
The energy density is then related to the $T\indices{^0_0}$ component where spatial gradients are neglected and the inverse ADM metric is used to find the second order density ,
\begin{equation} \label{eq:EnergyDensityFromT_ab}
	\rho = - T\indices{^0_0} = \frac{1}{2N^2} \dot{\phi}^I \dot{\phi}_I + V,
\end{equation}
where at zeroth order, $\rho = \frac{1}{2} \dot{\phi}^I \dot{\phi}_I + V$, as expected. Eqs.~\eqref{eq:Xepans1}, \eqref{eq:Xepans2} and \eqref{eq:LapseShiftExpans} are then used to perturb Eq.~\eqref{eq:EnergyDensityFromT_ab} to second-order and find the density perturbation, $\delta \rho$,
\begin{align} \label{eq:RhoPertUnsimplified}
	\begin{split}
		\delta \rho &= \dot{\phi}^I D_t Q_I + V_I Q^I + \frac{1}{2} \left( 3\alpha_1^2 - 2\alpha_2 - 2\alpha_1 \right) \dot{\phi}^I \dot{\phi}_I \\
		&+ \frac{1}{2} V_{IJ} Q^I Q^J + \frac{1}{2} D_t Q^I D_t Q_I - 2\alpha_1 \dot{\phi}^I D_t Q_I + \frac{1}{2} R_{IJKL} Q^I \dot{\phi}^J \dot{\phi}^K Q^L.
	\end{split}
\end{align}
The Hamiltonian constraint given in Eq.~\eqref{eq:HamiltonianConstraint} can then be used to reduce this expression to
\begin{equation} \label{eq:RhoPert_HamConstrnt}
	\delta \rho = 3H^2 \left( 3\alpha_1^2 - 2\alpha_2 - 2\alpha_1 \right)
\end{equation}
Then the lapse perturbations given in equations \eqref{eq:Alpha_1} \& \eqref{eq:Alpha_2} can be used to find the density perturbation in terms of fields only,
\begin{align}
	\begin{split} \label{eq:RhoPert_Final}
		\delta \rho = &-3H \dot{\phi}^I Q_I \\
		&+\frac{3}{2} \dot{\phi}^I \dot{\phi}^J Q_I Q_J - 3H \partial^{-2} \left[ \partial_i D_t Q^I \partial_i Q_I + D_t Q^I \partial^2 Q_I \right],
	\end{split}
\end{align}
where the first- and second-order terms are on the first and second lines respectively and the spatial derivatives have been neglected for the large scales we're interested in. Eq.~\eqref{eq:EnergyDensityFromT_ab} can be used to find $\dot{\rho}$ and $\ddot{\rho}$ and those results can be used with Eq.~\eqref{eq:RhoPert_Final} in Eq.~\eqref{eq:zeta(rho,delta_rho)} to find the uniform-density curvature perturbation, $\zeta$,
\begin{equation} \label{eq:zeta_1}
	\zeta_1 = - \frac{\dot{\phi}^I Q_I}{2H\epsilon},
\end{equation}
and
\begin{align} \label{eq:zeta_2}
	\begin{split}
		\zeta_2 = \frac{1}{6H^2 \epsilon} \Bigg\{ &\dot{\phi}_I \dot{\phi}_J \left( -\frac{3}{2} + \frac{9}{2\epsilon} + \frac{3}{4\epsilon^2} \frac{\dot{\phi}^K V_K}{H^3} \right) Q^I Q^J + \\
		&\frac{3}{\epsilon H} \dot{\phi}_I \dot{\phi}_J Q^I D_t Q^J - 3H \partial^{-2} \left( \partial_i D_t Q^I \partial_i Q_I + D_t Q^I \partial^2 Q_I \right) \Bigg\},
	\end{split}
\end{align}
where $\zeta_1$ and $\zeta_2$ are the first- and second-order terms respectively. These results are identical to the canonical case as given in \cite{Dias:2014msa} but it was important to check no curvature terms were introduced for the non-canonical field space here.

\subsubsection{Power spectra and $N$ tensors}
\label{subsubsec:PowSpecNtensors}

We now need to use equations \eqref{eq:zeta_1} \& \eqref{eq:zeta_2} to find the statistics of $\zeta$ in order to find the power spectrum and bispectrum for a multi-field inflation theory. For this we write $\zeta$ in Fourier space,
\begin{equation} \label{eq:zeta-Ntensors}
	\zeta(\vect{k}) = N_{\FouInd{a}} X^\FouInd{a} + \frac{1}{2} N_{\FouInd{{ab}}} X^\FouInd{a} X^\FouInd{b},
\end{equation}
where the $N$ tensors are
\begin{subequations}
	\begin{align}
		\label{eq:N_a def}
		N_\FouInd{a} (\vect{k}) &= (2\pi)^3 \delta(\vect{k} - \vect{k}_a) N_a, \\
		\label{eq:N_ab def}
		N_\FouInd{ab} (\vect{k}) &= (2\pi)^3 \delta(\vect{k} - \vect{k}_a - \vect{k}_b) N_{ab},
	\end{align}
\end{subequations}
and $X^\FouInd{a} = (Q^I, P^J)$. We can now Fourier transform equations \eqref{eq:zeta_1} \& \eqref{eq:zeta_2} to see that the coefficient matrices $N_a$ and $N_{ab}$ are
\begin{subequations}
\begin{align}
	\label{eq:N_a solution}
	N_a &= - \frac{\dot{\phi}_I}{2H \epsilon}
	\begin{pmatrix}
		1 \\
      	0 \\
    \end{pmatrix}, \\
    \label{eq:N_ab solution}
    N_{ab} &= \frac{1}{3H^2 \epsilon}
    \begin{pmatrix}
    \dot{\phi}_I \dot{\phi}_J \Big[ -\frac{3}{2} + \frac{9}{2\epsilon} + \frac{3}{4\epsilon^2} \frac{V_\gamma \pi^\gamma}{H^3} \Big] & \frac{3}{H\epsilon} \dot{\phi}_I \dot{\phi}_J - \frac{3H}{k^2} \Big[ \vect{k}_a \cdot \vect{k}_b + k_a^2 \Big] G_{IJ} \\
    \frac{3}{H\epsilon} \dot{\phi}_I \dot{\phi}_J - \frac{3H}{k^2} \Big[ \vect{k}_a \cdot \vect{k}_b + k_b^2 \Big] G_{IJ} & 0
    \end{pmatrix}.
\end{align}
\end{subequations}
The spectrum and bispectrum are given from the two and three point correlations of $\zeta$. They are defined by
\begin{subequations}
	\begin{align}
		\label{eq:PowSpecFrom2pnt}
		\braket{\zeta(\vect{k}_1)\zeta(\vect{k}_2)} &= (2\pi)^3 \delta(\vect{k}_1 + \vect{k}_2) P(k) \\
		\label{eq:BiSpecFrom3pnt}
		\braket{\zeta(\vect{k}_1)\zeta(\vect{k}_2)\zeta(\vect{k}_3)} &= (2\pi)^3 \delta(\vect{k}_1 + \vect{k}_2 + \vect{k}_3) B(k_1, k_2, k_3),
	\end{align}
\end{subequations}
with the power spectrum, $P(k)$, given by
\begin{equation} \label{eq:PowSpecDef}
	P(k) = N_a N_b \braket{X^a(k_a) X^b(k_b)},
\end{equation}
and the bispectrum $B(k_1, k_2, k_3)$ given by
\begin{align}
	\begin{split} \label{eq:BiSpecDef}
		B(k_1, k_2, k_3) =& N_a N_b N_c \braket{X^a(k_a) X^b(k_b) X^c(k_c)} + \\
		\big( &N_a N_b N_{cd} \braket{X^a(k_a) X^c(k_c)} \braket{X^b(k_b) X^d(k_d)} + 2 \ \mathrm{cyclic} \big),
	\end{split}
\end{align}
where `2 cyclic' indicates that there are 2 extra terms that are cyclic permutations of the indices.

\bibliographystyle{JHEP}
\bibliography{bibliography}

\end{document}